\PassOptionsToPackage{linktocpage}{hyperref}
\PassOptionsToPackage{colorlinks=true, linkcolor=red, citecolor=blue, urlcolor=blue}{hyperref}
\documentclass[12pt,onecolumn,floatfix,amsmath,nofootinbib,amssymb,floatfix,aps,prd,showpacs,letterpaper]{revtex4}
\usepackage{slashed}
\usepackage[level]{datetime}
\usepackage{graphicx,color,dcolumn,booktabs,bm}
\usepackage{longtable,lscape}
\usepackage{threeparttable}
\usepackage{booktabs}
\usepackage{txfonts}
\usepackage{subfigure}
\usepackage{float}
\usepackage{makecell}
\usepackage{tabularx}
\usepackage{array}
\usepackage{ragged2e}
\usepackage{tikz}
\usetikzlibrary{trees}
\usetikzlibrary{decorations.pathmorphing}
\usetikzlibrary{decorations.markings}

  \definecolor{jblue}  {RGB}{20,50,100}
  \definecolor{npurple}  {RGB} {153, 51, 204}
  \definecolor{wred}   {RGB}{217,0,56}
  \definecolor{white}   {RGB}{255,255,255}
   \definecolor{korange}   {RGB}{235, 80,  43}
  \definecolor{korange2}   {RGB}{245, 100,  63}
  \definecolor{kyelloworange}   {RGB}{255, 210,  110}
  \definecolor{kyelloworange2}   {RGB}{240, 170,  90}
  \definecolor{kred}   {RGB}{204,  102, 153}
  \definecolor{kpurple}   {RGB}{153,  61, 190}
  \definecolor{kpurplelight}   {RGB}{213,  161, 230}

	\tikzset{
	  photon/.style={decorate, decoration={snake}, draw=npurple,very thick},
	  boson/.style={decorate, decoration={snake}, draw=npurple,very thick},
	  electron/.style={draw=jblue,very thick, postaction={decorate},
	           decoration={markings,mark=at position .55 with {\arrow[draw=jblue]{>}}}
	  },
	  electron2/.style={draw=jblue,very thick, postaction={decorate},
	           decoration={markings,mark=at position .55 with {\arrow[draw=jblue]{<}}}
	  },
	  fermion/.style={draw=jblue,very thick, postaction={decorate},
	            decoration={markings,mark=at position .55 with {\arrow[draw=jblue]{}}}
	  },
	  gluon/.style={decorate, draw=korange,very thick, %kred
	    decoration={coil,amplitude=4pt, segment length=6pt}},
	  higgs/.style={draw=wred,very thick, postaction={decorate},
	           decoration={markings,mark=at position .55 with {\arrow[draw=wred]{>}}}
	  },
	  nothing/.style={draw=white,very thick}
	}
\newcolumntype{P}[1]{>{\centering\hspace{0pt}}p{#1}}
\newcolumntype{Z}{>{\centering\let\newline\\\arraybackslash\hspace{0pt}}x}

\usepackage{booktabs}
\usepackage{overpic}
\usepackage{amssymb}
\usepackage{indentfirst}
\usepackage{feynmf}   %{feynmp}
\usepackage{slashed}  %for Feynman symbols
\usepackage{cases}
\usepackage{color}
\usepackage{multirow}
\usepackage{epstopdf}
\usepackage{longtable}
\usepackage{indentfirst}
\usepackage{graphicx,color,dcolumn,booktabs,bm}
\graphicspath{{Figures/}} %
\allowdisplaybreaks
\usepackage{CJK}
\begin{document}
\title{Transition form factors for $D/B$ to $a_{0}(980)$ from light-cone QCD sum rules}
\author{Di Gao$^{1,2}$}
\email{digao@impcas.ac.cn}
\author{Jiangshan Lan$^{2,4,5}$}
\email{jiangshanlan@impcas.ac.cn}
\author{Duojie Jia$^{3,6}$ \thanks{}}
\email{jiadj@nwnu.edu.cn; Corresponding author}
\author{Xingbo Zhao$^{2,4,5}$}
\email{xbzhao@impcas.ac.cn}
\author{Yanjun Sun$^{1,6}$}
\email{sunyanjun@nwnu.edu.cn}
\affiliation{$^1$Institute of Theoretical Physics, College of Physics and Electronic Engineering, Northwest Normal University, Lanzhou 730070, China \\
$^2$Institute of Modern Physics, Chinese Academy of Sciences, Lanzhou 730000, China  \\
$^3$General Education Center, Qinghai Institute of Technology, Xining 810000, China \\
$^4$School of Nuclear Science and Technology, University of Chinese Academy of Sciences, Beijing 100049, China \\
$^5$CAS Key Laboratory of High Precision Nuclear Spectroscopy, Institute of Modern Physics, Chinese Academy of Sciences,  Lanzhou 730000, China\\
$^6$Lanzhou Center for Theoretical Physics, Lanzhou University, Lanzhou 730070, China  \\
}

\begin{abstract}
We apply the light-cone QCD sum rules with chiral currents to compute transition form factors of the semileptonic decay of the charmed and bottom scalar mesons $D/B$ to $a_{0}(980)l^{+}\nu_{l}$ $(l=e$, $\mu)$, where the scalar meson $a_{0}(980)$ are firstly regarded as two quark states, and contributions from the tetraquark component are then added via proper modulation of four distribution amplitudes considered. The obtained transition form factors and branching fraction are free of the contribution from the light-cone distribution amplitudes at the level of twist-three and two simple relations  connecting their form factors are obtained. Our computations indicates that the decay branch fractions are on the margin of the measurements reported by Beijing Spectrometer \uppercase\expandafter{\romannumeral3} in the pure two-quark scenario and are in good agreement with observations when tetraquark component is considered.
\end{abstract}

\maketitle
\date{\today }

\section{INTRODUCTION}

Exclusive semileptonic decays of mesons are essential for determining the couplings of the weak interaction in the Standard Model, owing to their simpler dynamics relative to nonleptonic decays. Till now, extensive research has been conducted on the semileptonic decays of heavy-light mesons to S-wave mesons, from both experimental \cite{BESIII:2017ylw,BaBar:2014xzf,CLEO:2011ab,CLEO:2009svp,Belle:2006idb} and theoretical \cite{Fu:2018yin,Faustov:2019mqr,Soni:2018adu,Jia:2017age,Li:2012gr,Khodjamirian:2009ys,Ball:2006yd,Fajfer:2005ug,Ball:1993tp,Khodjamirian:2000ds,Huang:2008sn} perspective. However, semileptonic decays involving P-wave mesons in the final states remain relatively unexplored. Over the past two decades, this has become a topics of growing interest with several exclusive decays to P-wave mesons being experimentally reported and measured
\cite{ParticleDataGroup:2014cgo,BaBar:2006wcl,BaBar:2006xju,BaBar:2007dwy,BaBar:2008ozy}.

From a theoretical perspective, the form factors (FFs) of a meson play a crucial role in quantitatively describing its exclusive decays. In the case of semileptonic decay of a charmed or bottom ($D/B$) meson, for instance, the FF can be computed by various methods like the large energy effective theory (LEET) \cite{Palmer:2013yia}, the lattice Quantum Chromodynamics (LQCD) \cite{Bernard:1991bz}, the QCD sum rule (QCDSR) \cite{FermilabLattice:2019ycs}, the heavy quark effective theory (HQET) \cite{Wang:2002zba} and the chiral perturbation theory (chPT) \cite{Wu:2006rd}. The applicability of these methods is divided into different regions based on the value of the momentum transfer squared, $q^{2}$, of the decay process: 
\begin{enumerate}
\item In the region where $q^{2}\rightarrow0$, the LEET \cite{Palmer:2013yia} applies.
\item In the region where $q^{2}$ is relatively large, the LQCD \cite{Bernard:1991bz} and the QCDSR \cite{FermilabLattice:2019ycs} apply.
\item In the region where $q^{2}$ is quite large, the HQET \cite{Wang:2002zba} and the chPT \cite{Wu:2006rd} apply.
 \end{enumerate}
\par
Regarding the decay processes of mesons such as the heavy-light mesons $Q\bar{q}$  shown in FIG.~\ref{fTds}, a notion of transition form factor (TFF) \cite{Dinh:2020inx,DAlise:2022ypp} is found to be necessary to describe exclusive decays quantitatively. The use of light-cone QCD sum rules(LCSR) without chiral currents \cite{Cheng:2017fkw}, commonly applied to correction function, to compute these FFs requires the inclusion of more distribution amplitudes(DAs)----namely, two twist-3 and one twist-2 light-cone DAs. This introduction of multiple DAs necessitates a more involved analysis of the LCSR parameters (Borel parameters and the thresholds), as these patrameters differ for each DA. Such complexity typically hampers accurate predictions of the transition FFs.

On the other hand, the physical properties of the scalar meson $a_{0}(980)$ is closely related to its light-cone DA. This meson is notably intriguing in its internal structure among the well-established scalar mesons. Since 2018, many attempts have been made to production of scalar mesons in decays of the $D$ mesons
\cite{Momeni:2022gqb,Feng:2022inv,BESIII:2021drk,Wang:2022vga,Cheng:2022vbw,DiCarlo:2021dzg,Chakraborty:2021qav,Huang:2021owr,Becirevic:2020rzi,Parrott:2020vbe,Gherardi:2020qhc,Achasov:2020qfx,Soni:2020sgn,Fleischer:2019wlx,Liu:2019tsi}. Regarding the meson $a_{0}(980)$, it is often interpreted as a multiquark state or a $K\bar{K}$ bound state \cite{ParticleDataGroup:2016lqr}. In a sense, the semileptonic $D$ decays provide a pristine environment where the $a_{0}(980)$ is produced by an isovector combination of the up/down quark ($u$ or $d$) and its anti-quark partner \cite{BESIII:2018sjg}. In 2010, a model-independent way was proposed in Ref. \cite{Wang:2009azc} to distinguish the quark components of light scalar mesons: a ratio R of partial decay widths is unity for the quark-antiquark (two-quark) picture of the scalar meson,  while it is $3$ for the four-quark description of the meson. This predictions has been confirmed by Beijing Spectrometer (BES\uppercase\expandafter{\romannumeral3}) in 2018 \cite{BESIII:2018sjg} and in 2021 \cite{BESIII:2021tfk}.

The purpose of this work is to apply the light-cone QCD sum rule to explore the transition FF associated with semileptonic decay of the charmed and bottom scalar mesons $D/B$ to $a_{0}(980)l^{+}\nu_{l}$ $(l=e$, $\mu)$, where the final state meson $a_{0}(980)$ is treated as a two-quark state firstly and further regarded to be tetraquark with $s\bar{s}$ insertion. In the two-quark scenaro of the $a_{0}(980)$, the method we use features itself in that it endows with chiral currents in the correlation functions and enable us to find the FFs free of the contributions from twist-3 light cone DAs of the scalar mesons. Four different light-cone DAs are explored for the scalar meson $a_{0}(980)$, to find its respective form factors, and thereby compute the decay widths of the semileptonic decays of the $D/B$ mesons. These DAs for the $a_{0}(980)$ include: (1) The light-cone DAs obtained by QCDSR at $z^{2}=0$ \cite{Cheng:2005nb}, denoted as DA-I; (2) The light-cone DAs obtained by QCDSR with a novel approximation technique in Ref. \cite{Hong:2024rhg}, denoted as DA-II; (3) The light-cone DAs obtained by basis light-front quantization (BLFQ) method \cite{Vary:2009gt,Xu:2021wwj,Zhao:2014xaa,Wiecki:2014ola}, denoted as DA-III; (4) The new light-cone DAs obtained through moments fitted with the DAs with the help of the BLFQ approach, denoted as DA-IV.

The paper is organized as follows: In Sect. II, we describe the flavor wave functions of the $a_{0}(980)$ and the light-cone DAs of the twist-2 and twist-3. In Sect. III, we discuss the four light-cone DAs obtained with four different methods. Sect. IV is devoted to computation of the FFs and branching ratios of the semileptonic decays on $D\rightarrow Sl^{+}\nu_{l}$ using the LCSR method. In Sect. V, we present numerical analysis of the FFs and branching ratios of the semileptonic decays of the $D/B$ mesons. A comparison is also given among the four predictions and experiments available.  In Sect. VI, we extend our computation to the tetraquark case of final state meson. The paper ends with summary in Sect. VII.

\section{PHYSICAL PROPERTIES OF SCALAR MESONS}

Two scalar mesons $f_{0}(980)$ and $a_{0}(980)$ were discovered in early 1990s \cite{Lockman:1989qa,LucioMartinez:1990uw,MARK-III:1990wgk,DM2:1990cwz,Heusch:1991sw,Tornqvist:1995kr} with their internal structure unsolved fully. There are two main scenarios about the structure of these two scalar mesons \cite{Jaffe:1976ig,Weinstein:1982gc,Weinstein:1983gd,Achasov:1987ts,Weinstein:1990gu,Scadron:2003yg,Jaffe:2004ph,Maiani:2004uc,Kalashnikova:2004ta,Cheng:2005nb,Sugiyama:2007sg,Kojo:2008hk,tHooft:2008rus,Alexandrou:2017itd}. One scenario \cite{Amsler:1995tu,Oller:1998zr,Lee:1999kv,Jamin:2000wn,Giacosa:2006tf,Albaladejo:2008qa,Gallas:2009qp,Zhou:2010ra,Parganlija:2012fy} suggests them to be the examples of novel mesons which differ only in isospin quantum number: the $f_{0}(980)$ is the isoscalar state of $\frac{u\bar{u}+d\bar{d}}{\sqrt{2}}$ while the $a_{0}(980)$ is the isovector state of  $\frac{u\bar{u}-d\bar{d}}{\sqrt{2}}$. The second scenario, regards them to be two members of a scalar tetraquark nonet fully composed of  light quarks \cite{Jaffe:1976ig,Jaffe:2004ph}, being degenerate approximately in mass due to isospin symmetry of the tetraquarks.

Experimentally measured mass of the $a_{0}(980)$ lies close to the opening of the $K\bar{K}$ channel, to which it strongly couples \cite{Abele:1998qd,OBELIX:2002lhi}, so this scalar meson must contains a large $K\bar{K}$ component in its wave function and might be described by a mixture of a $q\bar{q}$ state and a tetraquark, both with $J^{P}=0^{+}$.  For the $q\bar{q}$ component,  its flavor wavefunction is given by 
\begin{equation}
\begin{aligned}
a_{0}^{0}(980)&=\frac{1}{\sqrt{2}}(u\bar{u}-d\bar{d}),\\
 a_{0}^{-}(980)&=d\bar{u},\\
  a_{0}^{+}(980)&=u\bar{d}.\notag
\end{aligned}
\end{equation}
% However, this simple quark-anti quark picture faces two major difficulties:
% \begin{enumerate}
% \item It cannot explain the near mass degeneracy between $f_0(980)$ and $a_0(980)$ mesons.
% \item It struggles to account for the fact that the
% $\sigma$ and $\kappa$ resonances are much broader in width than $f_0(980)$ and $a_0(980)$.
% \end{enumerate}
% \par
% These issues can be naturally resolved in the tetraquark framework, where the light scalar mesons are considered as four-quark states. In this picture, the flavor wave functions take the form \cite{Jaffe:1976ig,Jaffe:2004ph}
For the tetraquark component strongly coupled to the $K\bar{K}$ channel, its flavor wavefucntion takes the form of 
\begin{equation}
\begin{aligned}
a_{0}^{0}(980)&=\frac{1}{\sqrt{2}}(u\bar{u}-d\bar{d})s\bar{s},\\
 a_{0}^{-}(980)&=d\bar{u}s\bar{s},\\
  a_{0}^{+}(980)&=u\bar{d}s\bar{s}.\notag
\end{aligned}
\end{equation}
% Ref. \cite{Cheng:2005nb} indicates that the distribution amplitude of scalar meson would be smaller in the four-quark model than in the two-quark picture. 
% As stated in Ref.~\cite{Cheng:2017fkw}, the light-cone DA of $a_0(980)$ made up of $q_2\bar{q_1}$ can be
% defined as

Estimation by QCD factorization \cite{Cheng:2005nb} indicates that distribution amplitude of the $a_0(980)$ tends to be smaller in the four-quark picture than in the $q\bar{q}$ picture of the $a_0(980)$. This can be understood that it is less likely for a light quark $q$ (thereby $\bar{q}$) to carry a  high momentum fraction $x$ (near endpoint $x\sim1$) in the tetraquark picture than in the two-quark picture because of partial momentum shared by the strange quark-antiquark ($s\bar{s}$) in the later case. Despite that $a_0(980)$ is most likely to be a four-quark bound state, a quantitative prediction based on the tetraquark picture turns out to be quite nontrivial or complex in practice. To avoid this complexity, we regard $a_0(980)$ as a mixture of a pure $q\bar{q}$ state and a tetraquark state given above, and apply the LCSR method, for the time being, to the $q\bar{q}$ component, that is, we compute its FFs based on the two-quark picture of the $a_0(980)$. Then, we introduce a proper modulation of the DA of the $a_0(980)$ to make smooth transition of the resulted computation to the four-quark scenario. For this, we begin with the light-cone DA of $a_0(980)$ which is made up of $q_2\bar{q_1}$ : 
\begin{equation}
\begin{aligned}
\langle S(p)|\bar{q}_{2}(y)\gamma_{\mu}q_{1}(x)|0\rangle&=
p_{\mu}\int_{0}^{1}due^{i(up\cdot y+\bar{u}p\cdot x)}\phi_{S}(u,\mu),\notag\\
\langle S(p)|\bar{q}_{2}(y)q_{1}(x)|0\rangle&=
m_{s}\int_{0}^{1}due^{i(up\cdot y+\bar{u}p\cdot x)}\phi^{s}_{S}(u,\mu),\notag\\
\langle S(p)|\bar{q}_{2}(y)\sigma_{\mu\nu}q_{1}(x)|0\rangle&=
-m_{s}(p_{\mu}z_{\nu}-p_{\nu}z_{\mu})\int_{0}^{1}due^{i(up\cdot y+\bar{u}p\cdot x)}\frac{\phi^{\sigma}_{S}(u,\mu)}{6},\notag\\
\end{aligned}
\end{equation}
where $\phi_{S}$ is the twist-2 light-cone DA while $\phi_{S}^{s}$ and $\phi_{S}^{\sigma}$ are the twist-3 light-cone DAs, $z=y-x$, and $u$($\bar{u}=1-u$) is the momentum fraction carried by the quark $q_{2}$($q_{1}$) in total momentum of the meson.

We consider only the twist-2 light-cone DA $\phi_{S}$ since the twist-3 light-cone DAs vanish given that chiral currents are employed in correlation function of the LCSR, as mentioned in Sect. I. For simplicity, the higher twist distribution amplitudes in LCSR are ignored in this work.

\section{The distribution amplitude of the scalar meson}
Till now, much devotion has been made to explore distribution amplitudes of the scalar mesons, see Ref. 
\cite{Fu:2018yin,Faustov:2019mqr,Soni:2018adu,Jia:2017age,Li:2012gr,Khodjamirian:2009ys,Ball:2006yd,Khodjamirian:2000ds,Huang:2008sn,Wu:2006rd,Wang:2009azc,Lu:2006fr,Fajfer:2005ug,Ball:1993tp} for a details. In this section, we typically consider four distribution amplitudes of the scalar meson
$a_{0}(980)$ which are obtained via four different methods and correspond to different
form factors, respectively, and apply them to compute respective decay widths of the semileptonic decay processes considered.

\subsection{The distribution amplitudes via Gegenbauer polynomial (DA-I)}
\par
For the meson with quark configuration $\bar{q}q$, one can generally write its light-cone DA  as:
\begin{equation}
\begin{aligned}&\langle S(p)|\bar{q}_{\beta}(y)q_{\alpha}(x)|0\rangle=\frac{1}{4}\int_{0}^{1}due^{i(up\cdot y+\bar{u}p\cdot x)}\Bigg\{\slashed{p}\phi_{S}(u)+m_{S}\Big[\phi^{s}_{S}(u)-\sigma_{\mu\nu}p^{\mu}z^{\nu}\frac{\phi^{\sigma}_{s}(u)}{6}\Big]\Bigg\}_{\beta\alpha}.
\label{1}
\end{aligned}
\end{equation}
where $S(p)$ stands for the scalar meson with four-momenta $p$. Here, we consider only the twist-2 light-cone DA $\phi_{S}$ thanks to the usage of chiral currents, as discussed in more details in Sec 4.2. For computation of the DA, one way is to expand it in terms of Gegenbauer polynomial  \cite{Chernyak:1983ej,Braun:2003rp}
\begin{equation}
\begin{aligned}
\phi_{S}(u,\mu)=&\bar{f_{S}}6u(1-u)[B_{0}(\mu)+\sum_{m=1}B_{m}(\mu)C_{m}^{3/2}(2u-1)],
\label{3}
\end{aligned}
\end{equation}
where the Gegenbauer coefficients $B_{m}$ are given by
\begin{equation}
\begin{aligned}
 B_{m}(\mu)=&\frac{1}{\bar{f_{S}}}\frac{2(2m+3)}{3(m+1)(m+2)}\int_{0}^{1}C_{m}^{3/2}(2u-1)\phi_{S}(u,\mu)du,
\end{aligned}
\end{equation}
with $C_{m}^{3/2}(2u-1)$ the Gegenbauer polynomials, and other parameters like $\bar{f_{S}}$ are defined by the following relations
\begin{gather*}
\langle S(p)|\bar{q}_{2}\gamma_{\mu}q_{1}|0\rangle=f_{S}p_{\mu},\notag\\
\langle S(p)|\bar{q}_{2}q_{1}|0\rangle=m_{S}\bar{f}_{S},\notag\\
\mu_{S}f_{S}=\bar{f}_{S},\notag\\
\mu_{S}=\frac{m_{S}}{m_{2}(\mu)-m_{1}(\mu)}.\notag
\end{gather*}
\par
Conformal invariance of QCD in chiral limit indicates that the partial waves in the expansion (\ref{3}) with different conformal spin cannot mix upon renormalization to a first approximation \cite{Braun:2003rp}. Consequently, one can renormalize the coefficients $B_{m}$ multiplicatively \cite{Cheng:2005nb}
\begin{equation}
%\begin{aligned}
B_{m}(\mu)=B_{m}(\mu_{0})\Big(\frac{\alpha_{s}(\mu_{0})}{\alpha_{s}(\mu)}\Big)^{-(\gamma_{(m)}+4)/b},\notag\\
\bar{f_{S}}(\mu)=\bar{f_{S}}(\mu_{0})\Big(\frac{\alpha_{s}(\mu_{0})}{\alpha_{s}(\mu)}\Big)^{4/b},
%\end{aligned}
\end{equation}
where the one-loop anomalous dimensions are given by \cite{Gross:1974cs,Shifman:1980dk}
\begin{equation}
\gamma_{(m)}=C_{F}\Big[1-\frac{2}{(m+1)(m+2)}+4\Big(\sum_{j=2}^{m+1}\frac{1}{j}\Big)\Big],
\label{g4}
\end{equation}
with $C_{F}=(N_{c}^{2}-1)/(2N_{c})$. At the energy scale $\mu=1\ \rm{GeV}$ associated with the $a_{0}(980)$ meson, the renormalized decay constant and Gegenbauer coefficients are found to be \cite{Cheng:2005nb} :
\begin{equation}
\begin{aligned}
\bar{f_{S}}(\mu=1\ \rm{GeV})&=0.365\ {\rm{GeV}},\\
 B_{1}(\mu=1\ {\rm{GeV}})=-0.93 &\pm 0.10,\
    B_{2}(\mu=1\ \rm{GeV})=0,\notag\\
 B_{3}(\mu=1\ {\rm{GeV}})=0.14 &\pm 0.08,\
     B_{4}(\mu=1\ {\rm{GeV}})=0.
\label{fBn}
\end{aligned}
\end{equation}

Generally, the Gegenbauer coefficients given above are related to moments of the scale meson with isospin $I=1$ through the following relations \cite{Cheng:2017fkw}:

\begin{equation}
\begin{aligned}
B_{0}&=\langle\xi^{0}\rangle,\\
B_{1}&=\frac{5}{3}\langle\xi^{1}\rangle,\\
B_{2}&=\frac{35}{12}\langle\xi^{2}\rangle-\frac{7}{12}\langle\xi^{0}\rangle,\\
B_{3}&=\frac{21}{4}\langle\xi^{3}\rangle-\frac{9}{4}\langle\xi^{1}\rangle,\\
B_{4}&=\frac{77}{8}\langle\xi^{4}\rangle-\frac{77}{12}\langle\xi^{2}\rangle+\frac{11}{24}\langle\xi^{0}\rangle,\\
\cdots
\label{Bn}
\end{aligned}
\end{equation}
In the approximation $z^{2}\simeq0$, the moments $\langle\xi_{S}^{n}\rangle$ can be computed in QCD sum rule, obtaining\cite{Cheng:2005nb}
\begin{equation}
\begin{aligned}
&\langle\xi_{S}^{n}\rangle m_{S}\bar{f}_{S}^{2}e^{-m_{s}^{2}/M^{2}}+\langle\xi_{S'}^{n}\rangle m_{S'}\bar{f}_{S'}^{2}e^{-m_{s'}^{2}/M^{2}}\\
&=-\frac{3}{16\pi^{2}}M^{2}\left(\frac{m_{q_{2}}+m_{q_{1}}}{n+2}+\frac{m_{q_{2}}-m_{q_{1}}}{n+1}\right)\left(1-e^{-\frac{s_{0}}{M^{2}}}\right) +\langle \bar{q}_{2}q_{2}\rangle  +\frac{10n-3}{24}\frac{\langle\bar{q}_{2}g_{s}\sigma\cdot Gq_{2}\rangle}{M^{2}}\\
&\quad+\frac{n(4n-5)}{36}\frac{\langle g_{s}^{2}G^{2}\rangle\langle\bar{q}_{2}q_{2}\rangle}{M^{4}}+(-1)^{n+1}\Bigg[-\frac{3}{16\pi^{2}}M^{2}\left(\frac{m_{q_{2}}+m_{q_{1}}}{n+2}+\frac{m_{q_{2}}-m_{q_{1}}}{n+1}\right)\left(1-e^{-\frac{s_{0}}{M^{2}}}\right)\\
&\quad+\langle \bar{q}_{1}q_{1}\rangle+\frac{10n-3}{24}\frac{\langle\bar{q}_{1}g_{s}\sigma\cdot Gq_{1}\rangle}{M^{2}}\frac{n(4n-5)}{36}\frac{\langle g_{s}^{2}G^{2}\rangle\langle\bar{q}_{1}q_{1}\rangle}{M^{4}}\Bigg].
\label{7}
\end{aligned}
\end{equation}
At $z^{2}=0$ and the threshold $s_{0}=3.1\ \rm{GeV}^{2}$, one can apply Eqs. (\ref{3}), (\ref{g4}) and (\ref{7}) to compute the light-cone DA. The result is shown in FIG.1.
\begin{figure}[H]
\centering

\centering
\includegraphics[height=6.5cm,width=8.5cm]{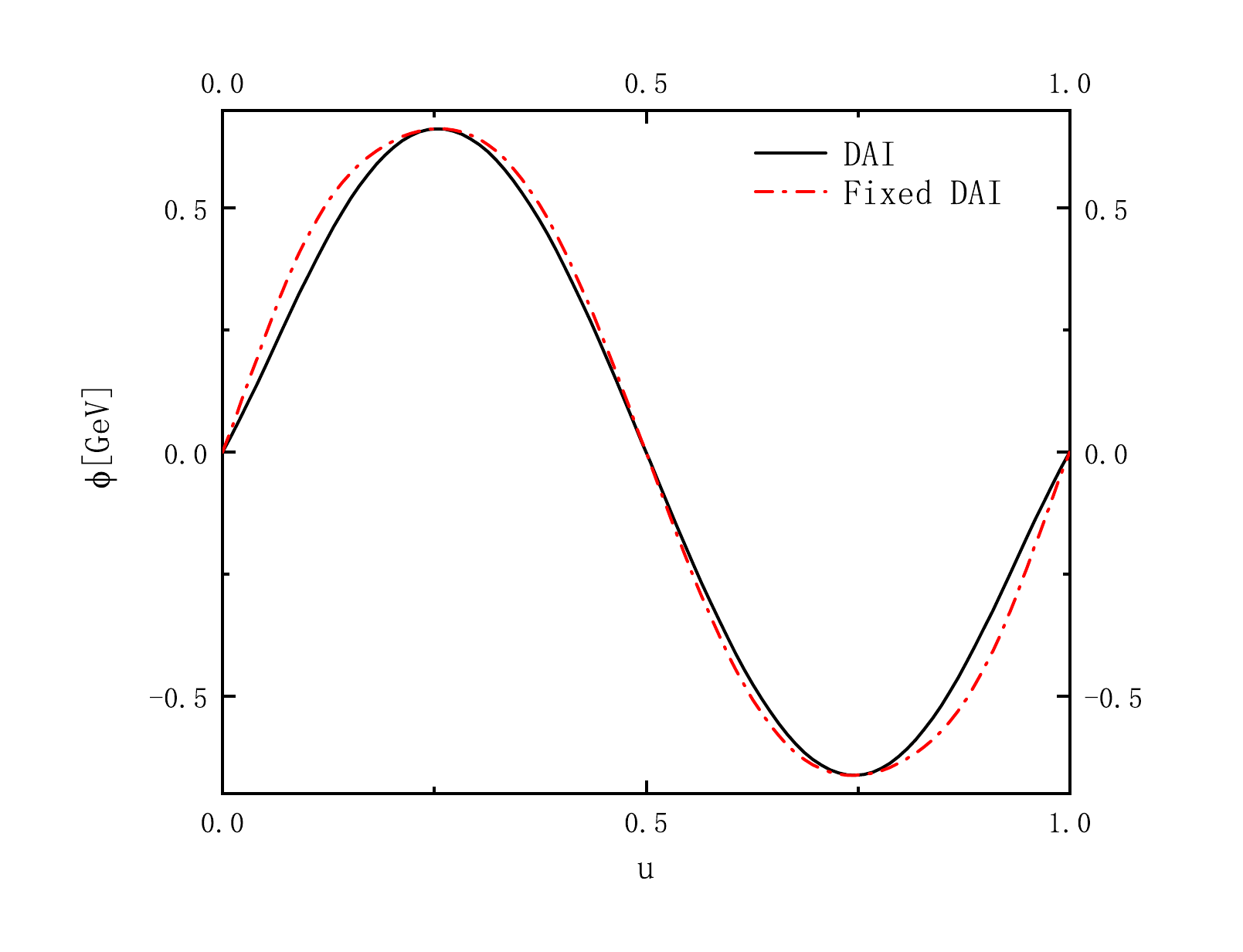}
\caption{The distribution amplitudes of the $a_{0}(980)$ obtained by QCD sum rule at $z^{2}=0$ for DA-I.}
\label{f1}
\end{figure}
\par
\subsection{The distribution amplitudes via Gegenbauer polynomial in the new approximation (DA-II)}
\par
\begin{figure}[H]
\centering
\centering
\includegraphics[height=6.5cm,width=8.5cm]{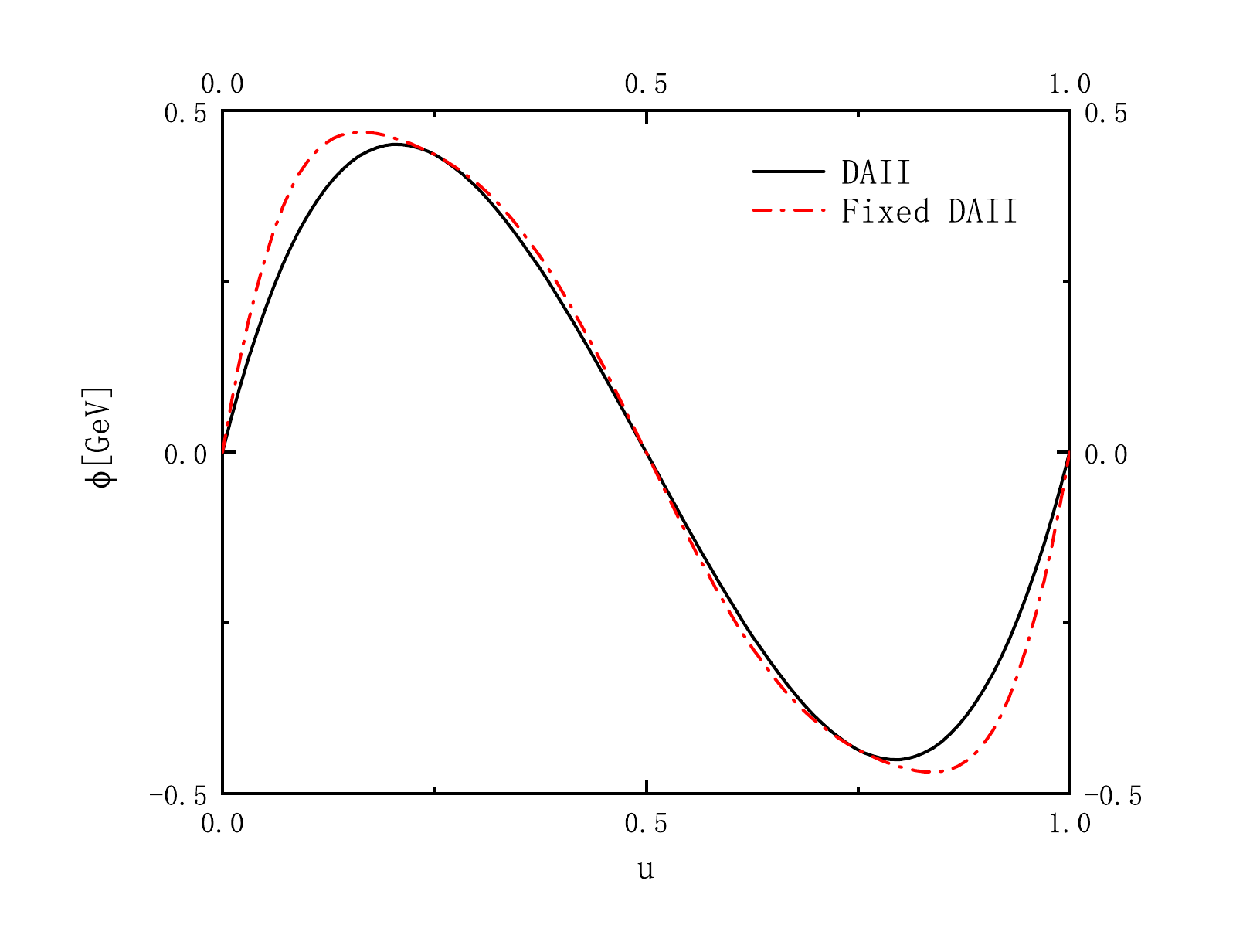}
\caption{The distribution amplitudes of the $a_{0}(980)$ obtained by QCD sum rule at a new approximation for DA-II.}
\label{f2}
\end{figure}

For the scalar meson with $I=1$, its moments include not only terms of order $z^{2}$, but also other terms of the same order, such as $x\cdot z$ and $x^{2}$. To compute moments of such a meson, Ref. \cite{Hong:2024rhg} introduces, differing from that in Eq. (\ref{7}), a new approximation, by which one can use the following relation to compute the moments,
\begin{equation}
\begin{aligned}
\langle\xi^{n}\rangle&=\frac{e^{m^{2}_{s}/M^{2}}}{m_{s}\bar{f}^{2}_{s}}\Bigg\{\frac{3}{16\pi^{2}}[3+(-1)^{n}+2n]\frac{1}{(n+2)(n+1)}\left[(-1)^{n+1}m_{1}+m_{2}\right]\left(1+e^{-\frac{s_{0}}{M^{2}}}\right)\\
&\quad+\langle0|\bar{q}_{2}q_{2}|0\rangle+(-1)^{n+1}\langle0|\bar{q}_{1}q_{1}|0\rangle+\frac{2n+1}{2M^{2}}
\left[m_{2}^{2}\langle0|\bar{q}_{2}q_{2}|0\rangle+(-1)^{n+1}m_{1}^{2}\langle0|\bar{q}_{1}q_{1}|0\rangle\right]\\
&\quad+\frac{m_{1}m_{2}}{2M^{2}}\left[\langle0|\bar{q}_{2}q_{2}|0\rangle+(-1)^{n+1}\langle0|\bar{q}_{1}q_{1}|0\rangle\right]\\
&\quad-\frac{2n}{3}\frac{1}{M^{2}}\left[\langle0|g\bar{q}_{2}TG\sigma q_{2}|0\rangle+(-1)^{n+1}\langle0|g\bar{q}_{1}TG\sigma q_{1}|0\rangle\right]\\
&\quad+\frac{8n\pi}{81}\frac{1}{M^{4}}\left[m_{1}\langle0|\sqrt{\alpha_{s}}\bar{q}_{2}q_{2}|0\rangle^{2}+(-1)^{n+1}m_{2}\langle0|\sqrt{\alpha_{s}}\bar{q}_{1}q_{1}|0\rangle^{2}\right]
\Bigg\}.
\label{xi}
\end{aligned}
\end{equation}
At the energy scale $\mu=1\ \rm{GeV}$, Eq. (\ref{xi}) and Eq. (\ref{Bn}) explicitly give rise to
\begin{gather}
 B_{1}(\mu=1\ \rm{GeV})=-0.718 ,\notag \quad
    \it{B}_{\rm{2}}(\mu=\rm{1}\ \rm{GeV})=0,\notag\\
 B_{3}(\mu=1\ \rm{GeV})=-0.011,\notag  \quad
     \it{B}_{\rm{4}}(\mu=\rm{1}\ \rm{GeV})=0. \notag
\end{gather}
The ensuing DA of the $a_{0}(980)$ by Eq. (\ref{3}) is shown in FIG.~\ref{f2}.

Using similar method, one can compute the higher moments and the corresponding Gegenbauer coefficients. Writing collectively, the computed results are
\begin{equation}
\begin{aligned}
\{\langle\xi^{1}\rangle, \langle\xi^{3}\rangle, \langle\xi^{5}\rangle, \langle\xi^{7}\rangle, \cdots\}_{\mu=1~\rm{GeV}}=\{-0.431, -0.187, -0.104, -0.066, \cdots \},\notag
\end{aligned}
\end{equation}
\begin{equation}
\begin{aligned}
\{B_{1}, B_{3}, B_{5}, B_{7}, \cdots\}_{\mu=1~\rm{GeV}}=\{-0.718, -0.011, -0.011, -0.009, \cdots\},\notag
\end{aligned}
\end{equation}
which exhibit a notable tendency of convergence associated with $n$. This is in contrast with that by Eq.~(\ref{7}).

\subsection{The distribution amplitudes based on BLFQ method (DA-III)}

In momentum space, the light-front wave functions (LFWFs) of a meson has the form of  \cite{Lan:2021wok,Bakker:2013cea,Brodsky:1997de,deTeramond:2005su,Brodsky:2006uqa,deTeramond:2008ht,Branz:2010ub,Brodsky:2014yha,Ahmady:2018muv,deTeramond:2021yyi,Ahmady:2021lsh,Ahmady:2021yzh}
\begin{equation}
\begin{aligned}
\Psi^{N,M_{J}}_{\{x_{i};\overrightarrow{p}_{\bot};\lambda_{i}\}}=\sum_{n_{i};m_{i}}\psi^{N}(\{\overline{\alpha_{i}}\})\prod_{i=1}^{N}\phi_{n_{i};m_{i}}(\overrightarrow{p}_{\bot};b),
\label{LFWF}
\end{aligned}
\end{equation}
where $\psi^{N=2}(\{\bar{\alpha_{i}}\})$ is the components of the mesonic eigenvectors associated with the Fock state $|q\bar{q}\rangle$, and $\psi^{N=3}(\{\bar{\alpha_{i}}\})$ is the components of the meson eigenvectors associated with $|q\bar{q}g\rangle$, $\{\bar{\alpha}\}\equiv \{x, n, m, \lambda\}$, in which $x$, $n$, $m$ and $\lambda$ are the longitudinal momentum fraction, the radial, angular and the spin quantum numbers, respectively,  and $\phi_{n_{i};m_{i}}(\overrightarrow{p}_{\bot};b)$ is the two-dimensional harmonic-oscillator basis functions, solved via diagonalizing the full Hamiltonian matrix in the basis light-front quantization (BLFQ) method \cite{Vary:2009gt,Xu:2021wwj,Zhao:2014xaa,Wiecki:2014ola}.

We note that in practical computation there are two up-limit numbers $N_{max}$ and $K$ fixing the total numbers of bases of the wave functions $\psi^{N}$ and $\phi_{n_{i};m_{i}}$, respectively. These two numbers, $N_{max}$ and $K$, are introduced to truncate the infinite bases of the respective Fock spaces. The first truncation via $N_{max}$ enables us to factorize out the transverse center of mass motion and manifest a natural ultra-violent (UV) regulator $\Lambda_{UV}\sim b\sqrt{N_{max}}$. Given the truncations {$N_{max}$=14, K=15}, one can fix the model parameters so that the ensuing eigenvalues of the full Hamiltonian matrix in BLFQ method reproduce the measured mass spectra of light mesons. Here, the model parameters include the quark mass $m_{q}$, the effective mass of gluon $m_{g}$, the harmonic oscillator
scale parameter $b$, the strength of the confinement $\kappa$, the two parameters $m_{f}$ and $g_{s}$ (independent of quark mass) in the vertex interaction and coupling constant. We summarize the resulted values of the parameters in Table I.
\begin{table}[H]
\centering
\caption{The model parameters for the truncation {$N_{max}$=14, K=15}. All are in units of $\rm{GeV}$ except $g_{s}$, which is dimensionless.}
    \begin{tabular}{c c c c c c}
    \hline
        \hline
        $m_{q}$ &  $m_{g}$&  $b$  & $\kappa$ & $m_{f}$ & $g_{s}$ \\
        % \hline
        0.39 & 0.60 & 0.29 & 0.65 & 5.69 & 1.92\\
        \hline
        \hline
    \end{tabular}
\end{table}
\noindent
\par
\begin{figure}[H]
\centering

\centering
\includegraphics[height=6.5cm,width=8.5cm]{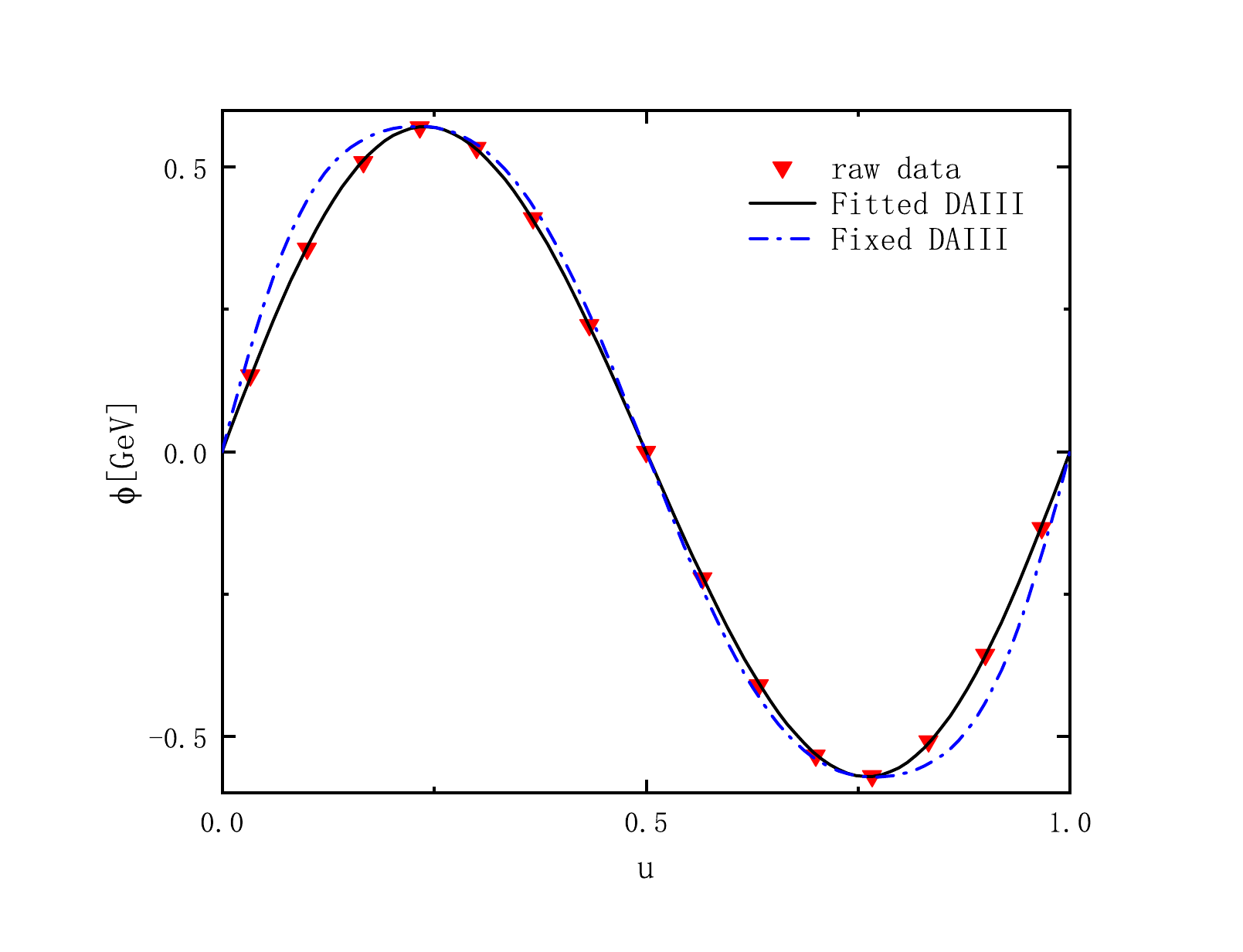}
\caption{The distribution amplitudes of the $a_{0}(980)$ obtained by BLFQ for DA-III.}
\label{f3}
\end{figure}

In the BLFQ method, the light-cone DAs of the scalar meson $q_{2}\bar{q}_{1}$  are
\begin{equation}
\begin{aligned}
\langle0|\bar{q}_{2}(z)\gamma^{+}&q_{1}(-z)|S(p)\rangle=ip^{+}\int_{0}^{1}dxe^{ip^{+}z^{-}(x-\frac{1}{2})}\phi_{S}(x).
\label{BLFQp}
\end{aligned}
\end{equation}
In terms of LFWFs, the light-cone DA in Eq.~(\ref{BLFQp}) can be solved to be\cite{Li:2015zda,Tang:2018myz,Jia:2018ary,Mondal:2019jdg,Lan:2019rba,Lan:2019vui,Qian:2020utg,Li:2021jqb,Lan:2021wok,Lan:2022blr}
\begin{equation}
\phi_{S}(x)=\bar{f}_{S}\frac{2\sqrt{2N_{C}}}{N_{F}}\int[d^{2}p_{\bot}]\Psi^{\uparrow\downarrow+\downarrow\uparrow}(x,p_{\bot}),
\label{fsp}
\end{equation}
with $N_{F}$ the renormalized constant,
\begin{equation}
N_{F}=\sum_{s_{1}s_{2}}\int[d^{3}p_{1}][d^{3}p_{2}]\Psi^{s_{1}s_{2}*}(p_{1},p_{2})\Psi^{s_{1}s_{2}}(p_{1},p_{2}),\notag
\end{equation}
where $p_{1,2}$ are the momentum of the quark $q_{1}$ and $\bar{q_{2}}$ and  $s_{1,2}$ are the spins of two quarks.
Given the LFWFs solved above, one can use Eq. (\ref{fsp}) and Eq. (\ref{LFWF}) to compute the distribution amplitude of the $a_{0}(980)$. The results are shown FIG.~\ref{f3}.
\par

\subsection{The distribution amplitudes as via Gegenbauer polynomial from BLFQ method (DA-IV)}
\par
\begin{figure}[H]
\centering

\centering
\includegraphics[height=6.5cm,width=8.5cm]{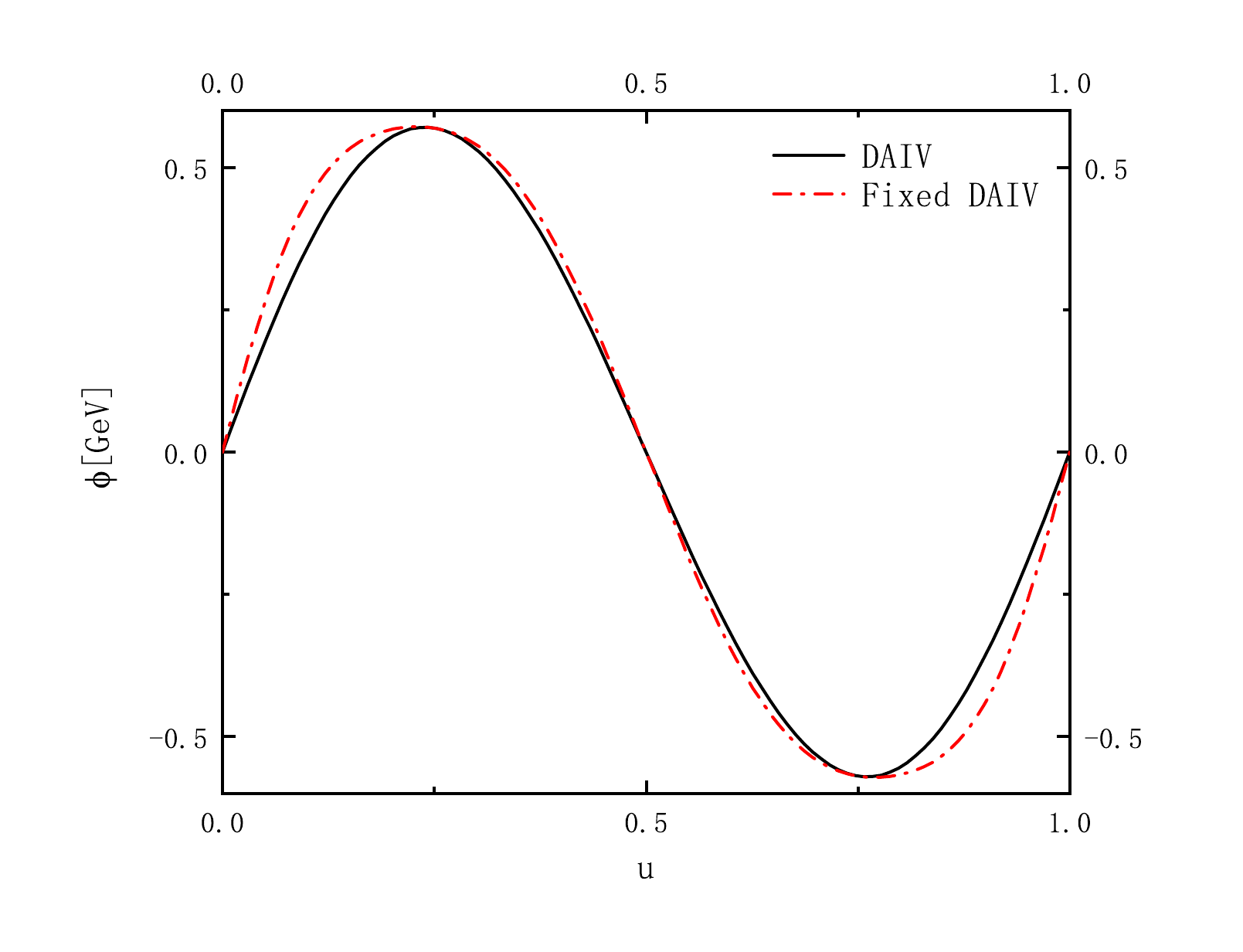}
\caption{The distribution amplitudes of the $a_{0}(980)$ obtained from BLFQ for DA-IV.}
\label{f4}
\end{figure}

Assuming distribution amplitudes in DA-III are known rigorously, one can obtain the following relation between moments and the DAs, without any approximation,
\begin{equation}
\langle \xi^{n}\rangle =\int_{0}^{1}dx (2x-1)^{n} \phi_{s}(x).\notag
\end{equation}
At the energy scale $\mu=1~\rm{GeV}$, this enable us to compute the moments and Gegenbauer coefficients. We list the results collectively
\begin{equation}
\begin{aligned}
&\{\langle\xi^{1}\rangle, \langle\xi^{3}\rangle, \langle\xi^{5}\rangle, \langle\xi^{7}\rangle, \cdots\}_{\mu=1~\rm{GeV}}=\{-0.510, -0.205, -0.109, -0.067, \cdots \},\notag
\end{aligned}
\end{equation}
\begin{equation}
\begin{aligned}
&\{B_{1}, B_{3}, B_{5}, B_{7}, \cdots\}_{\mu=1~\rm{GeV}}=\{-0.849, 0.073, -0.004, -0.001, \cdots\}.\notag
\end{aligned}
\end{equation}
One sees that moments and Gegenbauer coefficients also shows a notable tendency of convergence, with respect to $n$. This follows the distribution amplitude in the BLFQ method. We plot the results in FIG.~\ref{f4}.
\par

\par
\begin{figure}[H]
\centering

\centering
\includegraphics[height=6.5cm,width=8.5cm]{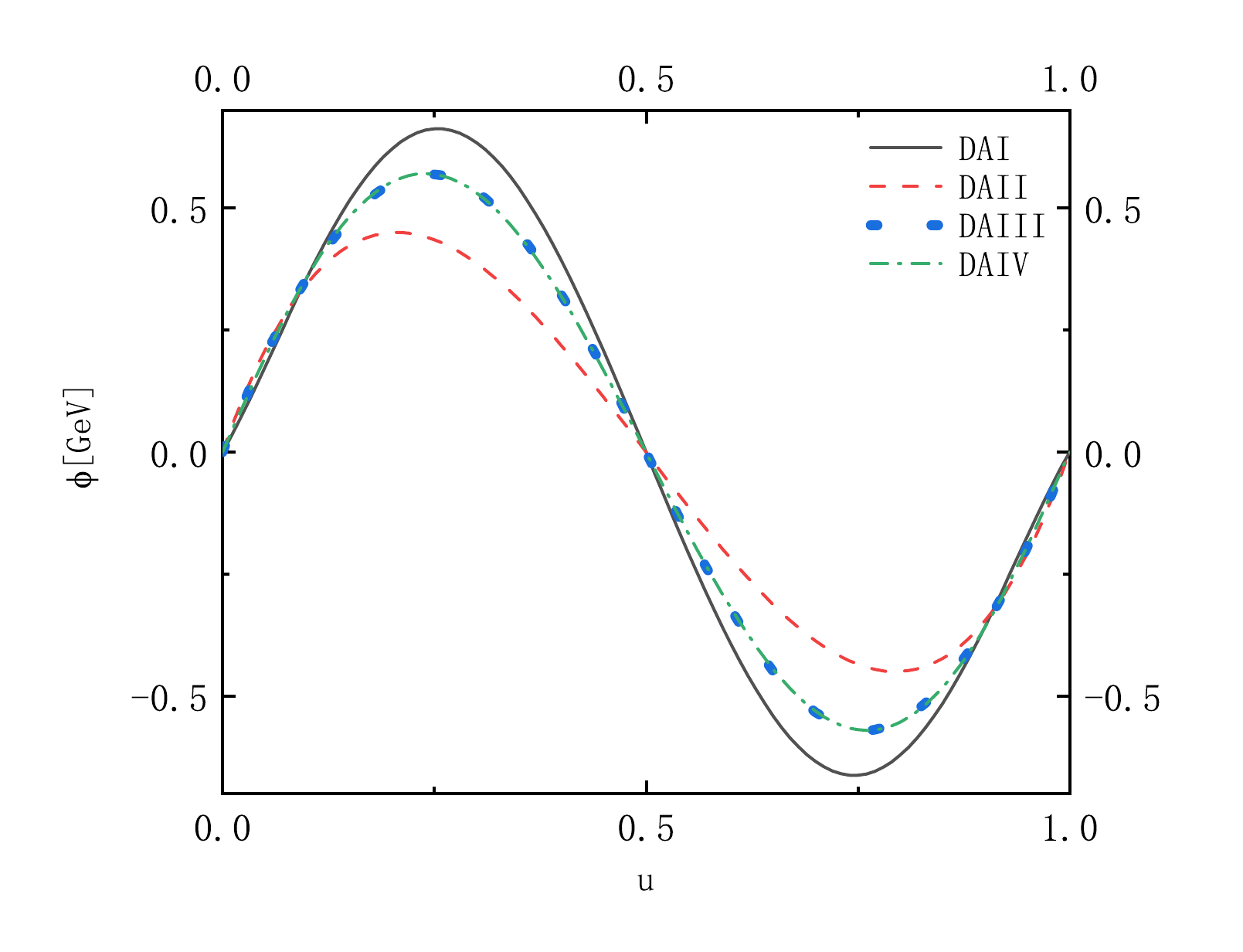}
\caption{The distribution amplitudes of the $a_{0}(980)$ shown with the solid (DA-I), the dash (DA-II), the dot (DA-III) and the dash dot (DA-IV) lines. Two lines corresponding to DA-III and DA-IV overlap exactly.}
\label{phi}
\end{figure}

Before ending of this Section, we briefly discuss the main features of four distribution amplitudes obtained from the different methods mentioned. We plot the four DAs associated with the respective methods collectively in FIG.~\ref{phi} for comparison.
Comparing the resulted light-cone DAs in FIG.~\ref{phi}, one can conclude :
\begin{enumerate}
\item The uncertainty of the distribution amplitudes computed via QCD sum rule comes mainly from the approximation on $z^{2}$.
\item The distribution amplitudes via the two methods, DA-III and DA-IV, are highly consistent with each others.
\end{enumerate}

\section{Form factors of semileptonic decays}

\begin{figure}[H]
\centering
\subfigbottomskip=1pt
\subfigure[]{
 \includegraphics[height=5cm,width=7cm]{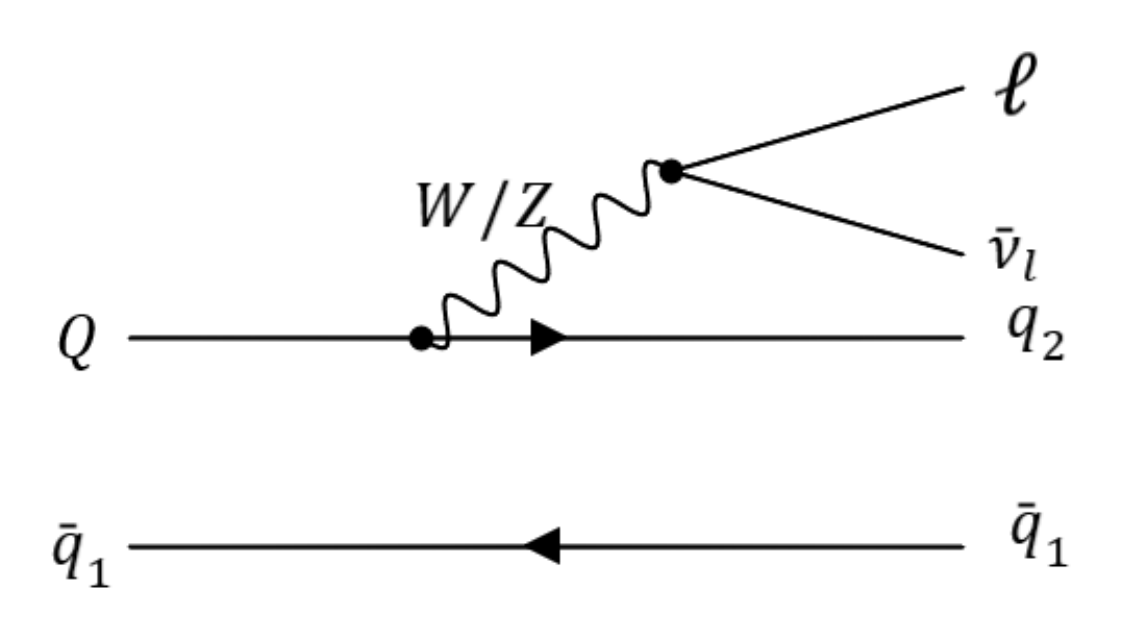}\label{fTds}}
\subfigure[]{
 \includegraphics[height=5cm,width=7cm]{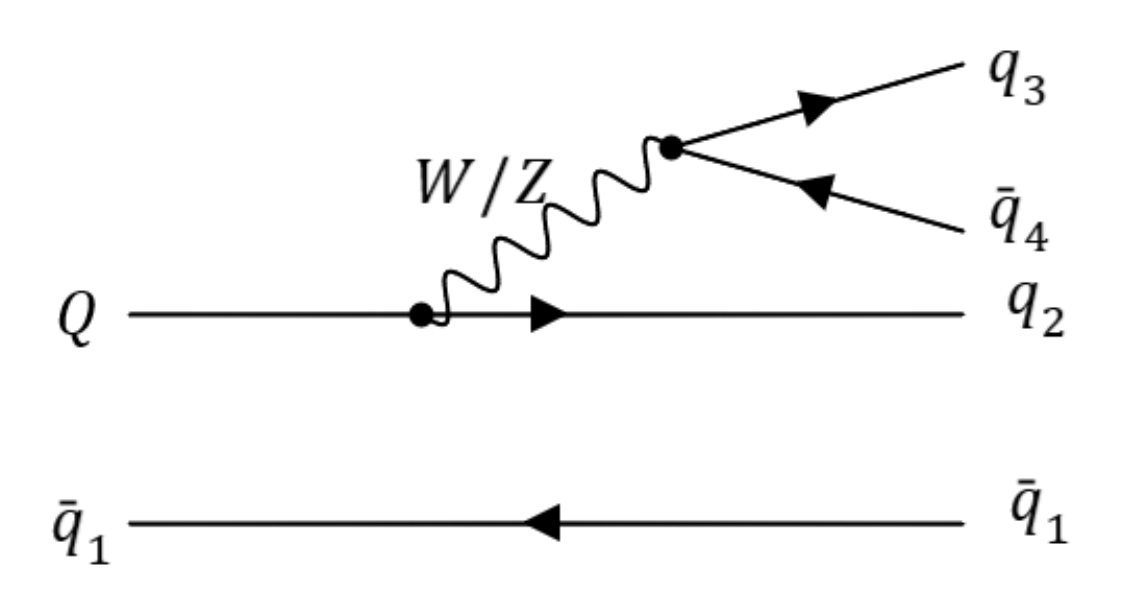}\label{fPds}}
\caption{(a) The processes associated with $f_{+}$ and $f_{-}$. The difference between $f_{+}$ and $f_{-}$ lies in the Lorentz structure. (b) The processes associated with $f_{T}$.}
\end{figure}

In the semileptonic decays, one can parameterize the hadron matrix elements, which implicitly encode strong interaction of quarks in hadrons, in terms of the computable FFs. As such, the FF plays an important role in the computation of the decay width. In this section, we employ LCSR to compute a set of (non-vanishing) independent FFs for the semileptonic decays of $D/B\rightarrow S$, which are referred as $f_{+}^{D/B\rightarrow S}$, $f_{-}^{D/B\rightarrow S}$ and $f_{T}^{D/B\rightarrow S}$. Generally, there are two kind of hadron transition, associated with two decays, which are of the vector and tensor types, for which the parameterizations takes the form \cite{Han:2013zg,Wang:2014vra,Cheng:2019tgh,Han:2023pgf}: 
\begin{equation}
\begin{aligned}
&\langle S(p)|\bar{q}_{2}\gamma_{\mu}\gamma_{5}Q|D/B(p+q)\rangle
&=-2if_{+}^{D/B\rightarrow S}(q^{2})p_{\mu}
&-i\Big[f_{+}^{D/B\rightarrow S}(q^{2})+f_{-}^{D/B\rightarrow  S}(q^{2})\Big]q_{\mu},\\
\label{9}
\end{aligned}
\end{equation}
\begin{equation}
\begin{aligned}
&\langle S(p)|\bar{q}_{2}\sigma_{\mu\nu}(1+\gamma_{5})q^{\nu}Q|D/B(p+q)\rangle
&=\Big[2p_{\mu}q^{2}-2q_{\mu}(p\cdot q)\Big]\frac{-f_{T}^{D/B\rightarrow S(q^2)}}{m_{D/B}+m_{S}},
\label{10}
\end{aligned}
\end{equation}
where $f_{+,-}^{D/B\rightarrow S}(q^{2})$ are the transition FFs associated with the transition processes of $Q\rightarrow q_{2}l\bar{\nu}_{l}$, plotted in FIG.~\ref{fTds}, and $f_{T}^{D/B\rightarrow S}(q^{2})$ is the penguin FFs associated with the penguin-induced transition processes of $Q\rightarrow q_{2} q_{3}\bar{q}_{4}$, shown in FIG.~\ref{fPds} (only the lowest order Feynman diagrams are presented here). The variable $q$ stands for the transfer momentum $q$ in transition, which should be distinguished from the same symbol used for the light quarks.

To compute the FFs above with LCSR, we construct two-point correlation functions
\begin{equation}
\begin{aligned}
\Pi_{1\mu}(p,q)&=i\int d^{4}xe^{iqx}\langle S(p)|T\{J_{1\mu}(x),J_{1}(0)\}|0\rangle,\\
\Pi_{2\mu}(p,q)&=i\int d^{4}xe^{iqx}\langle S(p)|T\{J_{2\mu}(x),J_{1}(0)\}|0\rangle,
\label{11}
\end{aligned}
\end{equation}
where $q=p_{D/B}-p$ stands for the transfer momentum, $p_{D/B}$ and $p$ are the four-momenta of the initial and final meson states, respectively. Here, $J_{1\mu}(x)$ stand for the chiral currents associated with the vector transition and $J_{2\mu}(x)$ for the chiral currents associated with the tensor transition, as given explicitly in Table II.
\begin{table*}
\caption{Chiral currents for different decay processes, where $Q$ stands for $b$ or $c$ quark}
    \begin{tabular}{c c c}
    \hline
        \hline
        Mode &  $J_{i\mu}(x)$&  $J_{i}(0)$  \\
%        \hline

        \multirow{2}*{$ D/B\rightarrow S $} & $J_{1\mu}(x)=\bar{q}_{2}(x)\gamma_{\mu}(1-\gamma_{5})Q(x)$ & $J_{1}(0)=\bar{Q}(0)i(1-\gamma_{5})q_{1}(0)$  \\
                                          & $J_{2\mu}(x)=\bar{q}_{2}(x)\sigma_{\mu\nu}(1+\gamma_{5})Q(x)$ & $J_{2}(0)=\bar{Q}(0)i(1-\gamma_{5})q_{1}(0)$  \\

        \hline
        \hline
    \end{tabular}
\end{table*}
\par
According to the method of LCSR, one can compute the correction functions given by Eq. (\ref{11}) separately from two viewpoints and then match ensuing results to extract the information of hadrons. The first viewpoint is that of phenomenological at hadronic level, in which one inserts a complete set of meson states between two currents appeared in the correlation function. The second viewpoint is that happens at the quark level, for which one applies the operator product expansion (OPE) to reformulate the correlation functions. During the computation via the method of LCSR, a complete set of the $D/B$ meson is going to be inserted in the matrix elements in Eq. (\ref{11}), which correspond to the initial state in the decay process.

\subsection{On the phenomenological side}
\par
Inserting the complete set of the initial ($D/B$ meson) states $|i\rangle$ in the correlation functions (\ref{11}) and isolating the ground-state components, which are referred also as $|D/B\rangle$, in the complete set, one has
\begin{equation}
\begin{aligned}
&\Pi_{1\mu}(p,q)=\frac{\langle S(p)|\bar{q}_{2}\gamma_{\mu}(1\pm \gamma_{5}) Q|D/B\rangle \langle D/B|\bar{Q}i\gamma_{5}q_{1}|0\rangle}{m^{2}_{D/B}-(p+q)^{2}}+\rm{higher \quad states},\\
&\Pi_{2\mu}(p,q)=\frac{\langle S(p)|\bar{q}_{2}\sigma_{\mu\nu}(1\pm \gamma_{5})q^{\nu}Q|D/B\rangle \langle D/B|\bar{Q}i\gamma_{5}q_{1}|0\rangle}{m^{2}_{D/B}-(p+q)^{2}}+\rm{higher \quad states},
\label{12}
\end{aligned}
\end{equation}
where the hadronic matrix elements $\langle S(p)|\bar{q}_{2}\gamma_{\mu}(1\pm \gamma_{5}) Q|D/B\rangle $ and $\langle S(p)|\bar{q}_{2}\sigma_{\mu\nu}(1\pm \gamma_{5})q^{\nu}Q|D/B\rangle $ are to be parameterized by the FFs, as shown in Eq. (\ref{9}) and Eq. (\ref{10}).

Besides Eq. (\ref{9}) and Eq. (\ref{10}), the other matrix element can be parameterized by
\begin{equation}
\begin{aligned}
\langle D/B|\bar{Q}i\gamma_{5}q_{1}|0\rangle&=\frac{m^{2}_{D/B}f_{D/B}}{m_{Q}+m_{q_{1}}}.
\label{fD}
\end{aligned}
\end{equation}
Substituting Eqs. (\ref{9}) and (\ref{10}) into Eq. (\ref{12}) and taking account of Lorentz structures, one can obtain the phenomenological formulation of two correlation functions. The results are, with the usage of Eq. (\ref{fD}),
\begin{equation}
\begin{aligned}
&\Pi_{1\mu}(p,q)=-\frac{2if_{+}^{D/B\rightarrow S}(q^{2})p_{\mu}+i\Big[f_{+}^{D/B\rightarrow S}(q^{2})+f_{-}^{D/B\rightarrow S}(q^{2})\Big]q_{\mu}}{m_{D/B}^{2}-(p+q)^{2}}\frac{m^{2}_{D/B}f_{D}}{m_{Q}+m_{q_{1}}}+\int_{s_{0}}^{\infty}\frac{\rho(s)ds}{s-(p+q)^{2}},\\
&\Pi_{2\mu}(p,q)=-\frac{q^{2}2p_{\mu}-2q_{\mu}(p\cdot q)}{m_{D/B}^{2}-(p+q)^{2}}\frac{if_{T}^{D/B\rightarrow S}}{m_{D/B}+m_{S}}\frac{m^{2}_{D/B}f_{D/B}}{m_{Q}+m_{q_{1}}}+\int_{s_{0}}^{\infty}\frac{\rho(s)ds}{s-(p+q)^{2}}.
\label{ps}
\end{aligned}
\end{equation}

To avoid the complexity of involving higher excited and continuum spectra on the phenomenological side, it is common to apply quark-hadron duality to simplify integrals involving higher excited states and continuum spectra~\cite{Shifman:1978by,Shifman:1978bw,Khodjamirian:1979fa,Shifman:1980dk,Khodjamirian:1983gd,Belyaev:1993wp,Khodjamirian:1998vk,Colangelo:2000dp}:
\begin{align}
\int_{s_{0}}^{\infty}ds\frac{\rho(s)}{s-(p+q)^{2}}\simeq \int_{s_{0}}^{\infty}ds\frac{1}{\pi}\frac{\rm{Im}\Pi_{\mu}^{pert}(\mathit{p,q})}{s-(p+q)^{2}},
\end{align}
where $s_{0}$ is the threshold near the squared mass of the first excited state of $D/B$ mesons, and $\rho(s)$ is the spectral density of the higher excited states and the continuum states.
\par
\subsection{On the theoretical side}
\par
At the quark level, which is relevant to the region of the large momentum, the computation of the correlation function will be based on the T-product expansion near the light-cone $x^{2}\approx 0$. By contracting quark fields $Q$ and $\bar{Q}$, one has
\begin{equation}
\begin{aligned}
&\Pi_{1\mu}(p,q)=i\int d^{4}xe^{iqx}\langle S(p)|\{\bar{q}_{2}(x)\gamma_{\mu}(1\pm\gamma_{5})S^{Q}(x,0)i(1\pm\gamma_{5})q_{1}(0)\}|0\rangle,\\
&\Pi_{2\mu}(p,q)=i\int d^{4}xe^{iqx}\langle S(p)|\{\bar{q}_{2}(x)\sigma_{\mu\nu}(1\pm\gamma_{5})q^{\nu}S^{Q}(x,0)i(1\pm\gamma_{5})q_{1}(0)\}|0\rangle,
\label{17}
\end{aligned}
\end{equation}
where $S^{Q}(x,0)$ is the propagator of the $Q$ quark,
\begin{align}
S^{Q}(x,0)&=-i\int\frac{d^{4}k}{(2\pi)^{4}}e^{-ik\cdot x}\frac{\slashed{k}+m_{Q}}{m^{2}_{Q}-k^{2}}.
\end{align}
Combining these relations, the correlation function become then
\begin{equation}
\begin{aligned}
\Pi_{1\mu}(p,q)=&i\int\frac{d^{4}xd^{4}k}{(2\pi )^{4}}\frac{e^{i(q-k)x}}{m_{c}^{2}-k^{2}}\rm{Tr}[\gamma_{\mu}(1\pm\gamma_{5})(\slashed{k}+\mathit{m_{Q}})(1\pm\gamma_{5})]_{\beta\alpha}\times\langle S(\mathit{p})|\bar{q}_{2\beta} (\emph{x}) q_{1\alpha}(0)|0\rangle,\\
\Pi_{2\mu}(p,q)=&i\int\frac{d^{4}xd^{4}k}{(2\pi )^{4}}\frac{e^{i(q-k)x}}{m_{c}^{2}-k^{2}}\rm{Tr}[\sigma_{\mu\nu}(1\pm\gamma_{5})q^{\nu}(\slashed{k}+\mathit{m_{Q}})(1\pm\gamma_{5})]_{\beta\alpha}\times\langle S(\mathit{p})|\bar{q}_{2\beta} (\emph{x}) q_{1\alpha}(0)|0\rangle.
\label{19}
\end{aligned}
\end{equation}
In the method of QCDSR, the non-vanishing matrix elements for scalar mesons are expanded in terms of distribution amplitudes, as shown in Eq. (\ref{1}).
\par
For the processes of $D/B\rightarrow Sl^{+}\nu_{l}$, one can find, upon tracing operation of Eq. (\ref{19}), that the contribution of twist-3 light cone distribution amplitudes vanishes, left merely with the contribution from the twist-2 LCDA. As a result, the theoretic side of correlation function become
\begin{equation}
\begin{aligned}
\Pi_{1\mu}=&2im_{Q}p_{\mu}\int^{1}_{0}du\frac{\phi_{S}(u)}{m^{2}_{Q}-(q+up)^{2}},\\
\Pi_{2\mu}=&-[2p_{\mu}q^{2}-2q_{\mu}(p\cdot q)]\int^{1}_{0}du\frac{\phi_{S}(u)}{m^{2}_{Q}-(q+up)^{2}}.
\label{ts}
\end{aligned}
\end{equation}
%It is clear that $\phi_{S}(u)$  plays an important role in the FFs of $D/B\rightarrow Sl\bar{\nu}_{l}$ processes.
\par
\subsection{Form factors}
Writing explicitly the correlations Eq.~(\ref{ps}) on the phenomenological side and that given by [Eq.~(\ref{ts})] on theoretical side and matching two sides, one gets
\begin{equation}
\begin{aligned}
-\frac{2if_{+}^{D/B\rightarrow S}(q^{2})p_{\mu}+i[f_{+}^{D/B\rightarrow S}(q^{2})+f_{-}^{D/B\rightarrow S}(q^{2})]q_{\mu}}{m_{D/B}^{2}-(p+q)^{2}}\frac{m^{2}_{D/B}f_{D/B}}{m_{Q}+m_{q_{1}}}=2im_{c}p_{\mu}\int^{1}_{\Delta}du\frac{\phi(u)}{m^{2}_{Q}-(q+up)^{2}},\\
-\frac{2[q^{2}p_{\mu}-q_{\mu}(p\cdot q)]}{m_{D/B}^{2}-(p+q)^{2}}\frac{if_{T}^{D/B\rightarrow S}}{m_{D/B}+m_{S}}\frac{m^{2}_{D/B}f_{D/B}}{m_{Q}+m_{q_{1}}}
=-[2p_{\mu}q^{2}-2q_{\mu}(p\cdot q)]\int^{1}_{\Delta}du\frac{\phi(u)}{m^{2}_{Q}-(q+up)^{2}},\\
\label{21}
\end{aligned}
\end{equation}
where $\Delta$ is one of the solution to the algebraic equation $us-m_{Q}^{2}-u(1-u)m_{S}^{2}+(1-u)q^{2}=0$,
\begin{equation}
\begin{aligned}
\Delta=&\frac{\sqrt{(s_{0}-m_{S}^{2}-q^{2})^{2}+4(m_{c}^{2}-q^{2})m_{S}^{2}}}{2m_{S}^{2}}-\frac{(s_{0}-m_{S}^{2}-q^{2})}{2m_{S}^{2}}.
\end{aligned}
\end{equation}

In the framework of LCSR, one has to suppress the contribution of the high excited as well as the continuum states in Eq. (\ref{21}). To do this, it is common to apply Borel transformation \cite{Colangelo:2000dp} on both sides of Eq. (\ref{21}).  The key relations of the Borel transformation consist of
\begin{equation}
\begin{aligned}
B_{M^2}\frac{1}{m_{D/B}^2-(q+p)^2}&=\frac{1}{M^2}e^{-\frac{m_{D/B}^2}{M^2}},\notag\\
B_{M^2}\frac{1}{m_{Q}^2-(q+up)^2}&=\frac{1}{uM^2}e^{-\frac{m_{Q}^2+u\bar{u}p^2-\bar{u}q^2}{uM^2}},
\end{aligned}
\end{equation}
where $M^2$ is the Borel parameter. Up the Borel transformation on Eq. (\ref{21}), we get the following FFs of $D/B\rightarrow Sl^{+}\nu_{l}$ processes,
\begin{equation}
\begin{aligned}
f_{+}^{D/B\rightarrow S}(q^2)=&-\frac{m_{Q}+m_{q_{1}}}{m_{D/B}^{2}f_{D/B}}m_{Q}\int^{1}_{\Delta}\frac{\phi(u)}{u}due^{FF},\\
f_{-}^{D/B\rightarrow S}(q^2)=&\frac{m_{Q}+m_{q_{1}}}{m_{D/B}^{2}f_{D/B}}m_{Q}\int^{1}_{\Delta}\frac{\phi(u)}{u}due^{FF},\\
f_{T}^{D/B\rightarrow S}(q^2)=&-(m_{D/B}+m_{S})\frac{m_{Q}+m_{q_{1}}}{m_{D/B}^{2}f_{D/B}}\int^{1}_{\Delta}\frac{\phi(u)}{u}due^{FF},
\end{aligned}
\end{equation}
where
\begin{align}
FF:&=-\frac{1}{uM^{2}}(m^{2}_{Q}+u\bar{u}p^{2}-\bar{u}q^{2})+\frac{m^{2}_{D/B}}{M^{2}}.
\end{align}

From the FFs obtained above, one can find a set of simple relation between them,
\begin{equation}
\begin{aligned}
f_{-}^{D/B\rightarrow S}(q^2)&=-f_{+}^{D/B\rightarrow S}(q^2),\\
f_{T}^{D/B\rightarrow S}(q^2)&=\frac{m_{D/B}+m_{S}}{m_{Q}}f_{+}^{D/B\rightarrow S}(q^2).\\
\end{aligned}
\end{equation}
\subsection{Branching fractions}
Given the above FFs, we are in the position to compute the differential decay widths of semileptonic decay~\cite{Huang:1998gp,Ball:1998kk,Yang:2005bv,Wang:2008da,Huang:2008sn,Li:2008tk,Ghahramany:2009zz,Zeng:2013nfa,Han:2013zg,Cheng:2013fba,Meissner:2013hya,Issadykov:2015iba,Zou:2016yhb,Huang:2022xny,Wang:2022yyn}.
% non-leptonic decay~\cite{Li:2012xy,Eeg:2010rk,Li:2013aca,Li:2015zra,Virto:2016fbw,Santorelli:2017fxm,Donini:2017rzi,Vos:2018zqi,Delepine:2019cpp,Zhou:2020bnm,Bediaga:2020qxg,Bell:2020qus,Oset:2020olh,Danilina:2021ukd,Corcella:2021upj,Han:2021gkl,Wang:2022fbk}.
For the process $D/B \rightarrow Sl^{+}\nu_{l}$, which we address in this work, the LCSR enables us to write the differential decay width $d\Gamma/dq^{2}$ in terms of the FFs \cite{Cheng:2017fkw}:
\begin{align}
\frac{d\Gamma (D/B \rightarrow Sl\bar {\nu}_{l})}{dq^{2}}&=\frac{G_{F}^{2}|V_{Qd}|^{2}}{768\pi^{3}m_{D/B}^{3}}\frac{(q^{2}-m_{l}^{2})^{2}}{q^{6}}\sqrt {(m_{D/B}^{2}+m_{S}^{2}-q^{2})^{2}-4m_{D/B}^{2}m_{S}^{2}}\notag\\
&\quad\times \Bigg\{(f_{+}(q^{2}))^{2}\Big[(q^{2}+m_{S}^{2}-m_{D/B}^{2})^{2}(q^{2}+2m_{l}^{2})-q^{2}m_{S}^{2}(4q^{2}+2m_{l}^{2})\Big]\notag\\
&\quad+6f_{+}(q^{2})f_{-}(q^{2})q^{2}m_{l}^{2}\left(m_{D/B}^{2}+m_{S}^{2}-q^{2}\right)+6(f_{-}(q^{2}))^{2}q^{4}m_{l}^{2} \Bigg\},
\label{25}
\end{align}
where $G_{F}$ is the Fermi coupling constant and $|V_{Qd}|$ is Cabibbo-Kabayashi-Maskawa matrix element. Integrating $q^{2}$ over a range of $m_{l}^{2}\leq q^{2}\leq (m_{D}-m_{S})^{2}$, one can obtain the decay width $\Gamma(D/B \rightarrow Sl^{+}\nu_{l}) $ and thereby the decay branching fractions
\begin{equation}
\mathcal {B} (D/B\rightarrow Sl\bar{\nu}_{e})=\frac{\Gamma (D/B \rightarrow Sl\bar {\nu}_{l})}{\Gamma_{total}}, \notag
\end{equation}
where $\Gamma_{total}$ is the total decay width and equal to the inverse lifetime of the $D/B$ mesons:
\begin{equation}
\Gamma_{total}=\frac{1}{\tau_{D/B}}. \notag
\end{equation}

\section{NUMERICAL ANALYSES AND DISCUSSIONS}

\subsection{Choices of input parameters}
\begin{figure}
\centering
\includegraphics[height=6.5cm,width=8.5cm]{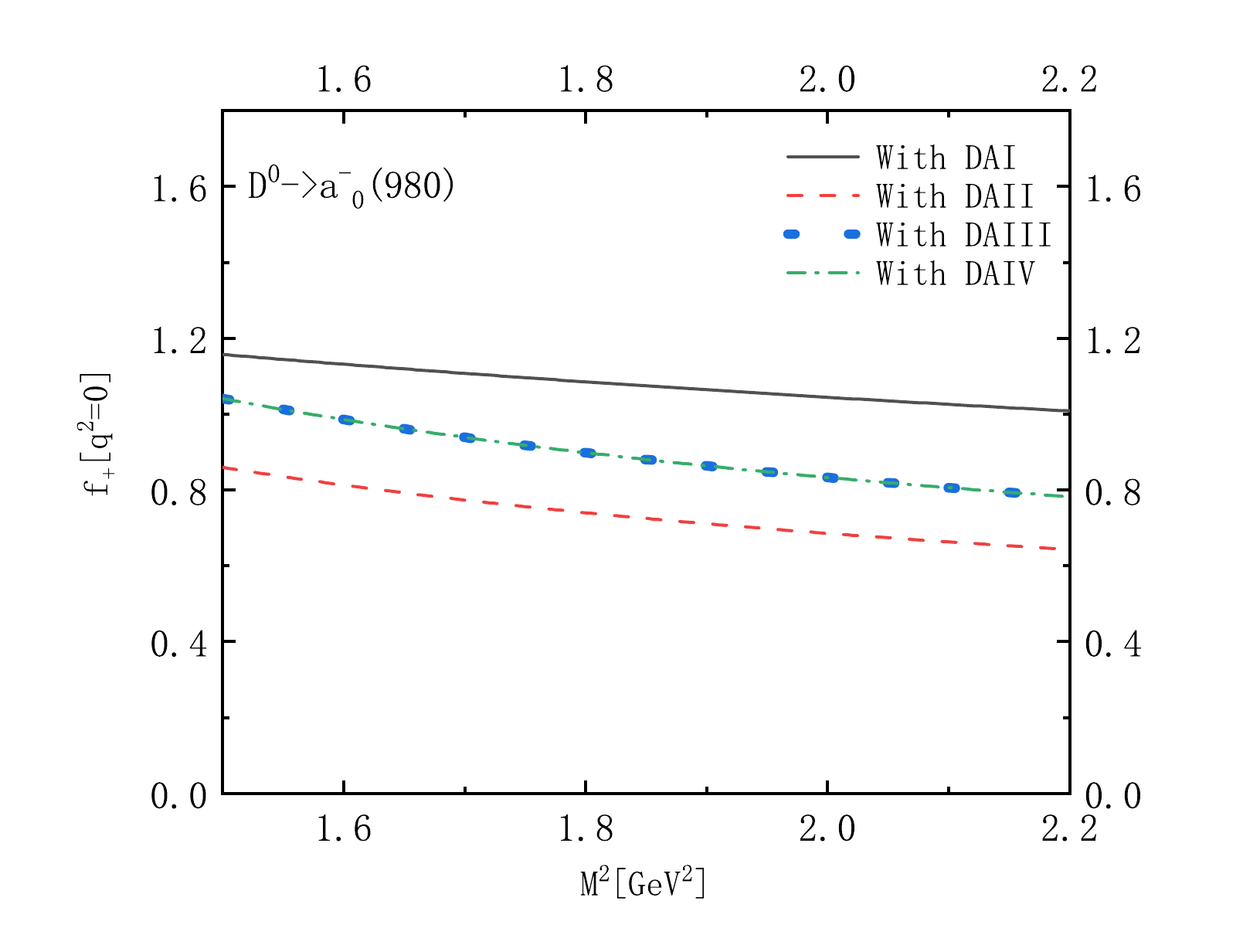}
\caption{The transition FF for $ D^{0}\rightarrow a^{-}_{0}(980)$ as a function of Borel parameter $M^{2}$ at threshold $s_{0}=5.18\ \rm{GeV}^{2}$ and $q^{2}=0$. The four lines of the solid (DA-I), dash (DA-II), dot (DA-III) and dash dot (DA-IV) correspond to the four distribution amplitudes given by different methods, with the two of later lines (DA-III and DA-IV) overlap exactly.}
\label{f5}
\end{figure}

\begin{figure}
\centering
\includegraphics[height=6.5cm,width=8.5cm]{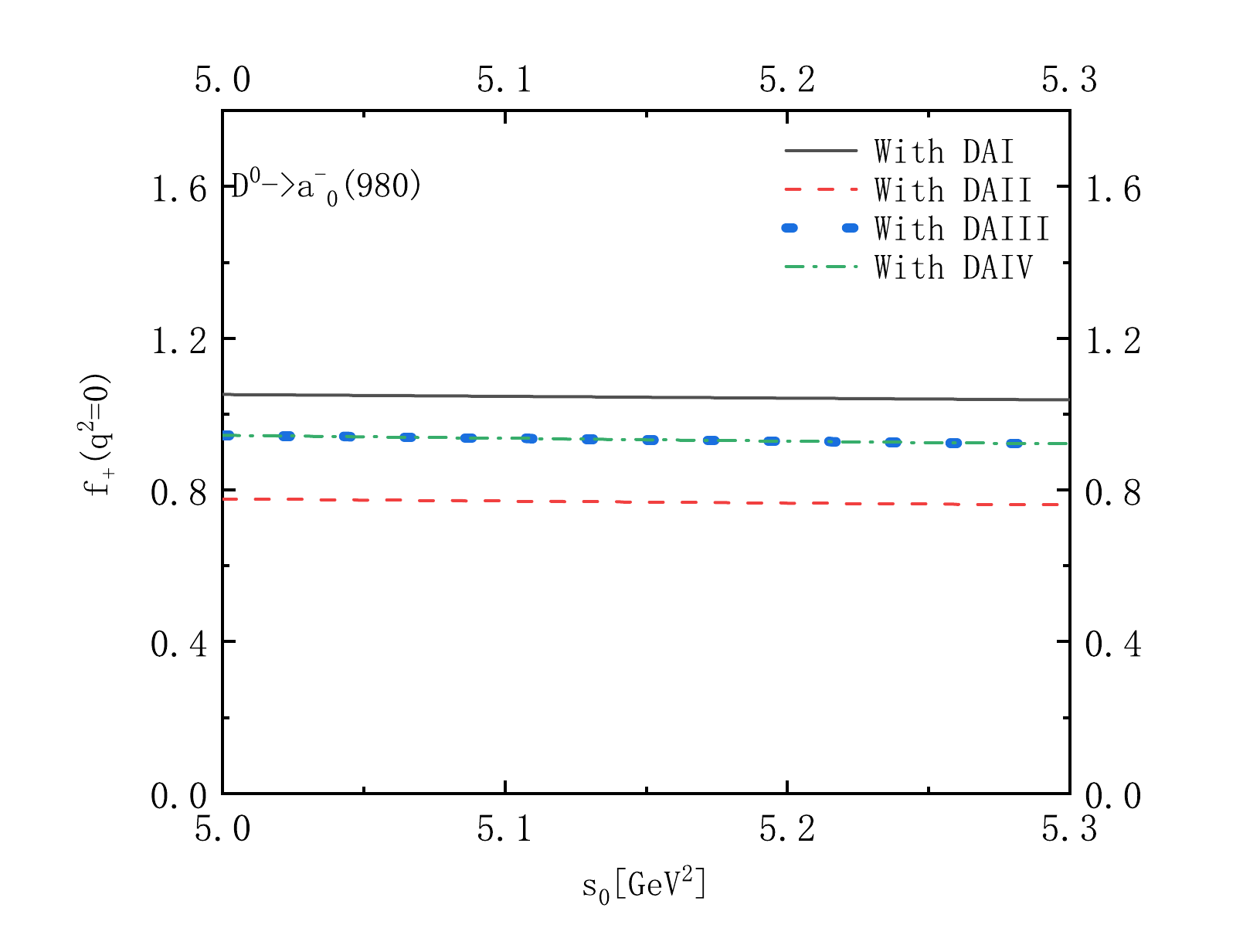}
\caption{The transition FF for $ D^{0}\rightarrow a^{-}_{0}(980)$ versus the threshold $s_{0}$ at the Borel parameter $M^{2}=1.72\ \rm{GeV^{2}}$ and $q^{2}=0$. The four lines of the solid (DA-I), dash (DA-II), dot (DA-III) and dash dot (DA-IV) correspond to the four distribution amplitudes given by different methods, with the two of later lines (DA-III and DA-IV) overlap exactly.}
\label{f6}
\end{figure}

In order to apply the relations in Sect. IV to numerically compute the FFs and branching fractions, one has to fix the parameters involved in the processes of semileptonic decay we consider. As numerical inputs, we employ the values of the parameters given in Refs. \cite{Cheng:2005nb,ParticleDataGroup:2024cfk,HFLAV:2019otj}, which are shown collectively in Table III.
\begin{table}[H]
    \centering
\caption{The values of the parameters for FFs and branching fractions \cite{Cheng:2005nb,ParticleDataGroup:2020ssz,HFLAV:2019otj}}
    \begin{tabular}{c c c c c}
        \hline
        \hline
         Method &  $m_{a_{0}(980)}$ [GeV] & $\bar{f}_{M}$[GeV]&$B_{1}$ $(\mu=1~\rm{GeV})$ &$B_{3}$ $(\mu=1~\rm{GeV}$) \\
        
         DA-I  & \multirow{4}*{0.980}  & \multirow{4}*{0.365}  & -0.93& 0.14 \\

         DA-II  & ~  & ~  & -0.72& -0.01\\

         DA-III  & ~  & ~  &   &  \\

       DA-IV  & ~  & ~  & -0.85& 0.07\\
        \hline
        
        $m_{D^{0}}$ [GeV] &$m_{D^{+}}$[GeV] &$|V_{cd}|$&$\tau_{B^{+}}$[$10^{-12}$s]  &$\tau_{B^{0}}$[$10^{-12}$s]\\
        
        $1.865$ & $1.870$ &$0.220$ &$1.638\pm0.004$ &$1.519\pm0.004$\\
        \hline
        
        $m_{b}$[GeV] & $m_{c}$[GeV] &$|V_{bu}|$ & $f_{D}$[GeV] & $f_{B}$[GeV]\\
        
        $4.18$ &$1.27$ &$0.004$ & $0.2054$ &$0.19$\\
        \hline
        
        $G_{F}[\rm{GeV}^{-2}]$ &$ m_{e}$[GeV] &$ m_{\mu}$[GeV]  &$ \tau_{D^{+}}$[$10^{-12}$s] &$\tau_{D^{0}}$[$10^{-13}$s] \\
        
        $1.1664\times 10^{-5}$ & $0.511\times 10^{-3}$ &$0.106$  &$1.040\pm0.007$  & $4.101\pm0.015$ \\
        \hline
        
        $m_{B^{0}}$[GeV] & $m_{B^{+}}$[GeV] &$m_{u}$[GeV] & $m_{d}$[GeV] & $-$\\
        
        $5.280$ & $5.280$ &$2.16\times 10^{-3}$ &$4.67\times 10^{-3}$ & $-$\\
        \hline
        \hline
    \end{tabular}
\end{table}
We note that the choices of the threshold and the Borel parameter are crucial to find the reasonable results within the LCSR. Normally, the threshold is chosen as squared mass of the first excited state of the $D$ or $B$ meson. However, the such-chosen values of the threshold are not universal for different processes and have to be determined individually. Since FFs should not depend on Borel parameter, one normally constraints the range of the Borel parameter so that
\begin{enumerate}
\item The contributions from the higher excited and continuum states are less than 30\%;
\item The dependence of the FF upon the Borel parameters is weak.
\end{enumerate}
Given the two constraints listed above, one can numerically search and determine the corresponding thresholds and Borel windows for the considered semileptonic decay processes. As such, one is able to use four different methods (DA-I,
DA-II,DA-III and DA-IV) to determine distribution amplitudes of the scalar
meson $a_{0}(980)$ and thereby one can compute respective form factors for the semileptonic decay we consider. The obtained results for the transition FFs for the decay $ D^{0}\rightarrow a^{-}_{0}(980)$ are plotted in FIG.~\ref{f5} and FIG.~\ref{f6}.

One sees from FIG.~\ref{f5} that $f_{+}$  is stable with variation of $M^{2}$ within the range of $1.5\ \rm{GeV}^{2}<M^{2}<2.2\ \rm{GeV}^{2}$. The dependence of the transition FFs for the decay $D^{0}\rightarrow a_{0}^{-}(980)$ on the threshold $s_{0}$ is plotted in FIG.~\ref{f6}. Following Ref \cite{Huang:1998sa}, we take the threshold $s_{0}$ to vary in the range $5\ \rm{GeV}^{2}\leq s_{0}<5.3\ \rm{GeV}{^2}$, which is somehow smaller than the measured threshold $ s_{0}=6.5\  \rm{GeV}{^2}$ reported in experiments \cite{ParticleDataGroup:2020ssz}.
\par
\begin{figure}[H]
\centering
\includegraphics[height=6.5cm,width=8.5cm]{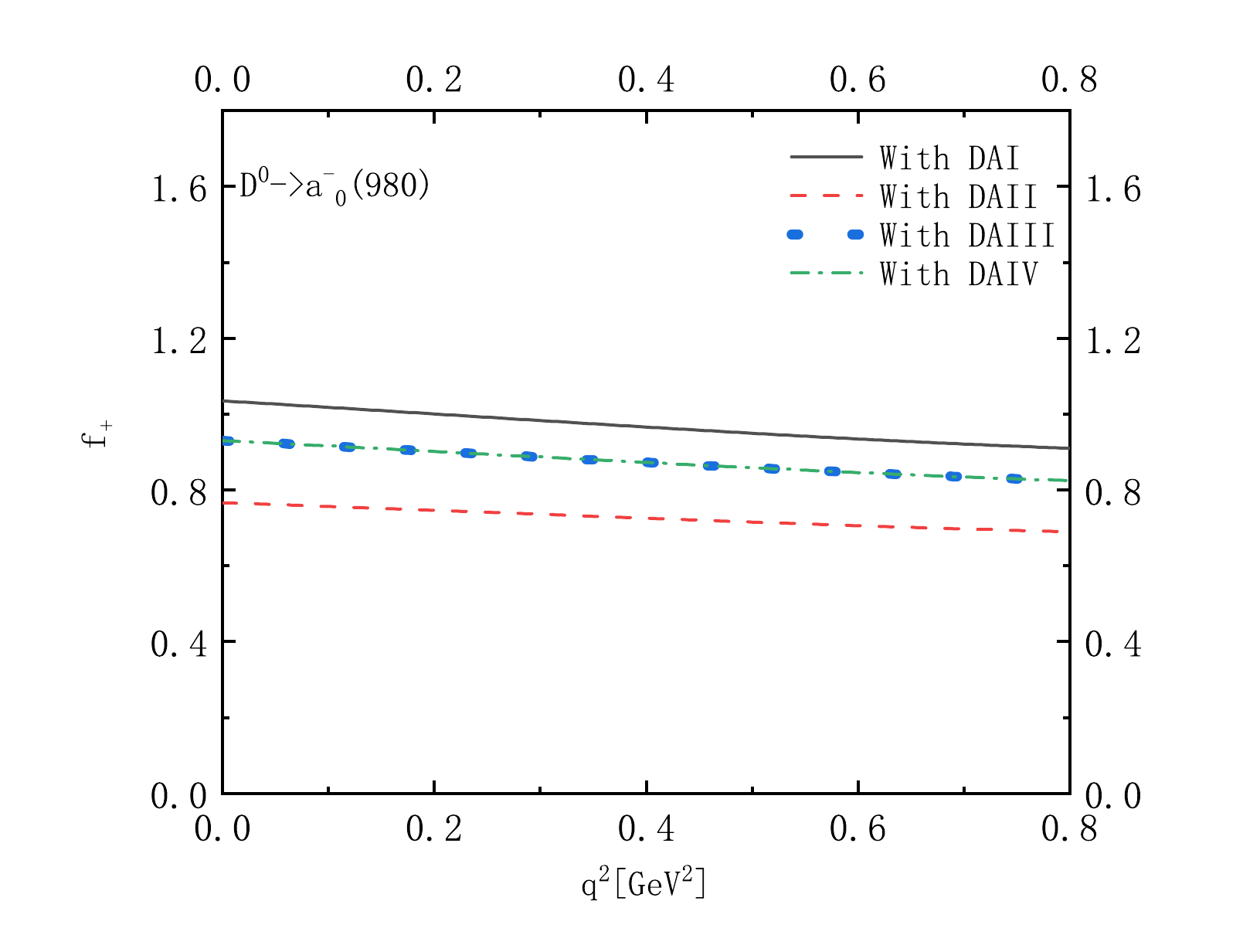}
\caption{The FFs of $D^{0}\rightarrow a_{0}^{-}(980)$ decay as a
function of $q^{2}$ at Borel parameter $M^{2}=1.72\ \rm{GeV}^{2}$ and the threshold $s_{0}=5.18\ \rm{GeV}^{2}$. The four lines of the solid (DA-I), dash (DA-II), dot (DA-III) and dash dot (DA-IV) correspond to the four distribution amplitudes given by different methods, with the two of later lines(DA-III and DA-IV) overlap exactly.}
\label{s1}
\end{figure}

Given the parameter windows for LCSR, one can fix the FFs at $q^{2}=0$ and list them in Table IV, where the obtained results are compared to that by other methods cited for the decays $D/B\rightarrow Sl^{+}\nu_{l}$.
\begin{table*}
\begin{center}
\caption{The form factors at $q^{2}=0$ for the transition $D/B\rightarrow S$ are compared with that via other theoretical methods cited. }
\begin{tabular}{c c c c c c}
\hline
\hline
\multicolumn{2}{c}{Processes}&  \multicolumn{2}{c}{$D\rightarrow a_{0}(980)$} & \multicolumn{2}{c}{$B\rightarrow a_{0}(980)$}  \\
\hline
\multicolumn{2}{c}{Form factor}&$f_{+}$&$f_{-}$&$f_{+}$&$f_{-}$\\
\hline
\multicolumn{2}{c}{LCSR with DA\uppercase\expandafter{\romannumeral1}}&1.03&-1.03&0.66&-0.66\\

\multicolumn{2}{c}{LCSR with DA\uppercase\expandafter{\romannumeral2}}&0.77&-0.77&0.53&-0.53\\

\multicolumn{2}{c}{LCSR with DA\uppercase\expandafter{\romannumeral3}}&0.93&-0.93&0.61&-0.61\\

\multicolumn{2}{c}{LCSR with DA\uppercase\expandafter{\romannumeral4}}&0.93&-0.93&0.61&-0.61\\

\multicolumn{2}{c}{QCDSR \cite{Sun:2010nv}}&$-$&$-$& 0.56&-0.56\\

\multicolumn{2}{c}{pQCD \cite{Li:2008tk}}&$-$&$-$&0.39&$-$\\

\multicolumn{2}{c}{LCSR \cite{Cheng:2017fkw}}&1.75&0.31&$-$&$-$\\

\multicolumn{2}{c}{The covariant confining
quark model \cite{Soni:2020sgn}}&0.55&$-$&$-$&$-$\\
\hline
\hline
\end{tabular}
\end{center}
\end{table*}
Further in FIG.~\ref{s1}, we show dependence of the FF on the transferred momentum $ q^{2}$ with $ m_{l}^{2}\leq q^{2}\leq(m_{D}-m_{S})^{2} $ for the $ D\rightarrow S $ at  $ s_{0}=5.18\ \rm{GeV}^{2} $ and $M^{2}=1.72\ \rm{GeV}^{2}$. As indicated from FIGs. ~\ref{f6} -~\ref{s1} the FFs are not sensitive to the threshold $s_{0}$ and the momentum transfer $q^{2}$.

\subsection{Branching fractions}
\begin{figure}[H]
\centering

\centering
\includegraphics[height=6.5cm,width=8.5cm]{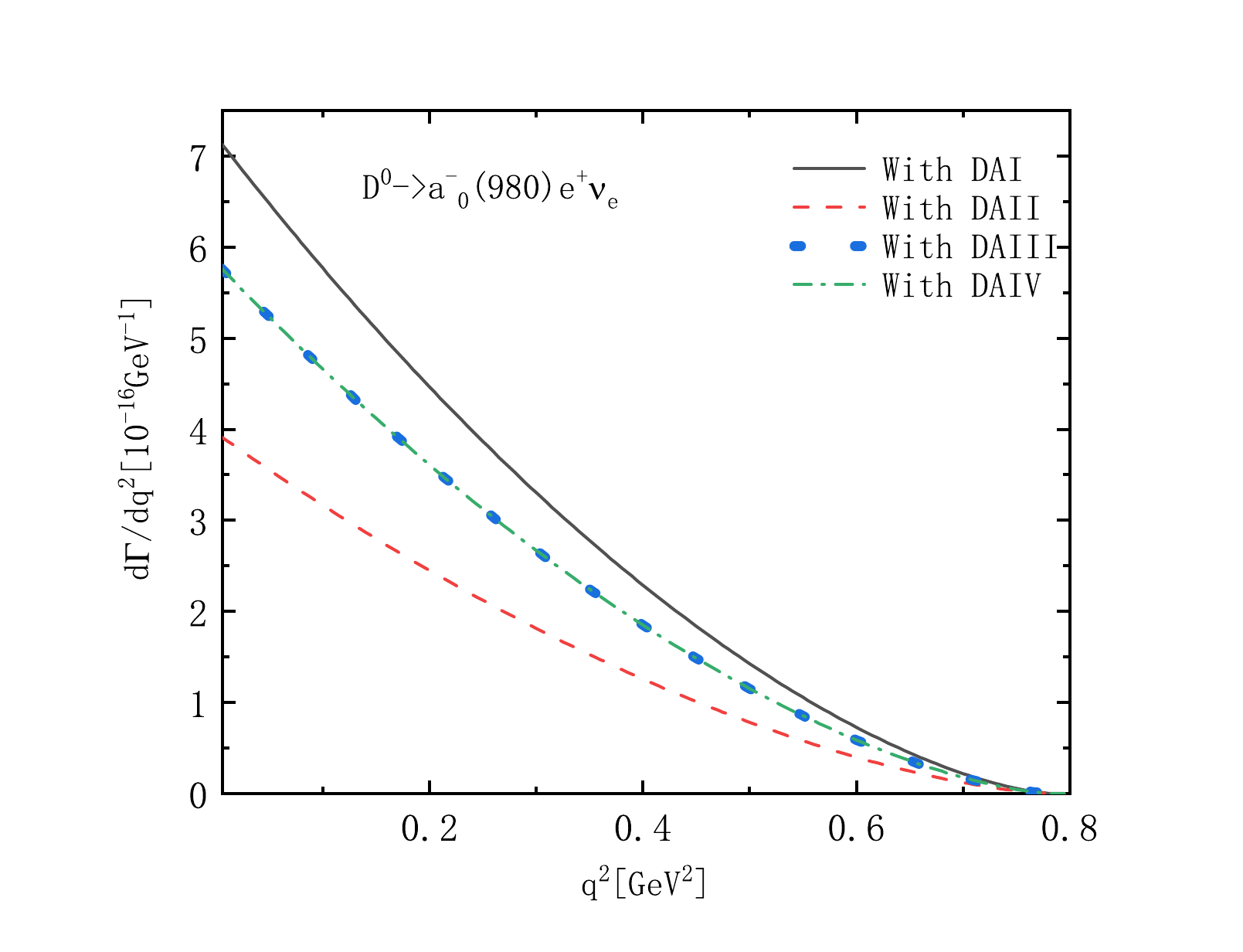}
\caption{The differential decay width of $ D^{0}\rightarrow a_{0}^{-}(980)e^{+}\nu_{e} $ for $m_{e}^{2}\leq q^{2}\leq (m_{D}-m_{a_{0}(980)})^{2}$ by four methods. The four lines of the solid (DA-I), dash (DA-II), dot (DA-III) and dash dot (DA-IV) correspond to the four distribution amplitudes given by different methods, with the two of later lines (DA-III and DA-IV) overlap exactly.}
\label{s2}
\end{figure}

\begin{figure}[H]
\centering

\centering
\includegraphics[height=6.5cm,width=8.5cm]{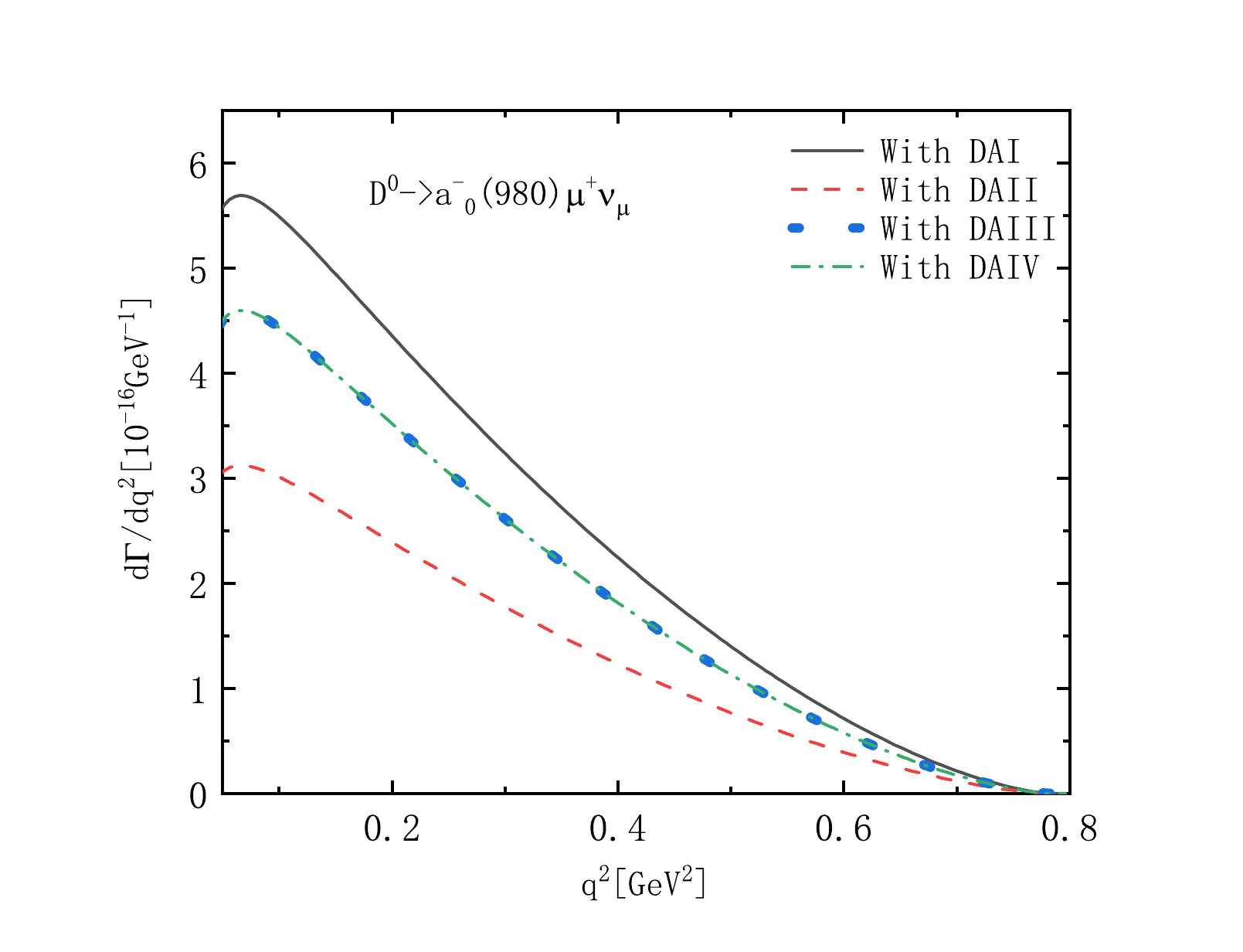}
\caption{$ D^{0}\rightarrow a_{0}^{-}(980)\mu^{+}\nu_{\mu} $ differential decay width at $m_{\mu}^{2}\leq q^{2}\leq (m_{D}-m_{a_{0}(980)})^{2}$ by four methods. The solid (DA-I), dash (DA-II), dot (DA-III) and dash dot (DA-IV) lines correspond to the four distribution amplitudes given by different methods, with the two of later lines(DA-III and DA-IV) overlap exactly.}
\label{s3}
\end{figure}

Application of the FFs obtained in subsect. V-A and the relations in Sect. IV leads to respective decay widths and corresponding branching fractions of the semileptonic decay processes considered in this work. In FIGs.~\ref{s2} and~\ref{s3}, we show the differential decay width versus transfer momentum $q^{2}$ .

\begin{figure}
\centering
\includegraphics[height=6.5cm,width=8.5cm]{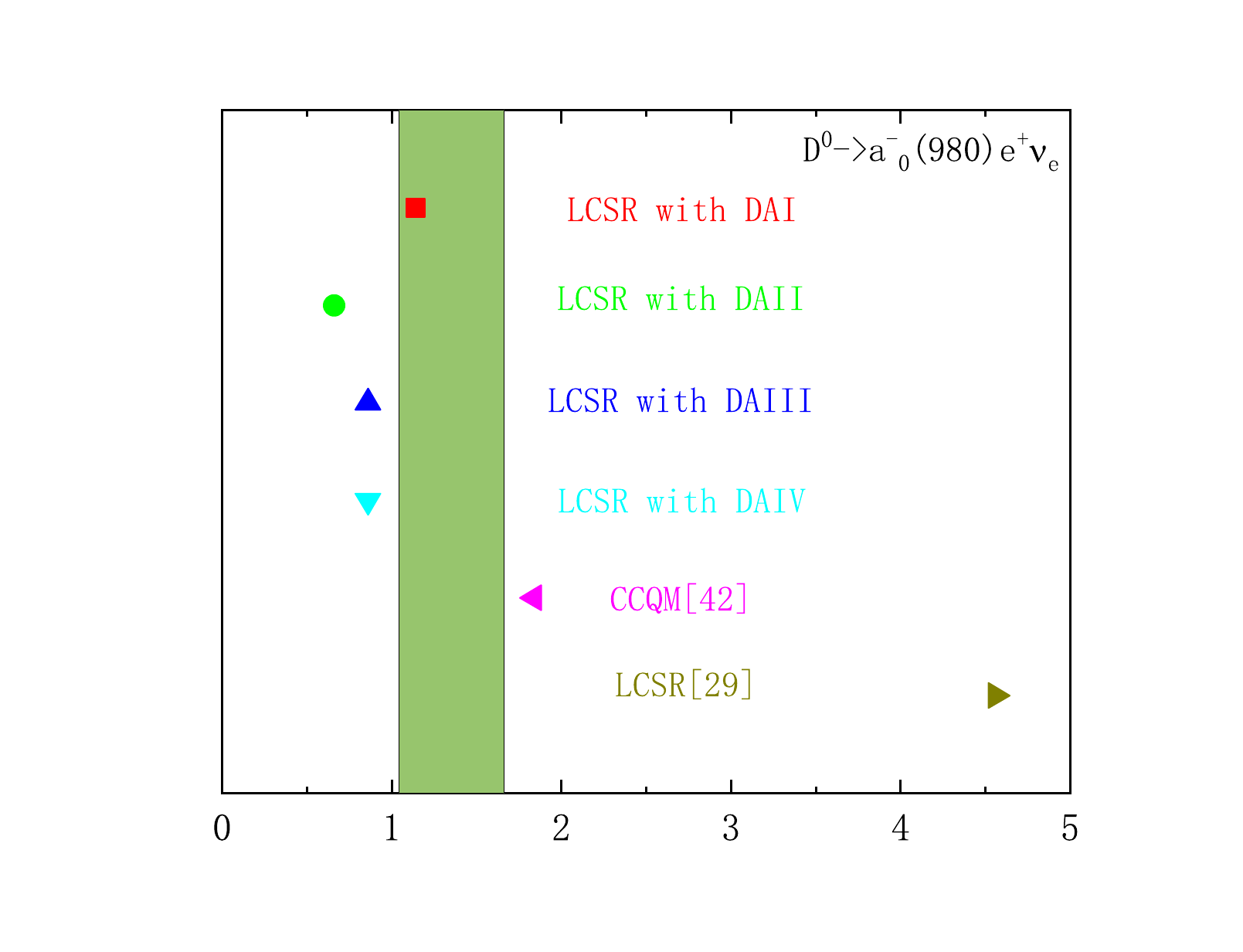}
\caption{Comparison of branching fractions for $ D^{0}\rightarrow a^{-}_{0}(980)e^{+}\nu_{e} $. The green band corresponds to the measurement of BES\uppercase\expandafter{\romannumeral3} \cite{BESIII:2018sjg}.}
\label{c1}
\end{figure}

\begin{figure}
\centering
\includegraphics[height=6.5cm,width=8.5cm]{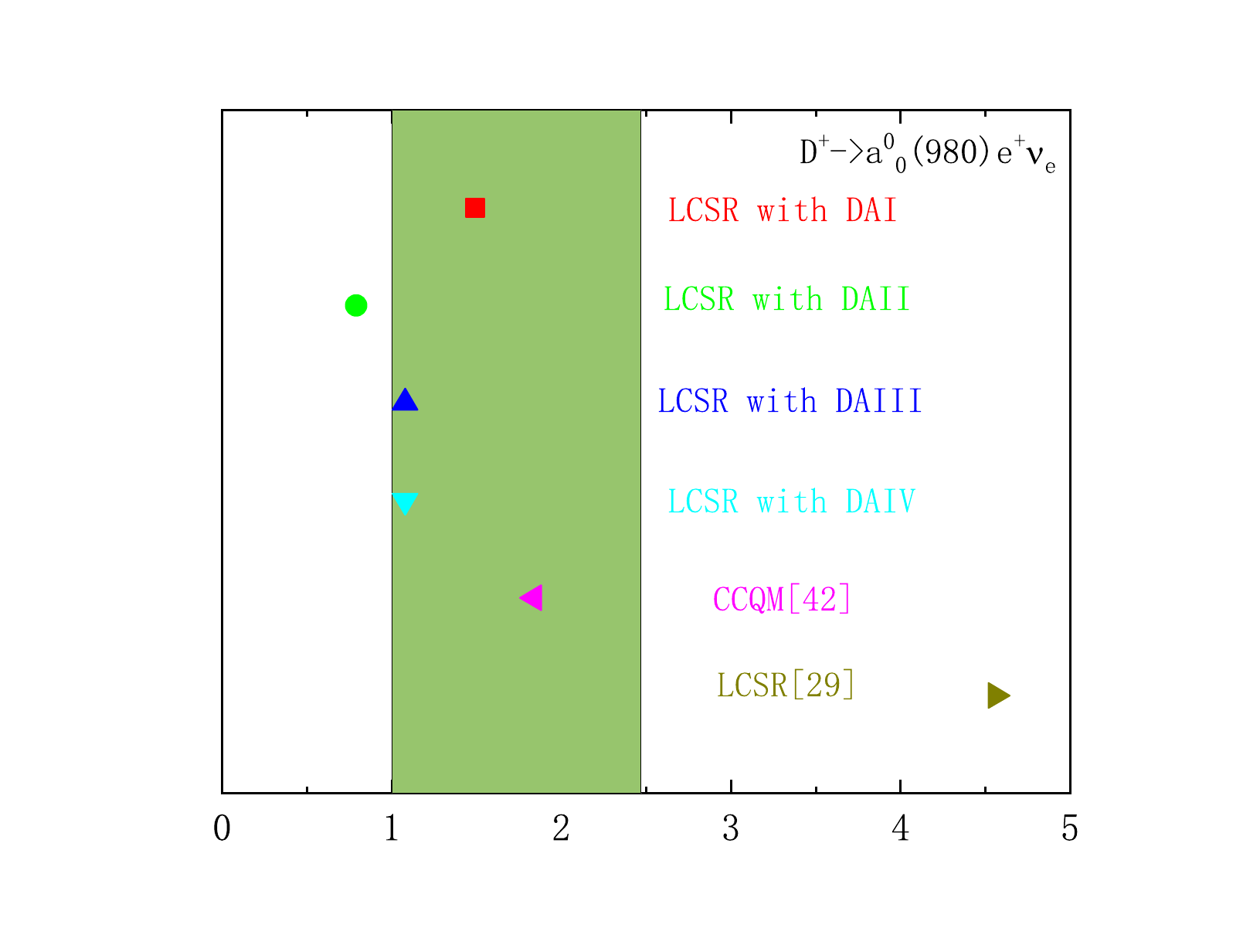}
\caption{Comparison of branching fractions for $ D^{+}\rightarrow a^{0}_{0}(980)e^{+}\nu_{e} $. The green band corresponds to the measurement of BESIII \cite{BESIII:2018sjg}.}
\label{c2}
\end{figure}

\begin{figure}
\centering
\includegraphics[height=6.5cm,width=8.5cm]{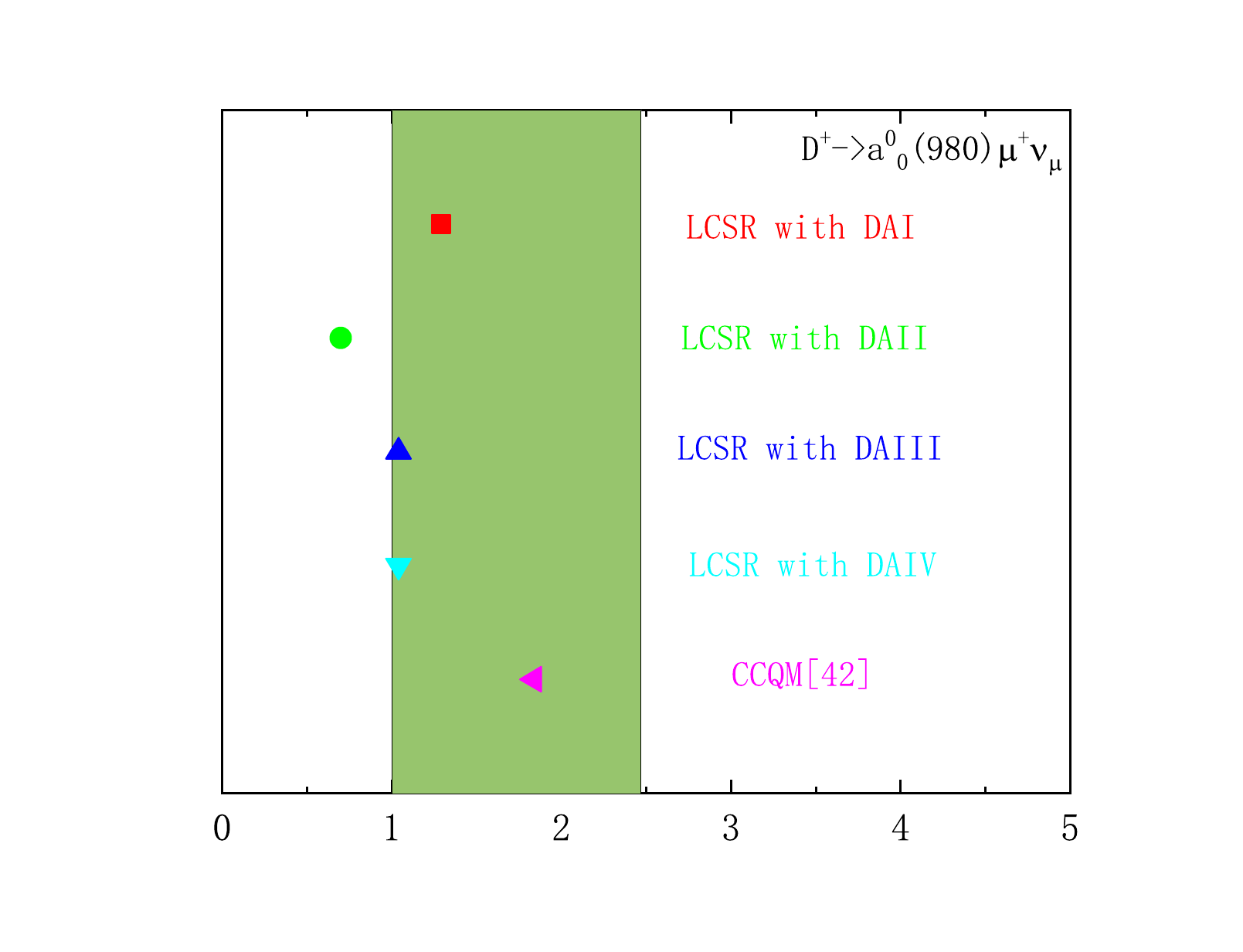}
\caption{Comparison of branching fractions for $ D^{+}\rightarrow a^{0}_{0}(980)\mu^{+}\nu_{\mu} $. The green band corresponds to the measurement of BESIII \cite{BESIII:2018sjg}.}
\label{c3}
\end{figure}

Integrating over a range of $m_{l}^{2}\leq q^{2}\leq (m_{D}-m_{S})^{2}$, we obtain the decay widths and thereby the decay branching fractions, with the numerical results shown in Table \uppercase\expandafter{\romannumeral5}. The branching fractions computed with different methods and their comparisons are plotted in FIG.~\ref{c1} for $ D^{0}\rightarrow a^{-}_{0}(980)e^{+}\nu_{e} $, FIG.~\ref{c2} for $ D^{+}\rightarrow a^{0}_{0}(980)e^{+}\nu_{e} $ and FIG.~\ref{c3} for $ D^{+}\rightarrow a^{0}_{0}(980)\mu^{+}\nu_{\mu} $. The measurement by BES\uppercase\expandafter{\romannumeral3} \cite{BESIII:2018sjg} yields the ratio of two partial widths, $\Gamma(D^{0}\rightarrow a_{0}^{-}(980)e^{+}\nu_{e})/\Gamma(D^{+}\rightarrow a_{0}^{0}(980)e^{+}\nu_{e})$, to be $2.03\pm0.95\pm0.06$, which is in consistent with our predictions for this ratio.
\begin{table*}
    \centering
    \caption{Branching fractions (in unit of $10^{-4}$) in this work compared with that via other theoretical methods cited as well as the experimental values for $D\rightarrow Sl^{+}\nu_{l}$. }
    \begin{tabular}{c|ccc}
    \hline
        \hline
       Process& Method & Branching fractions& $\mathcal {B} (D^{0(+)}\rightarrow Sl\bar{\nu}_{e})\times \mathcal {B} (S\rightarrow \eta\pi)$\\
       \hline
        \multirow{7}*{$ D^{0}\rightarrow a_{0}^{-}(980)e^{+}\nu_{e} $ } &LCSR with DA\uppercase\expandafter{\romannumeral1} & $1.35$&$1.14$\\
        
        ~ &LCSR with DA\uppercase\expandafter{\romannumeral2} & $0.74$&$0.6$\\
        
        ~ &LCSR with DA\uppercase\expandafter{\romannumeral3} & $1.02$&$0.86$\\
        
        ~ &LCSR with DA\uppercase\expandafter{\romannumeral4} & $1.02$&$0.86$\\
        
        ~ & CCQM \cite{Soni:2020sgn} &  $1.68\pm 0.15$&$1.84$\\
        
        ~ & LCSR \cite{Cheng:2017fkw} &  $4.08^{+1.37}_{-1.22}$&$4.56$\\
        
        ~ & BES\uppercase\expandafter{\romannumeral3} \cite{BESIII:2018sjg} & $- $&$1.33^{+0.33}_{-0.29} $\\
       \hline
        \multirow{5}*{$ D^{0}\rightarrow a_{0}^{-}(980)\mu^{+}\nu_{\mu} $ } &LCSR with DA\uppercase\expandafter{\romannumeral1} & $1.20$&$1.01$\\
        
        ~ &LCSR with DA\uppercase\expandafter{\romannumeral2} & $0.66$&$0.55$\\
        
        ~ &LCSR with DA\uppercase\expandafter{\romannumeral3} & $0.97$&$0.82$\\
        
        ~ &LCSR with DA\uppercase\expandafter{\romannumeral4} & $0.97$&$0.82$\\
        
        ~ & CCQM \cite{Soni:2020sgn} &  $1.63\pm 0.14$&$1.84$\\
        \hline
        \multirow{7}*{$ D^{+}\rightarrow a_{0}^{0}(980)e^{+}\nu_{e} $ } &LCSR with DA\uppercase\expandafter{\romannumeral1} & $1.77$&$1.49$\\
        
        ~ &LCSR with DA\uppercase\expandafter{\romannumeral2} & $0.94$&$0.79$\\
        
        ~ &LCSR with DA\uppercase\expandafter{\romannumeral3} & $1.29$&$1.08$\\
        
        ~ &LCSR with DA\uppercase\expandafter{\romannumeral4} & $1.29$&$1.08$\\
        
        ~ & CCQM \cite{Soni:2020sgn} &  $2.18\pm 0.38$&$1.84$\\
        
        ~ & LCSR \cite{Cheng:2017fkw} &  $5.40^{+1.78}_{-1.59}$&$4.56$\\
        
        ~ & BES\uppercase\expandafter{\romannumeral3} \cite{BESIII:2018sjg} & $- $&$1.66^{+0.81}_{-0.66} $\\
        \hline
        \multirow{6}*{$ D^{+}\rightarrow a_{0}^{0}(980)\mu^{+}\nu_{\mu} $ } &LCSR with DA\uppercase\expandafter{\romannumeral1} & $1.51$&$1.29$\\
        
        ~ &LCSR with DA\uppercase\expandafter{\romannumeral2} & $0.83$&$0.70$\\
        
        ~ &LCSR with DA\uppercase\expandafter{\romannumeral3} & $1.22$&$1.04$\\
        
        ~ &LCSR with DA\uppercase\expandafter{\romannumeral4} & $1.22$&$1.04$\\
        
        ~ & CCQM \cite{Soni:2020sgn} &  $2.12\pm 0.37$&$1.84$\\
        
        ~ & BES\uppercase\expandafter{\romannumeral3} \cite{BESIII:2018sjg} & $- $&$1.66^{+0.81}_{-0.66} $\\
    \hline
        \hline
    \end{tabular}\\
\end{table*}
For the decay processes of $ D^{0(+)}/B^{0(+)}\rightarrow a_{0}^{-(0)}(980)l^{+}\nu_{l} $ $(l=e,\mu)$ in Table V, one can easily show that our predictions of branching fractions are well consistent with the measured values by the BESIII \cite{BESIII:2018sjg} within the error range, where we have used the following branching ratio \cite{Cheng:2013fba}
\begin{align}
\mathcal {B} (a_{0}(980)\rightarrow \eta\pi)&= 0.845 \pm 0.017.\notag
\end{align}
Note that the difference of branching fractions between our and other theoretical methods comes mainly from the differences of the FF $f_{+}(q^{2})$, which makes the main contribution to the differential decay width.
\par
\section{Extending the TFF to tetraquark}
The DA is known to quantify the longitudinal momentum fraction carried by valence quarks (or partons) within a meson in the infinite-momentum frame, encoding essential non-perturbative information about its light-cone wavefunction. For conventional two-quark ($q\bar{q}$) configurations  of of a scalar meson, its distribution amplitude exhibits a typical double extremum, which substantially implies the underlying symmetry of the momentum carried by constituent (quark $q$) and anti-consittuent ($\bar{q}$) in meson.  The neson $a_{0}(980)$ typically conforms to this feature, two extremum of its DA happened at the symmetric positions of the center $u_c=0.5$, namely, $u\sim0.25$ and $0.75$, as explicitly demonstrated in FIGs.~\ref{f1}-\ref{phi}. 
\par
In Ref. \cite{Cheng:2005nb}, the DAs of the light mesons are analyzed with QCD factorization and is found to undergo some deformation, that is, the DA of a light meson made of tetraquark (with $q\bar{q}$ insertion) manifests a suppression near the endpoints $u=0$ and $u=1$ and enhances around the mid-region, relative to that for the two-quark mesons. This may be explained by tendency of momentum sharing that, in a tetraquark, constituent (e.g., quark $q$ or antiquark $\bar{q}$) of tetraquark is hard to carry a high momentum (near $u=1$ or $u=0$) but easy to have a averaged momentum due to the sharing of the momentum among all constituents.    
\par
Given the above features in DA, it is very useful to introduce, for simplicity, an extra multiplicative
``modulation function”, to parametrize the additional effect of a tetraquark component mixed in the $a_{0}(980)$ wavefunction. One simple function which demonstrates the deformation mentioned above is $1+0.5(1-e^{-k\frac{(u-0.25)^{2}(u-0.75)^{2}}{0.02}})$, with prefactor $k$ as a parameter. We multiply this function to the four DAs (modulate them) that we have used in sect. II through sect. V, and choose the deformed  DAs as the DA inputs for the tetraquarks with $s\bar{s}$ insertion. Given four modulated DAs, denoted also as DA-I through DA-IV, we apply the LCSR endowed with chiral currents to recompute the FFs of the same semileptonic decays of the $B/D$ mesons, as well for the different values of $k$. The fixed DAs are shown with dash and dot line in Fig.~\ref{f1}-\ref{f4}.  Furthermore, using these modulated DAs as input, we recompute the transition form factors and branching fractions for the semileptonic decays. The results for the branching fractions across a range of $k$ values are summarized in Table VI.

\begin{table}[H]
    \centering
    \caption{Branching fractions (in unit of $10^{-4}$) in this work compared for different values of $k$ for $\mathcal {B} (D\rightarrow a_{0}(980)l\nu_{l} )\times \mathcal {B} (a_{0}\rightarrow \eta\pi)$, with experiment included for comparison. }
    \begin{tabular}{c|ccccc}
\hline
\hline
    Method&Value of $k$ & $D^{0}\rightarrow a_{0}^{-}(980)e^{+}\nu_{e} $& $D^{0}\rightarrow a_{0}^{-}(980)\mu^{+}\nu_{\mu} $&$ D^{+}\rightarrow a_{0}^{0}(980)e^{+}\nu_{e} $ &$ D^{+}\rightarrow a_{0}^{0}(980)\mu^{+}\nu_{\mu} $ \\
    \hline
     \multirow{5}*{DA\uppercase\expandafter{\romannumeral1}}&
     $k=0$ & $1.14$&$1.01$&$1.49$&$1.29$\\
     
    % $k=0.5$ & $1.23$\\
    % \hline
    ~&$k=1$ & $1.29$&$1.25$&$1.69$&$1.46$\\
    
    ~&$k=1.3$ & $1.33$&$1.28$&$1.73$&$1.50$\\
     
    % $k=1.4$ & $1.34$\\
    %  \hline
    % $k=1.5$ & $1.35$\\
    % \hline
    ~&$k=2$ & $1.40$&$1.35$&$1.82$&$1.58$\\
    
    % $k=4.5$ & $1.56$\\
    % \hline
    ~&$k=7$ & $1.66$&$1.60$&$2.17$&$1.88$\\
    \hline
    \multirow{5}*{DA\uppercase\expandafter{\romannumeral2}}&
     $k=0$ & $0.66$&$0.55$&$0.79$&$0.70$\\
     
    % $k=0.5$ & $1.23$\\
    % \hline
    ~&$k=1$ & $0.78$&$0.65$&$0.86$&$0.83$\\
    
    ~&$k=1.3$ & $0.80$&$0.67$&$0.89$&$0.86$\\
     
    % $k=1.4$ & $1.34$\\
    %  \hline
    % $k=1.5$ & $1.35$\\
    % \hline
    ~&$k=2$ & $0.85$&$0.71$&$0.94$&$0.91$\\
    
    % $k=4.5$ & $1.56$\\
    % \hline
    ~&$k=7$ & $1.01$&$0.85$&$1.12$&$1.08$\\
    \hline
    \multirow{5}*{DA\uppercase\expandafter{\romannumeral3}/\uppercase\expandafter{\romannumeral4}}&
     $k=0$ & $0.86$&$0.82$&$1.08$&$1.04$\\
     
    % $k=0.5$ & $1.23$\\
    % \hline
    ~&$k=1$ & $1.00$&$0.94$&$1.24$&$1.20$\\
    
    ~&$k=1.3$ & $1.02$&$0.97$&$1.28$&$1.23$\\
     
    % $k=1.4$ & $1.34$\\
    %  \hline
    % $k=1.5$ & $1.35$\\
    % \hline
    ~&$k=2$ & $1.08$&$1.02$&$1.34$&$1.30$\\
    
    % $k=4.5$ & $1.56$\\
    % \hline
    ~&$k=7$ & $1.29$&$1.22$&$1.61$&$1.55$\\
    
    \hline
    Exp.&BES\uppercase\expandafter{\romannumeral3} \cite{BESIII:2018sjg} & $1.33^{+0.33}_{-0.29} $&-&$1.66^{+0.81}_{-0.66} $&$1.66^{+0.81}_{-0.66} $\\
    \hline
    \hline
    \end{tabular}
\end{table}
\par
% Our results offer strong, albeit indirect, support for the exotic nature of the $a_{0}(980)$. The significant improvement in the agreement with data when moving from the pure two-quark picture to the modulated DA picture indicates that the four-quark component plays a crucial role in its wave function. The LCSR method with chiral currents, combined with a phenomenological model for the tetraquark DA, thus serves as a powerful tool for probing the internal structure of exotic hadrons. Then, the fixed branching fractions computed with different methods and their comparisons are plotted in FIG.~\ref{c1m} for $ D^{0}\rightarrow a^{-}_{0}(980)e^{+}\nu_{e} $, FIG.~\ref{c2m} for $ D^{+}\rightarrow a^{0}_{0}(980)e^{+}\nu_{e} $ and FIG.~\ref{c3m} for $ D^{+}\rightarrow a^{0}_{0}(980)\mu^{+}\nu_{\mu} $.

\begin{figure}
\centering
\includegraphics[height=6.5cm,width=8.5cm]{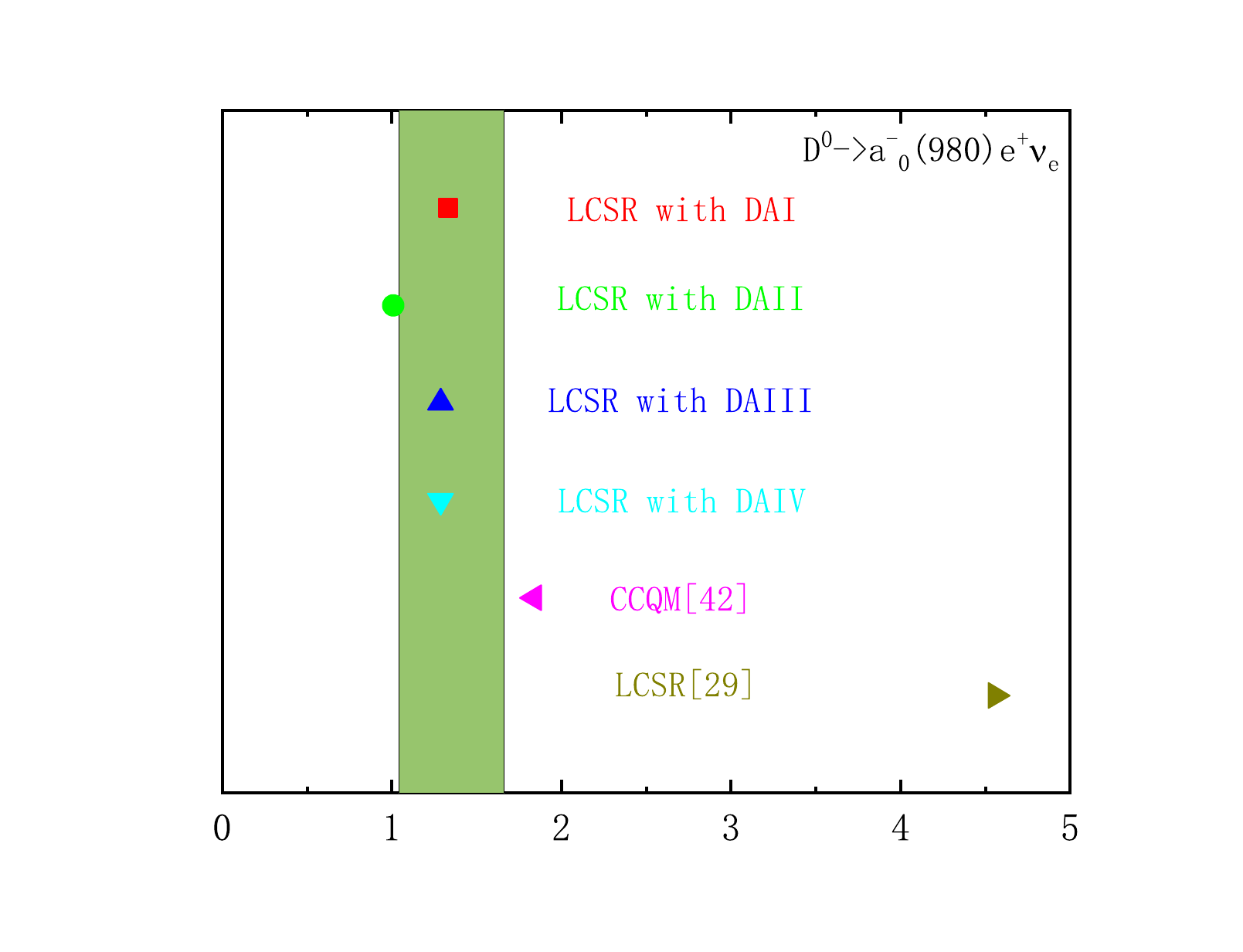}
\caption{Branching fractions for $ D^{0}\rightarrow a^{-}_{0}(980)e^{+}\nu_{e} $ with modulated DAs featuring a four-quark component, with $k=1.3$ for DA I and $k=7$ for DA II-IV. The green band corresponds to the measurement of BES\uppercase\expandafter{\romannumeral3} \cite{BESIII:2018sjg}.}
\label{c1m}
\end{figure}

\begin{figure}
\centering
\includegraphics[height=6.5cm,width=8.5cm]{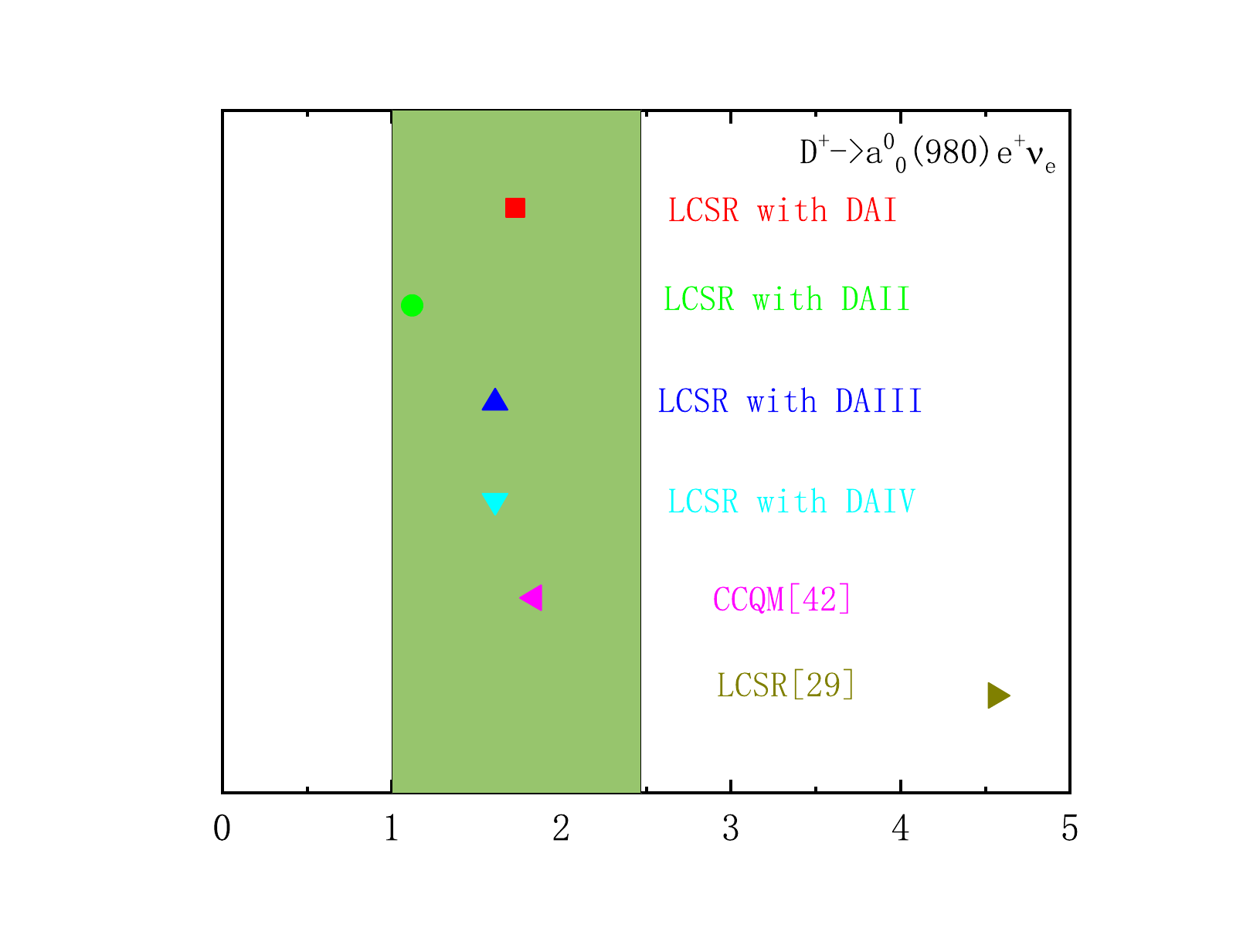}
\caption{Branching fractions for $ D^{+}\rightarrow a^{0}_{0}(980)e^{+}\nu_{e} $  with modulated DAs featuring a four-quark component, with $k=1.3$ for DA I and $k=7$ for DA II-IV. The green band corresponds to the measurement of BESIII \cite{BESIII:2018sjg}.}
\label{c2m}
\end{figure}

\begin{figure}
\centering
\includegraphics[height=6.5cm,width=8.5cm]{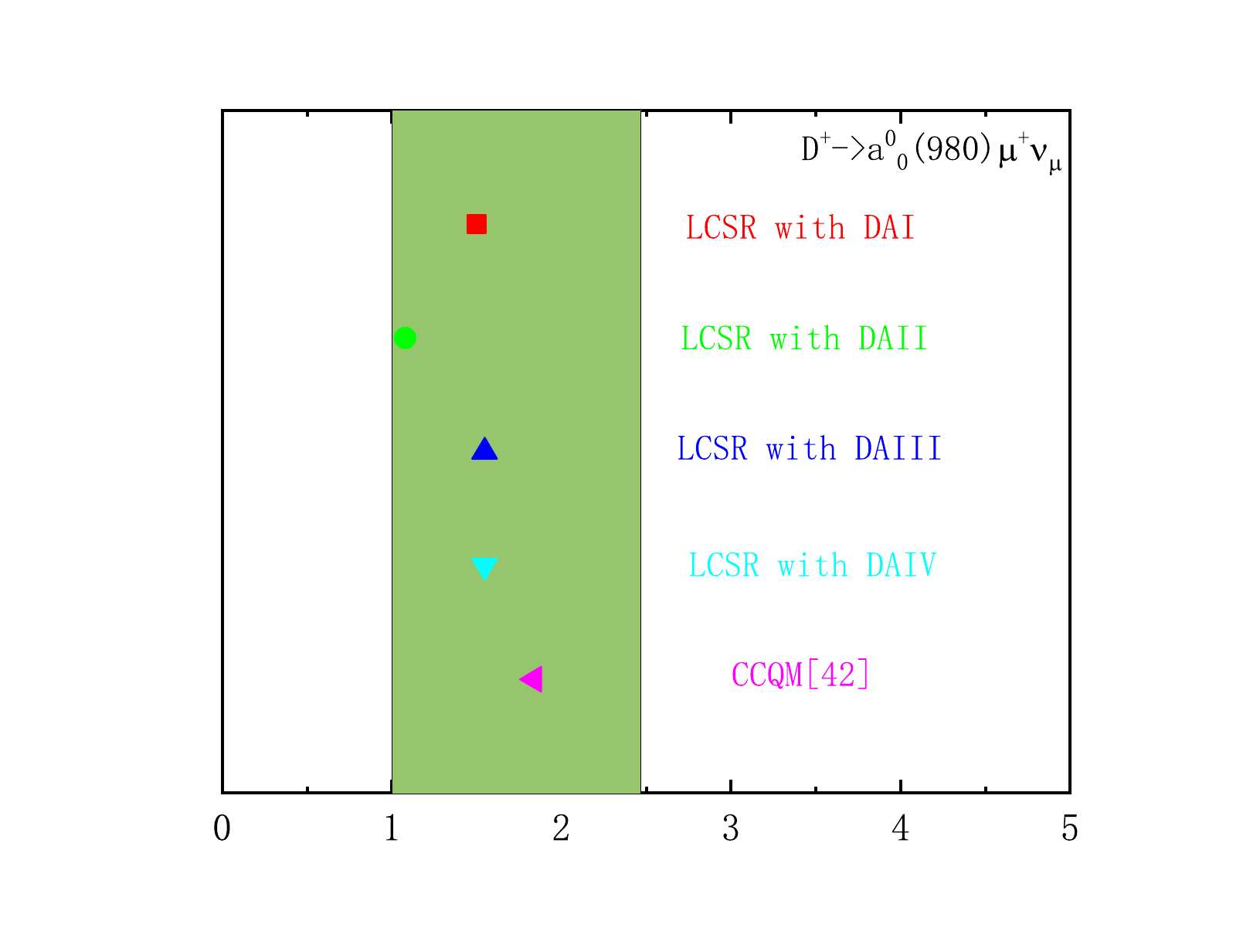}
\caption{Branching fractions for $ D^{+}\rightarrow a^{0}_{0}(980)\mu^{+}\nu_{\mu} $ with modulated DAs featuring a four-quark component, with $k=1.3$ for DA I and $k=7$ for DA II-IV. The green band corresponds to the measurement of BESIII \cite{BESIII:2018sjg}.}
\label{c3m}
\end{figure}

 From Table VI, one observes a continuous transition from DA of the (original) pure $q\bar{q}$ to that of tetraquark mixed configuration and ensuing branching fractions of the $B/D$ decays which increase monotonically with $k$. The obtained branching fractions computed with different methods and their comparisons are plotted in FIG.~\ref{c1m} for $ D^{0}\rightarrow a^{-}_{0}(980)e^{+}\nu_{e} $, FIG.~\ref{c2m} for $ D^{+}\rightarrow a^{0}_{0}(980)e^{+}\nu_{e} $ and FIG.~\ref{c3m} for $ D^{+}\rightarrow a^{0}_{0}(980)\mu^{+}\nu_{\mu} $. From these figures, one sees that a considerable improvement is achieved for extended computation of the TFFs and decay rates applied to the tetraquark-dominated meson in the sense that it captures main features of the BESIII measurements. By the way, our analysis gives strong support for the dominate four-quark structure of the $a_{0}(980)$  which agrees with its measured strong coupling to the $K\bar{K}$.

\section{SUMMARY}

In this work, we systematically study the semileptonic decay process of $ D/B\rightarrow a_{0}l^{+}\nu_{l} $ firstly in the two-quark picture and then in the four-quark pictures of the final state meson $a_0(980)$, with emphasis on the decay form factors that exclusively incorporate the contributions of the leading distribution amplitudes. We eliminate the contributions arising from twist-3 light-cone DAs  with the help of the light-cone QCD sum rules with chiral currents and  find a simple relationships among the form factors associated with the process of $ D/B\rightarrow Sl^{+}\nu_{l} $. An explicit expression of the form factors for $ D/B\rightarrow Sl^{+}\nu_{l} $ is presented in terms of $q^{2}$. The obtained FFs are utilized to predict the branching fractions of the relevant semileptonic decay process and to compare with that via the LCSR without chiral currents  \cite{Cheng:2017fkw} and via other methods. Our numerical computations for the pure $q\bar{q}$ picture suggest that the DAs in the four methods yields nearly same transition FFs and decay rates, with that by the LCSR with DAII smaller about $20-30\%$. The obtained branching fractions are tested against with the observed data of BES III, with the agreement achieved only marginally for the two-quark picture. 

% Furthermore, comparison among four methodologies of determining the FFs is made , which implies that the distribution amplitude based on the BLFQ method (DA\uppercase\expandafter{\romannumeral3}) aligns with the DA expanded into a sum of Gegenbauer polynomials from BLFQ method (DA\uppercase\expandafter{\romannumeral4}). Numerical analysis based on moment calculations implies that all distribution amplitudes expanded in Gegenbauer polynomials turns out to be convergent for scalar mesons. Meanwhile, it is of interest to further explore the semileptonic decays to examine if the improved approaches like BLFQ method are applicable to more processes of the semileptonic decays.
\par
% Finally, we introduce the modulated DA to make smooth transition of the resulted computation to the four-quark scenario. With the introduction of the tetraquark component via DA modulation, the predicted branching fractions increase and achieve agreement with the central values of the BESIII measurements.
We further extend our analysis for the two-quark meson to that of tetraquark configuration with the help of proper shape modulation of the DAs of the $a_0(980)$. The DA of the two-quark meson is deformed to become more steep as that for tetraquark should be under the modulation. As a results,  the transition FFs and branching fraction for the $D/B$ decays are re-computed with the improved  DAs via proper modulation  and found to be  in good agreement with the central data of the experiment by BES III. This behavior provides evidence for a significant four-quark component in the $a_0(980)$, lending further support to its interpretation as a state strongly coupled to the $K\bar{K}$ channel. This gives a further support for the dominate four-quark structure of the $a_0(980)$ which is indicated by its  measured strong coupling to the $K\bar{K}$. We hope future experiments by updated Beijing Electron Positron Collider (BEPC) and other faculties test our predictions of the TFFs and decay branching fractions in this work.

% We note that the explicit $K\bar{K}$ tetraquark mixing is possible in the $a_{0}(980)$ wave function, since the mass of the $a_{0}(980)$ lies very close to the opening of the $K\bar{K}$ channel, to which it strongly couples \cite{Abele:1998qd,OBELIX:2002lhi}. This
% possibility of the $K\bar{K}$ mixing is ignored in our
% computation. In spite of this limitation, our calculation reveals the key
% features of the transition form factors of the $D/B\rightarrow a_{0}(980)$
% and some intrinsic relations between meson form factors, which remains to be valid
% qualitatively, in the sense that the tetraquark mixing effects can be
% partially included in coupling channel effects \cite{Lipkin:1968woy,Tornqvist:1979hx,Kalashnikova:2005ui,Barnes:2007xu,Pennington:2007xr} of the $K\bar{K}$
% intermediate process during formation of the final state $a_{0}(980)$, while the explicit mixing effects beyond the $K\bar{K}$
% coupling channel is expected to be minor. For such a mixing in the pure $q\bar{q} $ state, a comprehensive analysis is
% required to improve the computations of the DA and transition form factors. 

\section*{ACKNOWLEDGMENTS}
D. G acknowledges the support of the China Scholarship Council program (Project ID: 202510530001). D. J is supported by National Natural Science Foundation of China under Grant No. 12165017. Y. S is supported by Natural Science Foundation of China under Grant Nos. 11365018, 11375240 and 11565023. JL is supported by Special Research Assistant Funding Project, Chinese Academy of Sciences, by the Natural Science Foundation of Gansu Province, China, Grant No.23JRRA631, and by National Natural Science Foundation of China, Grant No. 12305095. XZ is supported by new faculty startup funding by the Institute of Modern Physics, Chinese Academy of Sciences, by Key Research Program of Frontier Sciences, Chinese Academy of Sciences, Grant No. ZDBS-LY-7020, by the Natural Science Foundation of Gansu Province, China, Grant No. 20JR10RA067, by the Foundation for Key Talents of Gansu Province, by the Central Funds Guiding the Local Science and Technology Development of Gansu Province and by the Strategic Priority Research Program of the Chinese Academy of Sciences, Grant No. XDB34000000. This research is supported by Gansu International Collaboration and Talents Recruitment Base of Particle Physics (2023-2027), and supported by the International Partnership Program of Chinese Academy of Sciences, Grant No.016GJHZ2022103FN. A portion of the computational resources were also provided by Gansu Computing Center and by Sugon Computing Center in Xi'an.

\section*{Data Availability Statement}
This manuscript has associated data in a data repository. [Authors’ comment: All data included in this manuscript are available up request by contacting the corresponding authors, group (e.g., PDG group) of collaboration or looking into the cited references.]

%\bibliographystyle{unsrt}
%\bibliography{ref}

\begin{thebibliography}{99}

\bibitem{BESIII:2017ylw}
M.~Ablikim \textit{et al.} [BESIII],
%``Analysis of $D^+\to\bar K^0e^+\nu_e$ and $D^+\to\pi^0e^+\nu_e$ semileptonic decays,''
Phys. Rev. D \textbf{96} (2017) no.1, 012002
doi:10.1103/PhysRevD.96.012002
[arXiv:1703.09084 [hep-ex]].


\bibitem{BaBar:2014xzf}
J.~P.~Lees \textit{et al.} [BaBar],
%``Measurement of the $D^0 \to \pi^- e^+ \nu_e$ differential decay branching fraction as a function of $q^2$ and study of form factor parameterizations,''
Phys. Rev. D \textbf{91} (2015) no.5, 052022
doi:10.1103/PhysRevD.91.052022
[arXiv:1412.5502 [hep-ex]].

\bibitem{CLEO:2011ab}
S.~Dobbs \textit{et al.} [CLEO],
%``First Measurement of the Form Factors in the Decays $D^0 \to \rho^- e^+ \nu_e$ and $D^+ \to \rho^0 e^+ \nu_e$,''
Phys. Rev. Lett. \textbf{110}, no.13, 131802 (2013)
doi:10.1103/PhysRevLett.110.131802
[arXiv:1112.2884 [hep-ex]].

%\cite{CLEO:2009svp}
\bibitem{CLEO:2009svp}
D.~Besson \textit{et al.} [CLEO],
%``Improved measurements of D meson semileptonic decays to pi and K mesons,''
Phys. Rev. D \textbf{80} (2009), 032005
doi:10.1103/PhysRevD.80.032005
[arXiv:0906.2983 [hep-ex]].
%200 citations counted in INSPIRE as of 19 Sep 2024

%\cite{Belle:2006idb}
\bibitem{Belle:2006idb}
L.~Widhalm \textit{et al.} [Belle],
%``Measurement of D0 ---\ensuremath{>} pi l nu (Kl nu) Form Factors and Absolute Branching Fractions,''
Phys. Rev. Lett. \textbf{97} (2006), 061804
doi:10.1103/PhysRevLett.97.061804
[arXiv:hep-ex/0604049 [hep-ex]].
%170 citations counted in INSPIRE as of 19 Sep 2024

%\cite{Fu:2018yin}
\bibitem{Fu:2018yin}
H.~B.~Fu, L.~Zeng, R.~L\"u, W.~Cheng and X.~G.~Wu,
%``The $D\to \rho$ semileptonic and radiative decays within the light-cone sum rules,''
Eur. Phys. J. C \textbf{80} (2020) no.3, 194
doi:10.1140/epjc/s10052-020-7758-4
[arXiv:1808.06412 [hep-ph]].
%21 citations counted in INSPIRE as of 19 Sep 2024

%\cite{Faustov:2019mqr}
\bibitem{Faustov:2019mqr}
R.~N.~Faustov, V.~O.~Galkin and X.~W.~Kang,
%``Semileptonic decays of $D$ and $D_s$ mesons in the relativistic quark model,''
Phys. Rev. D \textbf{101} (2020) no.1, 013004
doi:10.1103/PhysRevD.101.013004
[arXiv:1911.08209 [hep-ph]].
%38 citations counted in INSPIRE as of 19 Sep 2024

%\cite{Soni:2018adu}
\bibitem{Soni:2018adu}
N.~R.~Soni, M.~A.~Ivanov, J.~G.~K\"orner, J.~N.~Pandya, P.~Santorelli and C.~T.~Tran,
%``Semileptonic $D_{(s)}$-meson decays in the light of recent data,''
Phys. Rev. D \textbf{98} (2018) no.11, 114031
doi:10.1103/PhysRevD.98.114031
[arXiv:1810.11907 [hep-ph]].
%50 citations counted in INSPIRE as of 19 Sep 2024

%\cite{Jia:2017age}
\bibitem{Jia:2017age}
D.~Jia, C.~Q.~Pang and A.~Hosaka,
%``Mass Formula for Light Nonstrange Mesons and Regge Trajectories in Quark Model,''
Int. J. Mod. Phys. A \textbf{32} (2017) no.25, 1750153
doi:10.1142/S0217751X17501536
[arXiv:1706.02788 [hep-ph]].
%11 citations counted in INSPIRE as of 19 Sep 2024

%\cite{Li:2012gr}
\bibitem{Li:2012gr}
Z.~H.~Li, N.~Zhu, X.~J.~Fan and T.~Huang,
%``Form Factors $f^{B\to \pi}_+(0)$ and $f^{D\to \pi}_+(0)$ in $QCD$ and Determination of $|V_{ub}|$ and $|V_{cd}|$,''
JHEP \textbf{05} (2012), 160
doi:10.1007/JHEP05(2012)160
[arXiv:1206.0091 [hep-ph]].
%28 citations counted in INSPIRE as of 19 Sep 2024

%\cite{Khodjamirian:2009ys}
\bibitem{Khodjamirian:2009ys}
A.~Khodjamirian, C.~Klein, T.~Mannel and N.~Offen,
%``Semileptonic charm decays D ---\ensuremath{>} pi l nu(l) and D ---\ensuremath{>} K l nu(l) from QCD Light-Cone Sum Rules,''
Phys. Rev. D \textbf{80} (2009), 114005
doi:10.1103/PhysRevD.80.114005
[arXiv:0907.2842 [hep-ph]].
%90 citations counted in INSPIRE as of 19 Sep 2024

%\cite{Ball:2006yd}
\bibitem{Ball:2006yd}
P.~Ball,
%``Testing QCD sum rules on the light-cone in D ---\ensuremath{>}(pi,K) l nu decays,''
Phys. Lett. B \textbf{641} (2006), 50-56
doi:10.1016/j.physletb.2006.08.038
[arXiv:hep-ph/0608116 [hep-ph]].
%50 citations counted in INSPIRE as of 19 Sep 2024

%\cite{Fajfer:2005ug}
\bibitem{Fajfer:2005ug}
S.~Fajfer and J.~F.~Kamenik,
%``Charm meson resonances and D ---\ensuremath{>} V semileptonic form-factors,''
Phys. Rev. D \textbf{72} (2005), 034029
doi:10.1103/PhysRevD.72.034029
[arXiv:hep-ph/0506051 [hep-ph]].
%65 citations counted in INSPIRE as of 19 Sep 2024

%\cite{Ball:1993tp}
\bibitem{Ball:1993tp}
P.~Ball,
%``The Semileptonic decays D ---\ensuremath{>} pi (rho) e neutrino and B ---\ensuremath{>} pi (rho) e neutrino from QCD sum rules,''
Phys. Rev. D \textbf{48} (1993), 3190-3203
doi:10.1103/PhysRevD.48.3190
[arXiv:hep-ph/9305267 [hep-ph]].
%200 citations counted in INSPIRE as of 19 Sep 2024

%\cite{Khodjamirian:2000ds}
\bibitem{Khodjamirian:2000ds}
A.~Khodjamirian, R.~Ruckl, S.~Weinzierl, C.~W.~Winhart and O.~I.~Yakovlev,
%``Predictions on B ---\ensuremath{>} pi anti-l nu(l), D ---\ensuremath{>} pi anti-l nu(l) and D ---\ensuremath{>} K anti-l nu(l) from QCD light cone sum rules,''
Phys. Rev. D \textbf{62} (2000), 114002
doi:10.1103/PhysRevD.62.114002
[arXiv:hep-ph/0001297 [hep-ph]].
%208 citations counted in INSPIRE as of 19 Sep 2024

%\cite{Huang:2008sn}
\bibitem{Huang:2008sn}
T.~Huang, Z.~H.~Li and F.~Zuo,
%``Heavy-to-light transition form factors and their relations in light-cone QCD sum rules,''
Eur. Phys. J. C \textbf{60} (2009), 63-71
doi:10.1140/epjc/s10052-008-0855-4
[arXiv:0809.0130 [hep-ph]].
%10 citations counted in INSPIRE as of 19 Sep 2024

%\cite{ParticleDataGroup:2014cgo}
\bibitem{ParticleDataGroup:2014cgo}
K.~A.~Olive \textit{et al.} [Particle Data Group],
%``Review of Particle Physics,''
Chin. Phys. C \textbf{38} (2014), 090001
doi:10.1088/1674-1137/38/9/090001
%9262 citations counted in INSPIRE as of 19 Sep 2024

%\cite{BaBar:2006wcl}
\bibitem{BaBar:2006wcl}
B.~Aubert \textit{et al.} [BaBar],
%``Measurements of CP-Violating Asymmetries in $B^0 \to$ a+-(1) (1260) $\pi^\mp$ decays,''
Phys. Rev. Lett. \textbf{98} (2007), 181803
doi:10.1103/PhysRevLett.98.181803
[arXiv:hep-ex/0612050 [hep-ex]].
%72 citations counted in INSPIRE as of 19 Sep 2024

%\cite{BaBar:2006xju}
\bibitem{BaBar:2006xju}
B.~Aubert \textit{et al.} [BaBar],
%``Observation of B0 Meson Decay to a+-(1)(1260) pi-+,''
Phys. Rev. Lett. \textbf{97} (2006), 051802
doi:10.1103/PhysRevLett.97.051802
[arXiv:hep-ex/0603050 [hep-ex]].
%70 citations counted in INSPIRE as of 19 Sep 2024

%\cite{BaBar:2007dwy}
\bibitem{BaBar:2007dwy}
B.~Aubert \textit{et al.} [BaBar],
%``Evidence for charged B meson decays to a+-(1)(1260) pi0 and a0(1)(1260) pi+-,''
Phys. Rev. Lett. \textbf{99} (2007), 261801
doi:10.1103/PhysRevLett.99.261801
[arXiv:0708.0050 [hep-ex]].
%23 citations counted in INSPIRE as of 19 Sep 2024


%\cite{BaBar:2008ozy}
\bibitem{BaBar:2008ozy}
B.~Aubert \textit{et al.} [BaBar],
%``Observation and Polarization Measurements of $B^\pm \to \phi K_{1}^\pm$ and $B^\pm \to \phi K_{2}^{*\pm}$,''
Phys. Rev. Lett. \textbf{101} (2008), 161801
doi:10.1103/PhysRevLett.101.161801
[arXiv:0806.4419 [hep-ex]].
%45 citations counted in INSPIRE as of 19 Sep 2024

%\cite{Palmer:2013yia}
\bibitem{Palmer:2013yia}
T.~Palmer and J.~O.~Eeg,
%``Form factors for semileptonic D decays,''
Phys. Rev. D \textbf{89} (2014) no.3, 034013
doi:10.1103/PhysRevD.89.034013
[arXiv:1306.0365 [hep-ph]].
%13 citations counted in INSPIRE as of 19 Sep 2024

%\cite{Bernard:1991bz}
\bibitem{Bernard:1991bz}
C.~W.~Bernard, A.~X.~El-Khadra and A.~Soni,
%``Lattice study of semileptonic decays of charm mesons into vector mesons,''
Phys. Rev. D \textbf{45} (1992), 869-874
doi:10.1103/PhysRevD.45.869
%100 citations counted in INSPIRE as of 19 Sep 2024

%\cite{FermilabLattice:2019ycs}
\bibitem{FermilabLattice:2019ycs}
R.~Li \textit{et al.} [Fermilab Lattice and MILC],
%``$D$ meson Semileptonic Decay Form Factors at $q^2 = 0$,''
PoS \textbf{LATTICE2018} (2019), 269
doi:10.22323/1.334.0269
[arXiv:1901.08989 [hep-lat]].
%16 citations counted in INSPIRE as of 19 Sep 2024

%\cite{Wang:2002zba}
\bibitem{Wang:2002zba}
W.~Y.~Wang, Y.~L.~Wu and M.~Zhong,
%``Heavy to light meson exclusive semileptonic decays in effective field theory of heavy quarks,''
Phys. Rev. D \textbf{67} (2003), 014024
doi:10.1103/PhysRevD.67.014024
[arXiv:hep-ph/0205157 [hep-ph]].
%49 citations counted in INSPIRE as of 19 Sep 2024

%\cite{Wu:2006rd}
\bibitem{Wu:2006rd}
Y.~L.~Wu, M.~Zhong and Y.~B.~Zuo,
%``B(s), D(s) ---\ensuremath{>} pi, K, eta, rho, K*, omega, phi Transition Form Factors and Decay Rates with Extraction of the CKM parameters |V(ub)|, |V(cs)|, |V(cd)|,''
Int. J. Mod. Phys. A \textbf{21} (2006), 6125-6172
doi:10.1142/S0217751X06033209
[arXiv:hep-ph/0604007 [hep-ph]].
%104 citations counted in INSPIRE as of 19 Sep 2024

%\cite{Dinh:2020inx}
\bibitem{Dinh:2020inx}
S.~Q.~Dinh and H.~M.~Tran,
%``Muon g-2 and semileptonic B decays in the B\'elanger-Delaunay-Westhoff model with gauge kinetic mixing,''
Phys. Rev. D \textbf{104} (2021) no.11, 115009
doi:10.1103/PhysRevD.104.115009
[arXiv:2011.07182 [hep-ph]].
%5 citations counted in INSPIRE as of 19 Sep 2024

%\cite{DAlise:2022ypp}
\bibitem{DAlise:2022ypp}
A.~D'Alise, G.~De Nardo, M.~G.~Di Luca, G.~Fabiano, D.~Frattulillo, G.~Gaudino, D.~Iacobacci, M.~Merola, F.~Sannino and P.~Santorelli, \textit{et al.}
%``Standard model anomalies: lepton flavour non-universality, g \ensuremath{-} 2 and W-mass,''
JHEP \textbf{08} (2022), 125
doi:10.1007/JHEP08(2022)125
[arXiv:2204.03686 [hep-ph]].
%19 citations counted in INSPIRE as of 19 Sep 2024

%\cite{Cheng:2017fkw}
\bibitem{Cheng:2017fkw}
X.~D.~Cheng, H.~B.~Li, B.~Wei, Y.~G.~Xu and M.~Z.~Yang,
%``Study of $\mathbf{\boldsymbol{D \rightarrow a_0 (980) e^+ \nu_e}}$ decay in the light-cone sum rules approach,''
Phys. Rev. D \textbf{96} (2017) no.3, 033002
doi:10.1103/PhysRevD.96.033002
[arXiv:1706.01019 [hep-ph]].
%18 citations counted in INSPIRE as of 19 Sep 2024

%\cite{Momeni:2022gqb}
\bibitem{Momeni:2022gqb}
S.~Momeni and M.~Saghebfar,
%``Semileptonic $D$ meson decays to the vector, axial vector and scalar mesons in Hard-Wall AdS/QCD correspondence,''
Eur. Phys. J. C \textbf{82} (2022) no.5, 473
doi:10.1140/epjc/s10052-022-10413-x
%7 citations counted in INSPIRE as of 19 Sep 2024

%\cite{Feng:2022inv}
\bibitem{Feng:2022inv}
J.~L.~Feng, F.~Kling, M.~H.~Reno, J.~Rojo, D.~Soldin, L.~A.~Anchordoqui, J.~Boyd, A.~Ismail, L.~Harland-Lang and K.~J.~Kelly, \textit{et al.}
%``The Forward Physics Facility at the High-Luminosity LHC,''
J. Phys. G \textbf{50} (2023) no.3, 030501
doi:10.1088/1361-6471/ac865e
[arXiv:2203.05090 [hep-ex]].
%262 citations counted in INSPIRE as of 19 Sep 2024

%\cite{BESIII:2021drk}
\bibitem{BESIII:2021drk}
M.~Ablikim \textit{et al.} [BESIII],
%``Study of light scalar mesons through $D^+_s \to \pi^0\pi^0e^+ \nu_e$ and $K^0_S K^0_S e^+ \nu_e$ decays,''
Phys. Rev. D \textbf{105} (2022) no.3, L031101
doi:10.1103/PhysRevD.105.L031101
[arXiv:2110.13994 [hep-ex]].
%7 citations counted in INSPIRE as of 19 Sep 2024

%\cite{Wang:2022vga}
\bibitem{Wang:2022vga}
Z.~Q.~Wang, X.~W.~Kang, J.~A.~Oller and L.~Zhang,
%``Analysis on the composite nature of the light scalar mesons f0(980) and a0(980),''
Phys. Rev. D \textbf{105} (2022) no.7, 074016
doi:10.1103/PhysRevD.105.074016
[arXiv:2201.00492 [hep-ph]].
%14 citations counted in INSPIRE as of 19 Sep 2024

%\cite{Cheng:2022vbw}
\bibitem{Cheng:2022vbw}
H.~Y.~Cheng, C.~W.~Chiang and Z.~Q.~Zhang,
%``Hadronic three-body D decays mediated by scalar resonances,''
Phys. Rev. D \textbf{105} (2022) no.3, 033006
doi:10.1103/PhysRevD.105.033006
[arXiv:2201.00460 [hep-ph]].
%14 citations counted in INSPIRE as of 19 Sep 2024

%\cite{DiCarlo:2021dzg}
\bibitem{DiCarlo:2021dzg}
M.~Di Carlo, G.~Martinelli, M.~Naviglio, F.~Sanfilippo, S.~Simula and L.~Vittorio,
%``Unitarity bounds for semileptonic decays in lattice QCD,''
Phys. Rev. D \textbf{104} (2021) no.5, 054502
doi:10.1103/PhysRevD.104.054502
[arXiv:2105.02497 [hep-lat]].
%45 citations counted in INSPIRE as of 19 Sep 2024

%\cite{Chakraborty:2021qav}
\bibitem{Chakraborty:2021qav}
B.~Chakraborty \textit{et al.} [(HPQCD Collaboration)\textsection{} and HPQCD],
%``Improved Vcs determination using precise lattice QCD form factors for D\textrightarrow{}K\ensuremath{\ell}\ensuremath{\nu},''
Phys. Rev. D \textbf{104} (2021) no.3, 034505
doi:10.1103/PhysRevD.104.034505
[arXiv:2104.09883 [hep-lat]].
%33 citations counted in INSPIRE as of 19 Sep 2024

%\cite{Huang:2021owr}
\bibitem{Huang:2021owr}
Q.~Huang, Y.~J.~Sun, D.~Gao, G.~H.~Zhao, B.~Wang and W.~Hong,
%``Study of form factors and branching ratios for $D\rightarrow S,Al\bar{\nu_{l}}$ with light-cone sum rules,''
[arXiv:2102.12241 [hep-ph]].
%6 citations counted in INSPIRE as of 19 Sep 2024

%\cite{Becirevic:2020rzi}
\bibitem{Becirevic:2020rzi}
D.~Be\v{c}irevi\'c, F.~Jaffredo, A.~Pe\~nuelas and O.~Sumensari,
%``New Physics effects in leptonic and semileptonic decays,''
JHEP \textbf{05} (2021), 175
doi:10.1007/JHEP05(2021)175
[arXiv:2012.09872 [hep-ph]].
%41 citations counted in INSPIRE as of 19 Sep 2024

%\cite{Parrott:2020vbe}
\bibitem{Parrott:2020vbe}
W.~G.~Parrott, C.~Bouchard, C.~T.~H.~Davies and D.~Hatton,
%``Toward accurate form factors for $B$-to-light meson decay from lattice QCD,''
Phys. Rev. D \textbf{103} (2021) no.9, 094506
doi:10.1103/PhysRevD.103.094506
[arXiv:2010.07980 [hep-lat]].
%9 citations counted in INSPIRE as of 19 Sep 2024

%\cite{Gherardi:2020qhc}
\bibitem{Gherardi:2020qhc}
V.~Gherardi, D.~Marzocca and E.~Venturini,
%``Low-energy phenomenology of scalar leptoquarks at one-loop accuracy,''
JHEP \textbf{01} (2021), 138
doi:10.1007/JHEP01(2021)138
[arXiv:2008.09548 [hep-ph]].
%100 citations counted in INSPIRE as of 19 Sep 2024

%\cite{Achasov:2020qfx}
\bibitem{Achasov:2020qfx}
N.~N.~Achasov, A.~V.~Kiselev and G.~N.~Shestakov,
%``Semileptonic decays $D\to\pi^+\pi^-e^+\nu_e$ and $D_s\to\pi^+\pi^-e^+\nu_e$ as the probe of constituent quark-antiquark pairs in the light scalar mesons,''
Phys. Rev. D \textbf{102} (2020) no.1, 016022
doi:10.1103/PhysRevD.102.016022
[arXiv:2005.06455 [hep-ph]].
%17 citations counted in INSPIRE as of 19 Sep 2024

%\cite{Soni:2020sgn}
\bibitem{Soni:2020sgn}
N.~R.~Soni, A.~N.~Gadaria, J.~J.~Patel and J.~N.~Pandya,
%``Semileptonic Decays of Charmed Mesons to Light Scalar Mesons,''
Phys. Rev. D \textbf{102} (2020) no.1, 016013
doi:10.1103/PhysRevD.102.016013
[arXiv:2001.10195 [hep-ph]].
%25 citations counted in INSPIRE as of 19 Sep 2024

%\cite{Fleischer:2019wlx}
\bibitem{Fleischer:2019wlx}
R.~Fleischer, R.~Jaarsma and G.~Koole,
%``Testing Lepton Flavour Universality with (Semi)-Leptonic $D_{(s)}$ Decays,''
Eur. Phys. J. C \textbf{80} (2020) no.2, 153
doi:10.1140/epjc/s10052-020-7702-7
[arXiv:1912.08641 [hep-ph]].
%19 citations counted in INSPIRE as of 19 Sep 2024

%\cite{Liu:2019tsi}
\bibitem{Liu:2019tsi}
K.~Liu [BESIII],
%``Semileptonic and leptonic $D$ decays at BESIII,''
PoS \textbf{LeptonPhoton2019} (2019), 046
doi:10.22323/1.367.0046
%0 citations counted in INSPIRE as of 19 Sep 2024

%\cite{ParticleDataGroup:2016lqr}
\bibitem{ParticleDataGroup:2016lqr}
C.~Patrignani \textit{et al.} [Particle Data Group],
%``Review of Particle Physics,''
Chin. Phys. C \textbf{40} (2016) no.10, 100001
doi:10.1088/1674-1137/40/10/100001
%7044 citations counted in INSPIRE as of 19 Sep 2024

%\cite{BESIII:2018sjg}
\bibitem{BESIII:2018sjg}
M.~Ablikim \textit{et al.} [BESIII],
%``Observation of the Semileptonic Decay $D^0 \to a_0(980)^- e^+ \nu_e$ and Evidence for $D^+ \to a_0(980)^0 e^+ \nu_e$,''
Phys. Rev. Lett. \textbf{121} (2018) no.8, 081802
doi:10.1103/PhysRevLett.121.081802
[arXiv:1803.02166 [hep-ex]].
%43 citations counted in INSPIRE as of 19 Sep 2024

%\cite{Wang:2009azc}
\bibitem{Wang:2009azc}
W.~Wang and C.~D.~Lu,
%``Distinguishing two kinds of scalar mesons from heavy meson decays,''
Phys. Rev. D \textbf{82} (2010), 034016
doi:10.1103/PhysRevD.82.034016
[arXiv:0910.0613 [hep-ph]].
%54 citations counted in INSPIRE as of 19 Sep 2024

%\cite{BESIII:2021tfk}
\bibitem{BESIII:2021tfk}
M.~Ablikim \textit{et al.} [BESIII],
%``Search for the decay $D_s^+\to a_0(980)^0e^+\nu_e$,''
Phys. Rev. D \textbf{103} (2021) no.9, 092004
doi:10.1103/PhysRevD.103.092004
[arXiv:2103.11855 [hep-ex]].
%12 citations counted in INSPIRE as of 19 Sep 2024

%\cite{Cheng:2005nb}
\bibitem{Cheng:2005nb}
H.~Y.~Cheng, C.~K.~Chua and K.~C.~Yang,
%``Charmless hadronic B decays involving scalar mesons: Implications to the nature of light scalar mesons,''
Phys. Rev. D \textbf{73} (2006), 014017
doi:10.1103/PhysRevD.73.014017
[arXiv:hep-ph/0508104 [hep-ph]].
%228 citations counted in INSPIRE as of 19 Sep 2024

%\cite{Hong:2024rhg}
\bibitem{Hong:2024rhg}
W.~Hong, D.~Gao and Y.~Sun,
%``Twist-2 distribution amplitudes of a0(980) and a0(1450),''
Nucl. Phys. A \textbf{1064} (2025), 123219
doi:10.1016/j.nuclphysa.2025.123219
[arXiv:2409.15776 [hep-ph]].
%1 citations counted in INSPIRE as of 23 Mar 2026

%\cite{Vary:2009gt}
\bibitem{Vary:2009gt}
J.~P.~Vary, H.~Honkanen, J.~Li, P.~Maris, S.~J.~Brodsky, A.~Harindranath, G.~F.~de Teramond, P.~Sternberg, E.~G.~Ng and C.~Yang,
%``Hamiltonian light-front field theory in a basis function approach,''
Phys. Rev. C \textbf{81} (2010), 035205
doi:10.1103/PhysRevC.81.035205
[arXiv:0905.1411 [nucl-th]].
%195 citations counted in INSPIRE as of 19 Sep 2024

%\cite{Xu:2021wwj}
\bibitem{Xu:2021wwj}
S.~Xu \textit{et al.} [BLFQ],
%``Nucleon structure from basis light-front quantization,''
Phys. Rev. D \textbf{104} (2021) no.9, 094036
doi:10.1103/PhysRevD.104.094036
[arXiv:2108.03909 [hep-ph]].
%58 citations counted in INSPIRE as of 19 Sep 2024

%\cite{Zhao:2014xaa}
\bibitem{Zhao:2014xaa}
X.~Zhao, H.~Honkanen, P.~Maris, J.~P.~Vary and S.~J.~Brodsky,
%``Electron g-2 in Light-Front Quantization,''
Phys. Lett. B \textbf{737} (2014), 65-69
doi:10.1016/j.physletb.2014.08.020
[arXiv:1402.4195 [nucl-th]].
%75 citations counted in INSPIRE as of 19 Sep 2024

%\cite{Wiecki:2014ola}
\bibitem{Wiecki:2014ola}
P.~Wiecki, Y.~Li, X.~Zhao, P.~Maris and J.~P.~Vary,
%``Basis Light-Front Quantization Approach to Positronium,''
Phys. Rev. D \textbf{91} (2015) no.10, 105009
doi:10.1103/PhysRevD.91.105009
[arXiv:1404.6234 [nucl-th]].
%91 citations counted in INSPIRE as of 19 Sep 2024

%\cite{Lockman:1989qa}
\bibitem{Lockman:1989qa}
W.~S.~Lockman [MARK-III],
%``PRODUCTION OF THE f0 (975) MESON IN J / psi DECAYS,''
Conf. Proc. C \textbf{890923} (1989), 109-118
SLAC-PUB-5139.
%3 citations counted in INSPIRE as of 19 Sep 2024

%\cite{LucioMartinez:1990uw}
\bibitem{LucioMartinez:1990uw}
J.~L.~Lucio Martinez and J.~Pestieau,
%``On the Phi ---\ensuremath{>} K0 anti-K0 gamma decay,''
Phys. Rev. D \textbf{42} (1990), 3253-3254
doi:10.1103/PhysRevD.42.3253
%79 citations counted in INSPIRE as of 19 Sep 2024

%\cite{MARK-III:1990wgk}
\bibitem{MARK-III:1990wgk}
Z.~Bai \textit{et al.} [MARK-III],
%``Partial wave analysis of J / psi ---\ensuremath{>} gamma K0(s) K+- pi-+,''
Phys. Rev. Lett. \textbf{65} (1990), 2507-2510
doi:10.1103/PhysRevLett.65.2507
%137 citations counted in INSPIRE as of 19 Sep 2024

%\cite{DM2:1990cwz}
\bibitem{DM2:1990cwz}
J.~E.~Augustin \textit{et al.} [DM2],
%``Partial wave analysis of DM2 data in the eta (1430) energy range,''
Phys. Rev. D \textbf{46} (1992), 1951-1958
doi:10.1103/PhysRevD.46.1951
%73 citations counted in INSPIRE as of 19 Sep 2024

%\cite{Heusch:1991sw}
\bibitem{Heusch:1991sw}
C.~A.~Heusch,
%``Status of gluonium searches,''
CERN-PPE-91-225.
%1 citations counted in INSPIRE as of 19 Sep 2024

%\cite{Tornqvist:1995kr}
\bibitem{Tornqvist:1995kr}
N.~A.~Tornqvist,
%``Understanding the scalar meson q anti-q nonet,''
Z. Phys. C \textbf{68} (1995), 647-660
doi:10.1007/BF01565264
[arXiv:hep-ph/9504372 [hep-ph]].
%490 citations counted in INSPIRE as of 19 Sep 2024

%\cite{Jaffe:1976ig}
\bibitem{Jaffe:1976ig}
R.~L.~Jaffe,
%``Multi-Quark Hadrons. 1. The Phenomenology of (2 Quark 2 anti-Quark) Mesons,''
Phys. Rev. D \textbf{15} (1977), 267
doi:10.1103/PhysRevD.15.267
%2246 citations counted in INSPIRE as of 19 Sep 2024

%\cite{Weinstein:1982gc}
\bibitem{Weinstein:1982gc}
J.~D.~Weinstein and N.~Isgur,
%``Do Multi-Quark Hadrons Exist?,''
Phys. Rev. Lett. \textbf{48} (1982), 659
doi:10.1103/PhysRevLett.48.659
%590 citations counted in INSPIRE as of 19 Sep 2024

%\cite{Weinstein:1983gd}
\bibitem{Weinstein:1983gd}
J.~D.~Weinstein and N.~Isgur,
%``The q q anti-q anti-q System in a Potential Model,''
Phys. Rev. D \textbf{27} (1983), 588
doi:10.1103/PhysRevD.27.588
%620 citations counted in INSPIRE as of 19 Sep 2024

%\cite{Achasov:1987ts}
\bibitem{Achasov:1987ts}
N.~N.~Achasov and V.~N.~Ivanchenko,
%``On a Search for Four Quark States in Radiative Decays of phi Meson,''
Nucl. Phys. B \textbf{315} (1989), 465-476
doi:10.1016/0550-3213(89)90364-7
%335 citations counted in INSPIRE as of 19 Sep 2024

%\cite{Weinstein:1990gu}
\bibitem{Weinstein:1990gu}
J.~D.~Weinstein and N.~Isgur,
%``K anti-K Molecules,''
Phys. Rev. D \textbf{41} (1990), 2236
doi:10.1103/PhysRevD.41.2236
%759 citations counted in INSPIRE as of 19 Sep 2024

%\cite{Scadron:2003yg}
\bibitem{Scadron:2003yg}
M.~D.~Scadron, G.~Rupp, F.~Kleefeld and E.~van Beveren,
%``Ground state scalar anti-q q nonet: SU(3) mass splittings and strong, electromagnetic, and weak decay rates,''
Phys. Rev. D \textbf{69} (2004), 014010
[erratum: Phys. Rev. D \textbf{69} (2004), 059901]
doi:10.1103/PhysRevD.69.014010
[arXiv:hep-ph/0309109 [hep-ph]].
%54 citations counted in INSPIRE as of 19 Sep 2024

%\cite{Jaffe:2004ph}
\bibitem{Jaffe:2004ph}
R.~L.~Jaffe,
%``Exotica,''
Phys. Rept. \textbf{409} (2005), 1-45
doi:10.1016/j.physrep.2004.11.005
[arXiv:hep-ph/0409065 [hep-ph]].
%625 citations counted in INSPIRE as of 19 Sep 2024

%\cite{Maiani:2004uc}
\bibitem{Maiani:2004uc}
L.~Maiani, F.~Piccinini, A.~D.~Polosa and V.~Riquer,
%``A New look at scalar mesons,''
Phys. Rev. Lett. \textbf{93} (2004), 212002
doi:10.1103/PhysRevLett.93.212002
[arXiv:hep-ph/0407017 [hep-ph]].
%429 citations counted in INSPIRE as of 19 Sep 2024

%\cite{Kalashnikova:2004ta}
\bibitem{Kalashnikova:2004ta}
Y.~S.~Kalashnikova, A.~E.~Kudryavtsev, A.~V.~Nefediev, C.~Hanhart and J.~Haidenbauer,
%``The Radiative decays phi ---\ensuremath{>} gamma a(0) / f(0) in the molecular model for the scalar mesons,''
Eur. Phys. J. A \textbf{24} (2005), 437-443
doi:10.1140/epja/i2005-10008-4
[arXiv:hep-ph/0412340 [hep-ph]].
%71 citations counted in INSPIRE as of 19 Sep 2024

%\cite{Sugiyama:2007sg}
\bibitem{Sugiyama:2007sg}
J.~Sugiyama, T.~Nakamura, N.~Ishii, T.~Nishikawa and M.~Oka,
%``Mixings of 4-quark components in light non-singlet scalar mesons in QCD sum rules,''
Phys. Rev. D \textbf{76} (2007), 114010
doi:10.1103/PhysRevD.76.114010
[arXiv:0707.2533 [hep-ph]].
%48 citations counted in INSPIRE as of 19 Sep 2024

%\cite{Kojo:2008hk}
\bibitem{Kojo:2008hk}
T.~Kojo and D.~Jido,
%``Sigma meson in pole-dominated QCD sum rules,''
Phys. Rev. D \textbf{78} (2008), 114005
doi:10.1103/PhysRevD.78.114005
[arXiv:0802.2372 [hep-ph]].
%41 citations counted in INSPIRE as of 19 Sep 2024

%\cite{tHooft:2008rus}
\bibitem{tHooft:2008rus}
G.~'t Hooft, G.~Isidori, L.~Maiani, A.~D.~Polosa and V.~Riquer,
%``A Theory of Scalar Mesons,''
Phys. Lett. B \textbf{662} (2008), 424-430
doi:10.1016/j.physletb.2008.03.036
[arXiv:0801.2288 [hep-ph]].
%275 citations counted in INSPIRE as of 19 Sep 2024

%\cite{Alexandrou:2017itd}
\bibitem{Alexandrou:2017itd}
C.~Alexandrou, J.~Berlin, M.~Dalla Brida, J.~Finkenrath, T.~Leontiou and M.~Wagner,
%``Lattice QCD investigation of the structure of the $a_0(980)$ meson,''
Phys. Rev. D \textbf{97} (2018) no.3, 034506
doi:10.1103/PhysRevD.97.034506
[arXiv:1711.09815 [hep-lat]].
%18 citations counted in INSPIRE as of 19 Sep 2024

%\cite{Amsler:1995tu}
\bibitem{Amsler:1995tu}
C.~Amsler and F.~E.~Close,
%``Evidence for a scalar glueball,''
Phys. Lett. B \textbf{353} (1995), 385-390
doi:10.1016/0370-2693(95)00579-A
[arXiv:hep-ph/9505219 [hep-ph]].
%319 citations counted in INSPIRE as of 19 Sep 2024

%\cite{Oller:1998zr}
\bibitem{Oller:1998zr}
J.~A.~Oller and E.~Oset,
%``N/D description of two meson amplitudes and chiral symmetry,''
Phys. Rev. D \textbf{60} (1999), 074023
doi:10.1103/PhysRevD.60.074023
[arXiv:hep-ph/9809337 [hep-ph]].
%655 citations counted in INSPIRE as of 19 Sep 2024

%\cite{Lee:1999kv}
\bibitem{Lee:1999kv}
W.~J.~Lee and D.~Weingarten,
%``Scalar quarkonium masses and mixing with the lightest scalar glueball,''
Phys. Rev. D \textbf{61} (2000), 014015
doi:10.1103/PhysRevD.61.014015
[arXiv:hep-lat/9910008 [hep-lat]].
%285 citations counted in INSPIRE as of 19 Sep 2024

%\cite{Jamin:2000wn}
\bibitem{Jamin:2000wn}
M.~Jamin, J.~A.~Oller and A.~Pich,
%``S wave K pi scattering in chiral perturbation theory with resonances,''
Nucl. Phys. B \textbf{587} (2000), 331-362
doi:10.1016/S0550-3213(00)00479-X
[arXiv:hep-ph/0006045 [hep-ph]].
%267 citations counted in INSPIRE as of 19 Sep 2024

%\cite{Giacosa:2006tf}
\bibitem{Giacosa:2006tf}
F.~Giacosa,
%``Mixing of scalar tetraquark and quarkonia states in a chiral approach,''
Phys. Rev. D \textbf{75} (2007), 054007
doi:10.1103/PhysRevD.75.054007
[arXiv:hep-ph/0611388 [hep-ph]].
%122 citations counted in INSPIRE as of 19 Sep 2024

%\cite{Albaladejo:2008qa}
\bibitem{Albaladejo:2008qa}
M.~Albaladejo and J.~A.~Oller,
%``Identification of a Scalar Glueball,''
Phys. Rev. Lett. \textbf{101} (2008), 252002
doi:10.1103/PhysRevLett.101.252002
[arXiv:0801.4929 [hep-ph]].
%181 citations counted in INSPIRE as of 19 Sep 2024

%\cite{Gallas:2009qp}
\bibitem{Gallas:2009qp}
S.~Gallas, F.~Giacosa and D.~H.~Rischke,
%``Vacuum phenomenology of the chiral partner of the nucleon in a linear sigma model with vector mesons,''
Phys. Rev. D \textbf{82} (2010), 014004
doi:10.1103/PhysRevD.82.014004
[arXiv:0907.5084 [hep-ph]].
%134 citations counted in INSPIRE as of 19 Sep 2024

%\cite{Zhou:2010ra}
\bibitem{Zhou:2010ra}
Z.~Y.~Zhou and Z.~Xiao,
%``The Origin of light $0^{+}$ scalar resonances,''
Phys. Rev. D \textbf{83} (2011), 014010
doi:10.1103/PhysRevD.83.014010
[arXiv:1007.2072 [hep-ph]].
%29 citations counted in INSPIRE as of 19 Sep 2024

%\cite{Parganlija:2012fy}
\bibitem{Parganlija:2012fy}
D.~Parganlija, P.~Kovacs, G.~Wolf, F.~Giacosa and D.~H.~Rischke,
%``Meson vacuum phenomenology in a three-flavor linear sigma model with (axial-)vector mesons,''
Phys. Rev. D \textbf{87} (2013) no.1, 014011
doi:10.1103/PhysRevD.87.014011
[arXiv:1208.0585 [hep-ph]].
%203 citations counted in INSPIRE as of 19 Sep 2024


% %\cite{Abele:1998qd}
\bibitem{Abele:1998qd}
A.~Abele, S.~Bischoff, P.~Blum, N.~Djaoshvili, D.~Engelhardt, A.~Herbstrith, C.~Holtzhaussen, M.~Tischhauser, J.~Adomeit and B.~Kammle, \textit{et al.}
%``p bar p annihilation at rest into K(L) K+ pi-+,''
Phys. Rev. D \textbf{57} (1998), 3860-3872
doi:10.1103/PhysRevD.57.3860
% %143 citations counted in INSPIRE as of 12 Sep 2025

% %\cite{OBELIX:2002lhi}
\bibitem{OBELIX:2002lhi}
M.~Bargiotti \textit{et al.} [OBELIX],
%``Coupled channel analysis of $\pi^+ \pi^ -\pi^0$, $K^+ K^- \pi^0$ and $K^{\pm} K^0_S \pi^{\mp}$ from $\bar p p$ annihilation at rest in hydrogen targets at three densities,''
Eur. Phys. J. C \textbf{26} (2003), 371-388
doi:10.1140/epjc/s2002-01080-7
% %42 citations counted in INSPIRE as of 12 Sep 2025


%\cite{Lu:2006fr}
\bibitem{Lu:2006fr}
C.~D.~Lu, Y.~M.~Wang and H.~Zou,
%``Twist-3 distribution amplitudes of scalar mesons from QCD sum rules,''
Phys. Rev. D \textbf{75} (2007), 056001
doi:10.1103/PhysRevD.75.056001
[arXiv:hep-ph/0612210 [hep-ph]].
%46 citations counted in INSPIRE as of 19 Sep 2024

%\cite{Chernyak:1983ej}
\bibitem{Chernyak:1983ej}
V.~L.~Chernyak and A.~R.~Zhitnitsky,
%``Asymptotic Behavior of Exclusive Processes in QCD,''
Phys. Rept. \textbf{112} (1984), 173
doi:10.1016/0370-1573(84)90126-1
%1447 citations counted in INSPIRE as of 19 Sep 2024

%\cite{Braun:2003rp}
\bibitem{Braun:2003rp}
V.~M.~Braun, G.~P.~Korchemsky and D.~M\"uller,
%``The Uses of conformal symmetry in QCD,''
Prog. Part. Nucl. Phys. \textbf{51} (2003), 311-398
doi:10.1016/S0146-6410(03)90004-4
[arXiv:hep-ph/0306057 [hep-ph]].
%359 citations counted in INSPIRE as of 19 Sep 2024

%\cite{Gross:1974cs}
\bibitem{Gross:1974cs}
D.~J.~Gross and F.~Wilczek,
%``ASYMPTOTICALLY FREE GAUGE THEORIES. 2.,''
Phys. Rev. D \textbf{9} (1974), 980-993
doi:10.1103/PhysRevD.9.980
%1485 citations counted in INSPIRE as of 19 Sep 2024

%\cite{Shifman:1980dk}
\bibitem{Shifman:1980dk}
M.~A.~Shifman and M.~I.~Vysotsky,
%``FORM-FACTORS OF HEAVY MESONS IN QCD,''
Nucl. Phys. B \textbf{186} (1981), 475-518
doi:10.1016/0550-3213(81)90023-7
%132 citations counted in INSPIRE as of 19 Sep 2024

%\cite{Lan:2021wok}
\bibitem{Lan:2021wok}
J.~Lan \textit{et al.} [BLFQ],
%``Light mesons with one dynamical gluon on the light front,''
Phys. Lett. B \textbf{825} (2022), 136890
doi:10.1016/j.physletb.2022.136890
[arXiv:2106.04954 [hep-ph]].
%49 citations counted in INSPIRE as of 19 Sep 2024

%\cite{Bakker:2013cea}
\bibitem{Bakker:2013cea}
B.~L.~G.~Bakker, A.~Bassetto, S.~J.~Brodsky, W.~Broniowski, S.~Dalley, T.~Frederico, S.~D.~Glazek, J.~R.~Hiller, C.~R.~Ji and V.~Karmanov, \textit{et al.}
%``Light-Front Quantum Chromodynamics: A framework for the analysis of hadron physics,''
Nucl. Phys. B Proc. Suppl. \textbf{251-252} (2014), 165-174
doi:10.1016/j.nuclphysbps.2014.05.004
[arXiv:1309.6333 [hep-ph]].
%67 citations counted in INSPIRE as of 19 Sep 2024

%\cite{Brodsky:1997de}
\bibitem{Brodsky:1997de}
S.~J.~Brodsky, H.~C.~Pauli and S.~S.~Pinsky,
%``Quantum chromodynamics and other field theories on the light cone,''
Phys. Rept. \textbf{301} (1998), 299-486
doi:10.1016/S0370-1573(97)00089-6
[arXiv:hep-ph/9705477 [hep-ph]].
%1576 citations counted in INSPIRE as of 19 Sep 2024

%\cite{deTeramond:2005su}
\bibitem{deTeramond:2005su}
G.~F.~de Teramond and S.~J.~Brodsky,
%``Hadronic spectrum of a holographic dual of QCD,''
Phys. Rev. Lett. \textbf{94} (2005), 201601
doi:10.1103/PhysRevLett.94.201601
[arXiv:hep-th/0501022 [hep-th]].
%462 citations counted in INSPIRE as of 19 Sep 2024

%\cite{Brodsky:2006uqa}
\bibitem{Brodsky:2006uqa}
S.~J.~Brodsky and G.~F.~de Teramond,
%``Hadronic spectra and light-front wavefunctions in holographic QCD,''
Phys. Rev. Lett. \textbf{96} (2006), 201601
doi:10.1103/PhysRevLett.96.201601
[arXiv:hep-ph/0602252 [hep-ph]].
%476 citations counted in INSPIRE as of 19 Sep 2024

%\cite{deTeramond:2008ht}
\bibitem{deTeramond:2008ht}
G.~F.~de Teramond and S.~J.~Brodsky,
%``Light-Front Holography: A First Approximation to QCD,''
Phys. Rev. Lett. \textbf{102} (2009), 081601
doi:10.1103/PhysRevLett.102.081601
[arXiv:0809.4899 [hep-ph]].
%351 citations counted in INSPIRE as of 19 Sep 2024

%\cite{Branz:2010ub}
\bibitem{Branz:2010ub}
T.~Branz, T.~Gutsche, V.~E.~Lyubovitskij, I.~Schmidt and A.~Vega,
%``Light and heavy mesons in a soft-wall holographic approach,''
Phys. Rev. D \textbf{82} (2010), 074022
doi:10.1103/PhysRevD.82.074022
[arXiv:1008.0268 [hep-ph]].
%183 citations counted in INSPIRE as of 19 Sep 2024

%\cite{Brodsky:2014yha}
\bibitem{Brodsky:2014yha}
S.~J.~Brodsky, G.~F.~de Teramond, H.~G.~Dosch and J.~Erlich,
%``Light-Front Holographic QCD and Emerging Confinement,''
Phys. Rept. \textbf{584} (2015), 1-105
doi:10.1016/j.physrep.2015.05.001
[arXiv:1407.8131 [hep-ph]].
%485 citations counted in INSPIRE as of 19 Sep 2024

%\cite{Ahmady:2018muv}
\bibitem{Ahmady:2018muv}
M.~Ahmady, C.~Mondal and R.~Sandapen,
%``Dynamical spin effects in the holographic light-front wavefunctions of light pseudoscalar mesons,''
Phys. Rev. D \textbf{98} (2018) no.3, 034010
doi:10.1103/PhysRevD.98.034010
[arXiv:1805.08911 [hep-ph]].
%36 citations counted in INSPIRE as of 19 Sep 2024

%\cite{deTeramond:2021yyi}
\bibitem{deTeramond:2021yyi}
G.~F.~de Teramond and S.~J.~Brodsky,
%``Longitudinal dynamics and chiral symmetry breaking in holographic light-front QCD,''
Phys. Rev. D \textbf{104} (2021) no.11, 116009
doi:10.1103/PhysRevD.104.116009
[arXiv:2103.10950 [hep-ph]].
%28 citations counted in INSPIRE as of 19 Sep 2024

%\cite{Ahmady:2021lsh}
\bibitem{Ahmady:2021lsh}
M.~Ahmady, H.~Dahiya, S.~Kaur, C.~Mondal, R.~Sandapen and N.~Sharma,
%``Extending light-front holographic QCD using the 't Hooft Equation,''
Phys. Lett. B \textbf{823} (2021), 136754
doi:10.1016/j.physletb.2021.136754
[arXiv:2105.01018 [hep-ph]].
%17 citations counted in INSPIRE as of 19 Sep 2024

%\cite{Ahmady:2021yzh}
\bibitem{Ahmady:2021yzh}
M.~Ahmady, S.~Kaur, S.~L.~MacKay, C.~Mondal and R.~Sandapen,
%``Hadron spectroscopy using the light-front holographic Schr\"odinger equation and the \textquoteright{}t Hooft equation,''
Phys. Rev. D \textbf{104} (2021) no.7, 074013
doi:10.1103/PhysRevD.104.074013
[arXiv:2108.03482 [hep-ph]].
%18 citations counted in INSPIRE as of 19 Sep 2024

%\cite{Li:2015zda}
\bibitem{Li:2015zda}
Y.~Li, P.~Maris, X.~Zhao and J.~P.~Vary,
%``Heavy Quarkonium in a Holographic Basis,''
Phys. Lett. B \textbf{758} (2016), 118-124
doi:10.1016/j.physletb.2016.04.065
[arXiv:1509.07212 [hep-ph]].
%126 citations counted in INSPIRE as of 19 Sep 2024

%\cite{Tang:2018myz}
\bibitem{Tang:2018myz}
S.~Tang, Y.~Li, P.~Maris and J.~P.~Vary,
%``$B_c$ mesons and their properties on the light front,''
Phys. Rev. D \textbf{98} (2018) no.11, 114038
doi:10.1103/PhysRevD.98.114038
[arXiv:1810.05971 [nucl-th]].
%52 citations counted in INSPIRE as of 19 Sep 2024

%\cite{Jia:2018ary}
\bibitem{Jia:2018ary}
S.~Jia and J.~P.~Vary,
%``Basis light front quantization for the charged light mesons with color singlet Nambu\textendash{}Jona-Lasinio interactions,''
Phys. Rev. C \textbf{99} (2019) no.3, 035206
doi:10.1103/PhysRevC.99.035206
[arXiv:1811.08512 [nucl-th]].
%74 citations counted in INSPIRE as of 19 Sep 2024

%\cite{Mondal:2019jdg}
\bibitem{Mondal:2019jdg}
C.~Mondal, S.~Xu, J.~Lan, X.~Zhao, Y.~Li, D.~Chakrabarti and J.~P.~Vary,
%``Proton structure from a light-front Hamiltonian,''
Phys. Rev. D \textbf{102} (2020) no.1, 016008
doi:10.1103/PhysRevD.102.016008
[arXiv:1911.10913 [hep-ph]].
%68 citations counted in INSPIRE as of 19 Sep 2024

%\cite{Lan:2019rba}
\bibitem{Lan:2019rba}
J.~Lan, C.~Mondal, S.~Jia, X.~Zhao and J.~P.~Vary,
%``Pion and kaon parton distribution functions from basis light front quantization and QCD evolution,''
Phys. Rev. D \textbf{101} (2020) no.3, 034024
doi:10.1103/PhysRevD.101.034024
[arXiv:1907.01509 [nucl-th]].
%87 citations counted in INSPIRE as of 19 Sep 2024

%\cite{Lan:2019vui}
\bibitem{Lan:2019vui}
J.~Lan, C.~Mondal, S.~Jia, X.~Zhao and J.~P.~Vary,
%``Parton Distribution Functions from a Light Front Hamiltonian and QCD Evolution for Light Mesons,''
Phys. Rev. Lett. \textbf{122} (2019) no.17, 172001
doi:10.1103/PhysRevLett.122.172001
[arXiv:1901.11430 [nucl-th]].
%97 citations counted in INSPIRE as of 19 Sep 2024

%\cite{Qian:2020utg}
\bibitem{Qian:2020utg}
W.~Qian, S.~Jia, Y.~Li and J.~P.~Vary,
%``Light mesons within the basis light-front quantization framework,''
Phys. Rev. C \textbf{102} (2020) no.5, 055207
doi:10.1103/PhysRevC.102.055207
[arXiv:2005.13806 [nucl-th]].
%32 citations counted in INSPIRE as of 19 Sep 2024

%\cite{Li:2021jqb}
\bibitem{Li:2021jqb}
Y.~Li and J.~P.~Vary,
%``Light-front holography with chiral symmetry breaking,''
Phys. Lett. B \textbf{825} (2022), 136860
doi:10.1016/j.physletb.2021.136860
[arXiv:2103.09993 [hep-ph]].
%31 citations counted in INSPIRE as of 19 Sep 2024

%\cite{Lan:2022blr}
\bibitem{Lan:2022blr}
J.~Lan \textit{et al.} [BLFQ],
%``Light mesons with one dynamical gluon within basis light-front quantization,''
[arXiv:2201.05987 [hep-ph]].
%1 citations counted in INSPIRE as of 19 Sep 2024

%\cite{Han:2013zg}
\bibitem{Han:2013zg}
H.~Y.~Han, X.~G.~Wu, H.~B.~Fu, Q.~L.~Zhang and T.~Zhong,
%``Twist-3 Distribution Amplitudes of Scalar Mesons within the QCD Sum Rules and Its Application to the $B \to S$ Transition Form Factors,''
Eur. Phys. J. A \textbf{49} (2013), 78
doi:10.1140/epja/i2013-13078-7
[arXiv:1301.3978 [hep-ph]].
%30 citations counted in INSPIRE as of 19 Sep 2024

%\cite{Wang:2014vra}
\bibitem{Wang:2014vra}
Z.~G.~Wang,
%``$B-S$ transition form-factors with the light-cone QCD sum rules,''
Eur. Phys. J. C \textbf{75} (2015) no.2, 50
doi:10.1140/epjc/s10052-015-3282-3
[arXiv:1409.6449 [hep-ph]].
%12 citations counted in INSPIRE as of 19 Sep 2024

%\cite{Cheng:2019tgh}
\bibitem{Cheng:2019tgh}
S.~Cheng and J.~M.~Shen,
%``$\bar{B}_s \rightarrow f_0(980)$ form factors and the width effect from light-cone sum rules,''
Eur. Phys. J. C \textbf{80} (2020) no.6, 554
doi:10.1140/epjc/s10052-020-8124-2
[arXiv:1907.08401 [hep-ph]].
%24 citations counted in INSPIRE as of 19 Sep 2024

%\cite{Han:2023pgf}
\bibitem{Han:2023pgf}
X.~Y.~Han, L.~S.~Lu, C.~D.~L\"u, Y.~L.~Shen and B.~X.~Shi,
%``Next-to-leading order QCD corrections to the form factors of $B$ to scalar meson decays,''
JHEP \textbf{11} (2023), 091
doi:10.1007/JHEP11(2023)091
[arXiv:2309.05631 [hep-ph]].
%4 citations counted in INSPIRE as of 19 Sep 2024

%\cite{Shifman:1978by}
\bibitem{Shifman:1978by}
M.~A.~Shifman, A.~I.~Vainshtein and V.~I.~Zakharov,
%``QCD and Resonance Physics: Applications,''
Nucl. Phys. B \textbf{147} (1979), 448-518
doi:10.1016/0550-3213(79)90023-3
%3209 citations counted in INSPIRE as of 19 Sep 2024

%\cite{Shifman:1978bw}
\bibitem{Shifman:1978bw}
M.~A.~Shifman, A.~I.~Vainshtein and V.~I.~Zakharov,
%``QCD and Resonance Physics. The rho-omega Mixing,''
Nucl. Phys. B \textbf{147} (1979), 519-534
doi:10.1016/0550-3213(79)90024-5
%887 citations counted in INSPIRE as of 19 Sep 2024

%\cite{Khodjamirian:1979fa}
\bibitem{Khodjamirian:1979fa}
A.~Y.~Khodjamirian,
%``Dispersion Sum Rules for the Amplitudes of Radiative Transitions in Quarkonium,''
Phys. Lett. B \textbf{90} (1980), 460-464
doi:10.1016/0370-2693(80)90974-0
%42 citations counted in INSPIRE as of 19 Sep 2024

%\cite{Khodjamirian:1983gd}
\bibitem{Khodjamirian:1983gd}
A.~Y.~Khodjamirian,
%``On the Calculation of $J/\psi \to \eta_c \gamma$ Width in {QCD},''
Sov. J. Nucl. Phys. \textbf{39} (1984), 614
EFI-651-41-83-YEREVAN.
%28 citations counted in INSPIRE as of 19 Sep 2024

%\cite{Belyaev:1993wp}
\bibitem{Belyaev:1993wp}
V.~M.~Belyaev, A.~Khodjamirian and R.~Ruckl,
%``QCD calculation of the B ---\ensuremath{>} pi, K form-factors,''
Z. Phys. C \textbf{60} (1993), 349-356
doi:10.1007/BF01474633
[arXiv:hep-ph/9305348 [hep-ph]].
%217 citations counted in INSPIRE as of 19 Sep 2024

%\cite{Khodjamirian:1998vk}
\bibitem{Khodjamirian:1998vk}
A.~Khodjamirian, R.~Ruckl and C.~W.~Winhart,
%``The Scalar B ---\ensuremath{>} pi and D ---\ensuremath{>} pi form-factors in QCD,''
Phys. Rev. D \textbf{58} (1998), 054013
doi:10.1103/PhysRevD.58.054013
[arXiv:hep-ph/9802412 [hep-ph]].
%61 citations counted in INSPIRE as of 19 Sep 2024

%\cite{Colangelo:2000dp}
\bibitem{Colangelo:2000dp}
P.~Colangelo and A.~Khodjamirian,
%``QCD sum rules, a modern perspective,''
doi:10.1142/9789812810458\_0033
[arXiv:hep-ph/0010175 [hep-ph]].
%779 citations counted in INSPIRE as of 19 Sep 2024

%\cite{Huang:1998gp}
\bibitem{Huang:1998gp}
T.~Huang and Z.~H.~Li,
%``B --\ensuremath{>} K* gamma in the light cone QCD sum rule,''
Phys. Rev. D \textbf{57} (1998), 1993-1996
doi:10.1103/PhysRevD.57.1993
%34 citations counted in INSPIRE as of 19 Sep 2024

%\cite{Ball:1998kk}
\bibitem{Ball:1998kk}
P.~Ball and V.~M.~Braun,
%``Exclusive semileptonic and rare B meson decays in QCD,''
Phys. Rev. D \textbf{58} (1998), 094016
doi:10.1103/PhysRevD.58.094016
[arXiv:hep-ph/9805422 [hep-ph]].
%515 citations counted in INSPIRE as of 19 Sep 2024

%\cite{Yang:2005bv}
\bibitem{Yang:2005bv}
M.~Z.~Yang,
%``Semileptonic decay of B and D ---\ensuremath{>} K*0 (1430) anti-l nu from QCD sum rule,''
Phys. Rev. D \textbf{73} (2006), 034027
[erratum: Phys. Rev. D \textbf{73} (2006), 079901]
doi:10.1103/PhysRevD.73.079901
[arXiv:hep-ph/0509103 [hep-ph]].
%42 citations counted in INSPIRE as of 19 Sep 2024

%\cite{Wang:2008da}
\bibitem{Wang:2008da}
Y.~M.~Wang, M.~J.~Aslam and C.~D.~Lu,
%``Scalar mesons in weak semileptonic decays of B(s),''
Phys. Rev. D \textbf{78} (2008), 014006
doi:10.1103/PhysRevD.78.014006
[arXiv:0804.2204 [hep-ph]].
%40 citations counted in INSPIRE as of 19 Sep 2024

%\cite{Li:2008tk}
\bibitem{Li:2008tk}
R.~H.~Li, C.~D.~Lu, W.~Wang and X.~X.~Wang,
%``B ---\ensuremath{>} S Transition Form Factors in the PQCD approach,''
Phys. Rev. D \textbf{79} (2009), 014013
doi:10.1103/PhysRevD.79.014013
[arXiv:0811.2648 [hep-ph]].
%89 citations counted in INSPIRE as of 19 Sep 2024

%\cite{Ghahramany:2009zz}
\bibitem{Ghahramany:2009zz}
N.~Ghahramany and R.~Khosravi,
%``Analysis of the rare semileptonic decays of Bs to f0 (980) and K-0* (1430) scalar mesons in QCD sum rules,''
Phys. Rev. D \textbf{80} (2009), 016009
doi:10.1103/PhysRevD.80.016009
%21 citations counted in INSPIRE as of 19 Sep 2024

%\cite{Zeng:2013nfa}
\bibitem{Zeng:2013nfa}
D.~M.~Zeng and Z.~Y.~Fang,
%``$B_{s} \to f_{0}(980)$ transition form factors within the $k_{T}$ factorization approach,''
Commun. Theor. Phys. \textbf{59} (2013), 457-461
doi:10.1088/0253-6102/59/4/12
%0 citations counted in INSPIRE as of 19 Sep 2024

%\cite{Cheng:2013fba}
\bibitem{Cheng:2013fba}
H.~Y.~Cheng, C.~K.~Chua, K.~C.~Yang and Z.~Q.~Zhang,
%``Revisiting charmless hadronic B decays to scalar mesons,''
Phys. Rev. D \textbf{87} (2013) no.11, 114001
doi:10.1103/PhysRevD.87.114001
[arXiv:1303.4403 [hep-ph]].
%65 citations counted in INSPIRE as of 19 Sep 2024

%\cite{Meissner:2013hya}
\bibitem{Meissner:2013hya}
U.~G.~Mei\ss{}ner and W.~Wang,
%``Generalized Heavy-to-Light Form Factors in Light-Cone Sum Rules,''
Phys. Lett. B \textbf{730} (2014), 336-341
doi:10.1016/j.physletb.2014.02.009
[arXiv:1312.3087 [hep-ph]].
%76 citations counted in INSPIRE as of 19 Sep 2024

%\cite{Issadykov:2015iba}
\bibitem{Issadykov:2015iba}
A.~Issadykov, M.~A.~Ivanov and S.~K.~Sakhiyev,
%``Form factors of the B-S-transitions in the covariant quark model,''
Phys. Rev. D \textbf{91} (2015) no.7, 074007
doi:10.1103/PhysRevD.91.074007
[arXiv:1502.05280 [hep-ph]].
%26 citations counted in INSPIRE as of 19 Sep 2024

%\cite{Zou:2016yhb}
\bibitem{Zou:2016yhb}
Z.~T.~Zou, Y.~Li and X.~liu,
%``Two-body charmed B(s) decays involving a light scalar meson,''
Phys. Rev. D \textbf{95} (2017) no.1, 016011
doi:10.1103/PhysRevD.95.016011
[arXiv:1609.06444 [hep-ph]].
%15 citations counted in INSPIRE as of 19 Sep 2024

%\cite{Huang:2022xny}
\bibitem{Huang:2022xny}
D.~Huang, T.~Zhong, H.~B.~Fu, Z.~H.~Wu, X.~G.~Wu and H.~Tong,
%``$K_0^*(1430)$ twist-2 distribution amplitude and $B_s,D_s \rightarrow K_0^*(1430)$ transition form factors,''
Eur. Phys. J. C \textbf{83} (2023) no.7, 680
doi:10.1140/epjc/s10052-023-11851-x
[arXiv:2211.06211 [hep-ph]].
%4 citations counted in INSPIRE as of 19 Sep 2024

%\cite{Wang:2022yyn}
\bibitem{Wang:2022yyn}
R.~M.~Wang, Y.~X.~Liu, M.~Y.~Wan, C.~Hua, J.~H.~Sheng and Y.~G.~Xu,
%``Semileptonic decays D \textrightarrow{} P/V/S\ensuremath{\ell}+\ensuremath{\nu}\ensuremath{\ell} with the SU(3) flavor symmetry/breaking,''
Nucl. Phys. B \textbf{995} (2023), 116349
doi:10.1016/j.nuclphysb.2023.116349
[arXiv:2301.00079 [hep-ph]].
%6 citations counted in INSPIRE as of 19 Sep 2024

%\cite{ParticleDataGroup:2024cfk}
\bibitem{ParticleDataGroup:2024cfk}
S.~Navas \textit{et al.} [Particle Data Group],
%``Review of particle physics,''
Phys. Rev. D \textbf{110} (2024) no.3, 030001
doi:10.1103/PhysRevD.110.030001
%133 citations counted in INSPIRE as of 19 Sep 2024

%\cite{HFLAV:2019otj}
\bibitem{HFLAV:2019otj}
Y.~S.~Amhis \textit{et al.} [HFLAV],
%``Averages of b-hadron, c-hadron, and $\tau $-lepton properties as of 2018,''
Eur. Phys. J. C \textbf{81} (2021) no.3, 226
doi:10.1140/epjc/s10052-020-8156-7
[arXiv:1909.12524 [hep-ex]].
%899 citations counted in INSPIRE as of 19 Sep 2024

%\cite{ParticleDataGroup:2020ssz}
\bibitem{ParticleDataGroup:2020ssz}
P.~A.~Zyla \textit{et al.} [Particle Data Group],
%``Review of Particle Physics,''
PTEP \textbf{2020} (2020) no.8, 083C01
doi:10.1093/ptep/ptaa104
%5999 citations counted in INSPIRE as of 19 Sep 2024

%\cite{Huang:1998sa}
\bibitem{Huang:1998sa}
T.~Huang and Z.~H.~Li,
%``The binding energy of the excited heavy light mesons in HQET,''
Phys. Lett. B \textbf{438} (1998), 159-164
doi:10.1016/S0370-2693(98)00952-6
%7 citations counted in INSPIRE as of 19 Sep 2024

%\cite{Sun:2010nv}
\bibitem{Sun:2010nv}
Y.~J.~Sun, Z.~H.~Li and T.~Huang,
%``$B_{(s)}\to S$ transitions in the light cone sum rules with the chiral current,''
Phys. Rev. D \textbf{83} (2011), 025024
doi:10.1103/PhysRevD.83.025024
[arXiv:1011.3901 [hep-ph]].
%36 citations counted in INSPIRE as of 19 Sep 2024






% %\cite{Lipkin:1968woy}
% \bibitem{Lipkin:1968woy}
% H.~J.~Lipkin,
% %``Interference, mixing, and angular correlations in decays of boson resonances,''
% Phys. Rev. \textbf{176} (1968), 1709-1714
% doi:10.1103/PhysRev.176.1709
% %43 citations counted in INSPIRE as of 12 Sep 2025

% %\cite{Tornqvist:1979hx}
% \bibitem{Tornqvist:1979hx}
% N.~A.~Tornqvist,
% %``The Meson Mass Spectrum and Unitarity,''
% Annals Phys. \textbf{123} (1979), 1
% doi:10.1016/0003-4916(79)90262-8
% %92 citations counted in INSPIRE as of 12 Sep 2025

% %\cite{Kalashnikova:2005ui}
% \bibitem{Kalashnikova:2005ui}
% Y.~S.~Kalashnikova,
% %``Coupled-channel model for charmonium levels and an option for X(3872),''
% Phys. Rev. D \textbf{72} (2005), 034010
% doi:10.1103/PhysRevD.72.034010
% [arXiv:hep-ph/0506270 [hep-ph]].
% %220 citations counted in INSPIRE as of 12 Sep 2025

% %\cite{Barnes:2007xu}
% \bibitem{Barnes:2007xu}
% T.~Barnes and E.~S.~Swanson,
% %``Hadron loops: General theorems and application to charmonium,''
% Phys. Rev. C \textbf{77} (2008), 055206
% doi:10.1103/PhysRevC.77.055206
% [arXiv:0711.2080 [hep-ph]].
% %156 citations counted in INSPIRE as of 12 Sep 2025

% %\cite{Pennington:2007xr}
% \bibitem{Pennington:2007xr}
% M.~R.~Pennington and D.~J.~Wilson,
% %``Decay channels and charmonium mass-shifts,''
% Phys. Rev. D \textbf{76} (2007), 077502
% doi:10.1103/PhysRevD.76.077502
% [arXiv:0704.3384 [hep-ph]].
% %112 citations counted in INSPIRE as of 12 Sep 2025






\end{thebibliography}

\end{document}